\newcommand{\kms}{\ensuremath{\mathrm{km\ s^{-1}}}}
\newcommand{\pPXF}{\textsc pPXF}
\newcommand{\pfrac}{{\tt yfrac}}
\newcommand{\cf}{{\it cf.}}

\documentclass[twocolumn,revtex4]{emulateapj}

\usepackage{subfigure}
\usepackage{upgreek}
\usepackage{amsmath}
\usepackage{xspace}
\usepackage{appendix}
\usepackage{pdflscape}
\usepackage{multirow}
\usepackage{hyperref}
\usepackage{natbib}
\usepackage{xcolor}
\usepackage{soul}

\begin{document}

\title{Stellar Population Synthesis with Distinct Kinematics:
  Multi-Age Asymmetric Drift in SDSS-IV MaNGA Galaxies}

\author{Shravan Shetty\altaffilmark{1}, 
Matthew A. Bershady\altaffilmark{2,3,4},
Kyle B. Westfall\altaffilmark{5},
Michele Cappellari\altaffilmark{6},
Niv Drory\altaffilmark{7},
David R. Law\altaffilmark{8},
Renbin Yan\altaffilmark{9}, 
Kevin Bundy\altaffilmark{5}
}

\altaffiltext{1}{Kavli Institute for Astronomy and Astrophysics, Peking University, Beijing 100871, China; shravan.shetty@pku.edu.cn}

\altaffiltext{2}{Department of Astronomy, University of Wisconsin-Madison,  475
N. Charter St., Madison, WI 53706, USA}

\altaffiltext{3}{South African Astronomical Observatory, PO Box 9, Observatory 7935, Cape Town, South Africa}

\altaffiltext{4}{Department of Astronomy, University of Cape Town, Private Bag X3, Rondebosch 7701, South Africa}
 
\altaffiltext{5}{University of California Observatories, University of
  California, Santa Cruz, 1156 High St., Santa Cruz, CA 95064, USA}

\altaffiltext{6}{Sub-Department of Astrophysics, Department of
  Physics, University of Oxford, Denys Wilkinson Building, Keble Road,
  Oxford OX1 3RH, UK}

\altaffiltext{7}{McDonald Observatory, The University of Texas at
  Austin, 1 University Station, Austin, TX 78712, USA}

\altaffiltext{8}{Space Telescope Science Institute, 3700 San Martin
  Drive, Baltimore, MD 21218, USA}

\altaffiltext{9}{Department of Physics and Astronomy, University of
  Kentucky, 505 Rose St., Lexington, KY 40506-0057, USA}


\begin{abstract}

We present the first asymmetric drift (AD) measurements for unresolved
stellar populations of different characteristic ages above and below
1.5 Gyr. These measurements sample the age-velocity relation (AVR) in
galaxy disks. In this first paper we develop two efficient algorithms
to extract AD on a spaxel-by-spaxel basis from optical integral-field
spectroscopic (IFS) data-cubes. The algorithms apply different
spectral templates, one using simple stellar populations and the other
a stellar library; their comparison allows us to assess systematic
errors in derived multi-component velocities, such as
template-mismatch. We test algorithm reliability using mock spectra
and Monte Carlo Markov Chains on real data from the MaNGA survey in
SDSS-IV. We quantify random and systematic errors in AD as a function
of signal-to-noise and stellar population properties with the aim of
applying this technique to large subsets of the MaNGA galaxy
sample. As a demonstration of our methods, we apply them to an initial
sample of seven galaxies with comparable stellar mass and color to the
Milky Way. We find a wide range of distinct AD radial profiles for
young and old stellar populations.

\end{abstract}


\section{Introduction}
\label{sec:intro}

It has long been known that there exists a correlation between stellar
population age and their vertical scale height in the Milky Way (MW)
solar neighborhood \citep{Stromberg1925, Wielen77}. The dynamical
linkage between scale-height to velocity dispersion and tangential
speed has led to this correlation being referred to as the
age--velocity or age--velocity-dispersion relation (AVR). This
relationship has been studied exhaustively in the Milky Way's Solar
neighborhood, but it still remains unclear whether disk populations
form a discrete or continuous dynamical and chemical distribution
\citep{Nordstrom04, Holmbergetal2007, SeabrokeGilmore2007,
  Aumer09}. AVR measurements are critical for understanding disk
evolution because AVR modulates in response to chemo-dynamical
processes. AVR measurements are essential for accurate dynamical
estimates of stellar disk mass because AVR encapsulates vertical
population gradients.

AVR's have been detected in M31
\citep{Collins2011,Dorman15,Quirketal2019}, M33 \citep{Beasley15},
several other Local Group (LG) dwarf galaxies \citep[compiled
  by][]{Leaman17}, and inferred indirectly via photometric means
in a handful of nearby, low-mass edge-on galaxies
\citep{Seth05a}. Despite a consistent general trend -- older disk
stellar populations are dynamically hotter than their younger cohort
-- the slope and normalization of AVR in other galaxies is found to be
significantly different from that seen in the Milky Way. No consensus
has been reached regarding the shape (or amplitude) of the
relationship and its dependence on other galaxies properties.

The origin of the AVR remains uncertain. Three processes have been
suggested. One invokes scattering with giant molecular clouds, spiral
arms or bars within galaxies \citep{Spitzer51, Spitzer53,
  KokuboIda1992, Carlberg85}, or by minor mergers
\citep{TothOstriker1992,Walkeretal1996,
  HuangCarlberg1997,Benson2004,House2011,Helmi2012,Few2012,RuizLara2016}. A
second suggests accretion of dynamically hot debris from mergers of
old and metal poor satellites \citep{Abadietal2003b,Pinna2019a}, raising the
possibility for discrete chemical enrichment patterns between thin and
thick disks. A third suggests older stellar populations formed in a
thicker, more turbulent star-forming gas layer, i.e., older
populations were dynamically hotter ab initio \citep{Brook04,
  Bournaud09, Forbesetal2012}. The observed velocity dispersions of
ionized gas in star-forming galaxy disks at higher redshift
\citep{Weiner06, Law07, Forster-Schreiber09, Wisnioski15} make this a
compelling if perhaps incomplete scenario. Simulations suggest several
mechanisms can be at play even in individual galaxies \citep[e.g.,][]
{Bird13,Martig14b}, while \cite{Leaman17} suggest the interplay
between heating mechanisms correlates with galaxy mass, based on eight
LG galaxies, only two of which have stellar masses above
10$^{10}M_\odot$.

The aim of this paper is to develop reliable methods toward measuring
AVR in the galaxy population at large, as observed by the MaNGA survey
\citep{Bundy15}. This work can potentially be extended to other
surveys such as SAMI \citep{Croom12} or CALIFA
\citep{CALIFA}. However, these surveys resolve physical scales of
$\sim$1~kpc and 70~\kms (stellar $\sigma$), and do not typically
  achieve signal-to-noise ratios (S/N) ample to recover scales below
  the instrumental resolution \citep[e.g.,][]{Toloba2011,Rys2013}.
For reference, MW vertical velocity dispersions for the older disk at
the solar radius are 30 to 50~\kms\ while the younger counterpart has
vertical velocity dispersions of 10 to 20~\kms. Hence, the boundary
conditions for such a method are that it cannot depend on resolved
stellar populations (available for only LG and very nearby galaxies),
nor can it depend on stellar velocity dispersions.

A compelling solution is found in the asymmetric drift (AD) that
relates in-plane stellar velocity dispersion components to the lag of
the stellar tangential speed from the circular speed of the
potential. The latter is well-estimated from ionized or neutral gas
velocities when these gas-phase components are observed to have very
small velocity dispersions. In early-type disk systems where sometimes
small corrections for pressure support are needed to bring ionized-gas
tangential speeds in line with circular speeds
\citep[e.g.,][]{Davis2013}, such corrections can be estimated
from gas density gradients without recourse to velocity dispersion
information \citep{Dalcanton2010}. Stellar and gas velocity centroids
are robust kinematic measures far below the instrumental
resolution, and the former are relatively immune (compared to
higher-order moments) to systematics from template mismatch. Hence
this paper develops methods to measure multi-component asymmetric
drift as a function of population age to determine AVR.

Using asymmetric drift is not without its challenges in disentangling
multiple kinematic components from spectra of integrated star-light.
Previous work attempting to disentangle kinematic components have
focused mainly on counter-rotating disks
\citep[e.g.,][]{Coccatoetal2011,Johnstonetal2013} or early-type
galaxies \citep[e.g.,][]{DeBruyne04, Tabor17,Poci2019}, although these
studies have availed themselves of higher-order moments. The
  work of \citet{Poci2019} combines detailed Schwarzchild dynamical
  modeling and stellar population synthesis to correlate the relative
  weights of stellar orbits and single stellar populations, thus
  generating a decomposed model for the galaxy structure. This
  technique provides a powerful approach to investigate the underlying
  structures in galaxies, but is sensitive to the reliable
  identification and modeling of all structures in a galaxy and hence
  may not be ideal for scaling to large galaxy samples. The work of
\cite{Tabor17} is of particular relevance because of its use of
disk-bulge decompositions as further constraints and its application
intended for MaNGA. In contrast to this work we use an age constraint
instead, and limit ourselves to a metric which depends on velocity
only.

This paper uses optical IFS of nearby galaxies from the MaNGA survey
\citep{Bundy15,Yan16b} in SDSS-IV \citep{Blanton17}. MaNGA uses fiber
integral-field units \citep{Drory15} with the BOSS spectrographs
\citep{Smee13} on the Sloan 2.5m telescope \citep{Gunn06}. The MaNGA
observing strategy, target selection, spectrophotometric calibration
and data reduction pipeline are described in \citet{Law15},
\citet{Wake17}, \citet{Yan16a}, and \citet{Lawetal2016_MaNGA_DRP},
respectively. Salient features of the spectroscopic data are given in
relevant sections below. We take advantage of existing gas and stellar
kinematic measurements from the Data Analysis Pipeline (Westfall et
al. submitted) to select targets and define geometries.

After defining our data sets and basic fitting methods in
Section~\ref{sec:methods}, we demonstrate that a differential
asymmetric-drift signal between young and old stellar populations can
be measured in MaNGA spectra (Section~\ref{sec:hypothesis}).  This
initial demonstration uses simple stellar populations (SSPs) from
stellar population synthesis models in a Markov-Chain Monte Carlo
(MCMC) implementation of the penalized pixel fitting code (\pPXF) code
of \cite{Cappellari2017}. This implementation is computationally
inefficient and, at low signal-to-noise, sensitive to degeneracies in
stellar population fitting. For these reasons, in the next three
sections we motivate and develop two efficient and robust
algorithms, also based on \pPXF, that do not depend on MCMC. These
algorithms avoid stellar population fitting degeneracies via a
combination of a local minimizer and priors. One of these
  algorithms employs SSPs, while the other employs an empirical
  stellar library; they have comparable performance, and together they
  provide a means to estimate systematics due to template mismatch.

Readers interested only in the performance and results of these 
two final algorithms may skip to Section~\ref{sec:results}. In this
Section we derive estimates of random and systematic errors for our
algorithms, and present our asymmetric drift measurements for seven
MaNGA galaxies with near-MW mass in this
context. Section~\ref{sec:conclude} summarizes our findings and
conclusions.

Reader interested in the algorithm development will find the
  presentation in the intervening sections: In
Section~\ref{sec:ssp_algo} we define the metrics used to
  determine algorithm performance.  We then construct an efficient
SSP-based algorithm using \pPXF\ that robustly measures the
differential asymmetric-drift signal between young and old stellar
populations. This algorithm is successful in avoiding local minima in
the likelihood space, while also avoiding global degeneracies in
population-synthesis fitting noted above. We identify the potential
for non-negligible systematics from template mismatch in {\it
  velocities} in Section~\ref{sec:mismatch}. Consequently we construct
a second algorithm to measure the differential asymmetric-drift signal
between young and old stellar populations using empirical stellar
libraries rather than SSPs (Section~\ref{sec:stellar_algo}).
Appendices provide details of our stellar library selection
(\ref{app:stellar_lib}), mock spectra (\ref{app:mocks}), MCMC analysis
(\ref{app:mcmc_ppd}), and kinematic maps for the studied galaxy sample
(\ref{sec:maps}).


\section{Sample, Data and Fitting Methods}
\label{sec:methods}

\subsection{Galaxy Sample}
\label{sec:sample}

\begin{deluxetable*}{llccccccccc}
\tablewidth{0pt}
\tabletypesize{\tiny}
\tablecaption{Galaxy Sample}
\tablehead{
  \colhead{MID} &
  \colhead{plate-IFU} &
  \colhead{redshift} &
  \colhead{b/a} &
  \colhead{$R_e$} &
  \colhead{$(g-r)_0$} &
  \colhead{${\rm M}_r$ } &
  \colhead{$({\rm NUV}-i)_0$} &
  \colhead{${\rm M}_i$ } &
  \colhead{M*} & 
  \colhead{n$_S$} \\
  \colhead{} &
  \colhead{} &
  \colhead{} &
  \colhead{} &
  \colhead{(kpc)} &
  \colhead{(mag)} &
  \colhead{(mag)} &
  \colhead{(mag)} &
  \colhead{(mag)} &
  \colhead{($10^{10} {\rm M}_\odot$)} &
  \colhead{}
}
\startdata
1-339041 & 8138-12704 & 0.031 & 0.69 & 6.7 & 0.67 & -21.73 & 4.06 & -22.09 & 6.62 & 4.14 \\
1-209537 & 8486-12701 & 0.038 & 0.63 & 8.6 & 0.74 & -21.44 & 4.52 & -21.79 & 5.27 & 6.00 \\
1-532459 & 8320-9102  & 0.052 & 0.67 & 7.6 & 0.60 & -21.63 & 3.07 & -21.90 & 4.77 & 2.72 \\
1-251279 & 8332-12705 & 0.033 & 0.81 & 3.6 & 0.69 & -21.16 & 4.10 & -21.52 & 3.88 & 3.33 \\
1-542358 & 8482-3702  & 0.040 & 0.93 & 3.9 & 0.63 & -21.25 & 3.62 & -21.58 & 3.74 & 3.19 \\
1-265988 & 8329-6103  & 0.031 & 0.64 & 4.0 & 0.62 & -20.30 & 3.68 & -20.63 & 1.65 & 1.47 \\
1-209199 & 8485-9102  & 0.026 & 0.72 & 4.1 & 0.56 & -20.33 & 3.30 & -20.68 & 1.61 & 2.26 
\enddata
\label{tab:sample}
\end{deluxetable*}

We selected seven galaxies from the MaNGA survey to develop and test
the algorithms in this study. The galaxies were chosen to have
exceptionally regular gas and stellar kinematics (in terms of
azimuthal symmetry, with little evidence for bar distortions),
moderate inclinations ($20^\circ<i<50^\circ$), and clear evidence for
asymmetric drift between their (single component) stellar velocities
and ionized gas (as reckoned by the MaNGA data analysis pipeline data
products; hereafter DAP). This list is by no means exhaustive of good
candidates.

These galaxies were also chosen to have stellar masses and colors
similar to the Milky Way (MW).\footnote{MW stellar mass is estimated
  to be $5.7\pm1.3\times10^{10} {\rm M_\odot}$ \citep{Licquia16}, with
  an $i$-band absolute magnitude of ${\rm M}_i = -21.27\pm0.37$ mag
  and $g-r$ color of $0.68\pm0.06$ mag \citep{Licquia15b}.}
Photometric properties of these galaxies, sorted by stellar mass, are
summarized in Table~\ref{tab:sample}, where redshift, b/a (minor to
major axis ratio), half-light radius, rest-frame $g-r$ and NUV-$i$
colors, SDSS $r$- and $i$-band absolute magnitude, stellar mass and
Sersic index (n$_S$) are from the NASA-Sloan Atlas
\citep[NSA,][]{Blanton11}.\footnote{http://www.nsatlas.org} Except for
n$_S$, all photometric quantities report their elliptical Petrosian
aperture measurements.  Magnitudes and colors have AB zero-points;
absolute magnitudes and masses are scaled to H$_0 =
100$~\kms\ Mpc$^{-1}$ throughout. A value of n$_S = 6$ is a hard upper
limit in the \cite{Blanton11} analysis. The NUV-$i$ color range is
blueward of the red sequence \citep[e.g., Figure 2 of][]{Yan16b} but
in the redder half of the blue cloud. As we will see, the sample has
peak rotation speeds between 200 to 300~\kms, with both slow- and
fast-rising rotation curves.

We refer to these galaxies by their plate-IFU (integral field unit)
designation throughout the paper. Table~\ref{tab:sample} provides
their corresponding, unique MaNGA ID (MID).

\subsection{MaNGA Data Cubes}
\label{sec:cubes}

For the spectral analysis throughout this paper we use the ``LOGCUBE''
datacube from the Data Reduction Pipeline (DRP) of MaNGA
\citep{Lawetal2016_MaNGA_DRP} from an internal release\footnote{MaNGA
  Product Launch (MPL) 5, v2\_0\_1.}  closely related to the versions
in Data Release (DR)14 \citep{SDSS_DR14}.  We spot-checked our
analysis to verify that subsequent subtle changes in the
data-processing through DR15 \citep{SDSS_DR15} and the next internal
data release\footnote{MPL 8, v2\_5\_3} do not alter the results in any
significant way.  Given the significant investment in MCMC analysis,
we have retained the results from the older release.

MaNGA datacubes contain the reduced and combined spectra from the
dithered IFU observations of a single galaxy. Spaxels have been
resampled to a logarithmic bin of 70~\kms\ and to a common wavelength
grid.\footnote{For reference, the median instrumental resolution (FWHM)
  varies between 250~\kms\ at the blue limit of 362~nm to 125~\kms\ at
  950~nm, with a median FWHM value close to 150~\kms\ characteristic of
  the performance between 500 and 750~nm
  \citet{Lawetal2016_MaNGA_DRP,Yan16b}.} Data cubes also contain
information on the mean and variance of the spectral line spread
function (LSF). We use this LSF to match the resolution of our
spectral templates to the data, spaxel-by-spaxel, before fitting. We
measure the S/N of individual spaxels in MaNGA
datacubes using the median of the spectra and inverse variance ({\tt
  IVAR} in the datacube) within the wavelength range of 360-940 nm. In
our analysis we ignore spaxels with S/N per \AA\ less than 1.

\subsection{General Considerations for \pPXF}
\label{sec:general}

\begin{figure}
  \centering
  \includegraphics[width=0.95\linewidth]{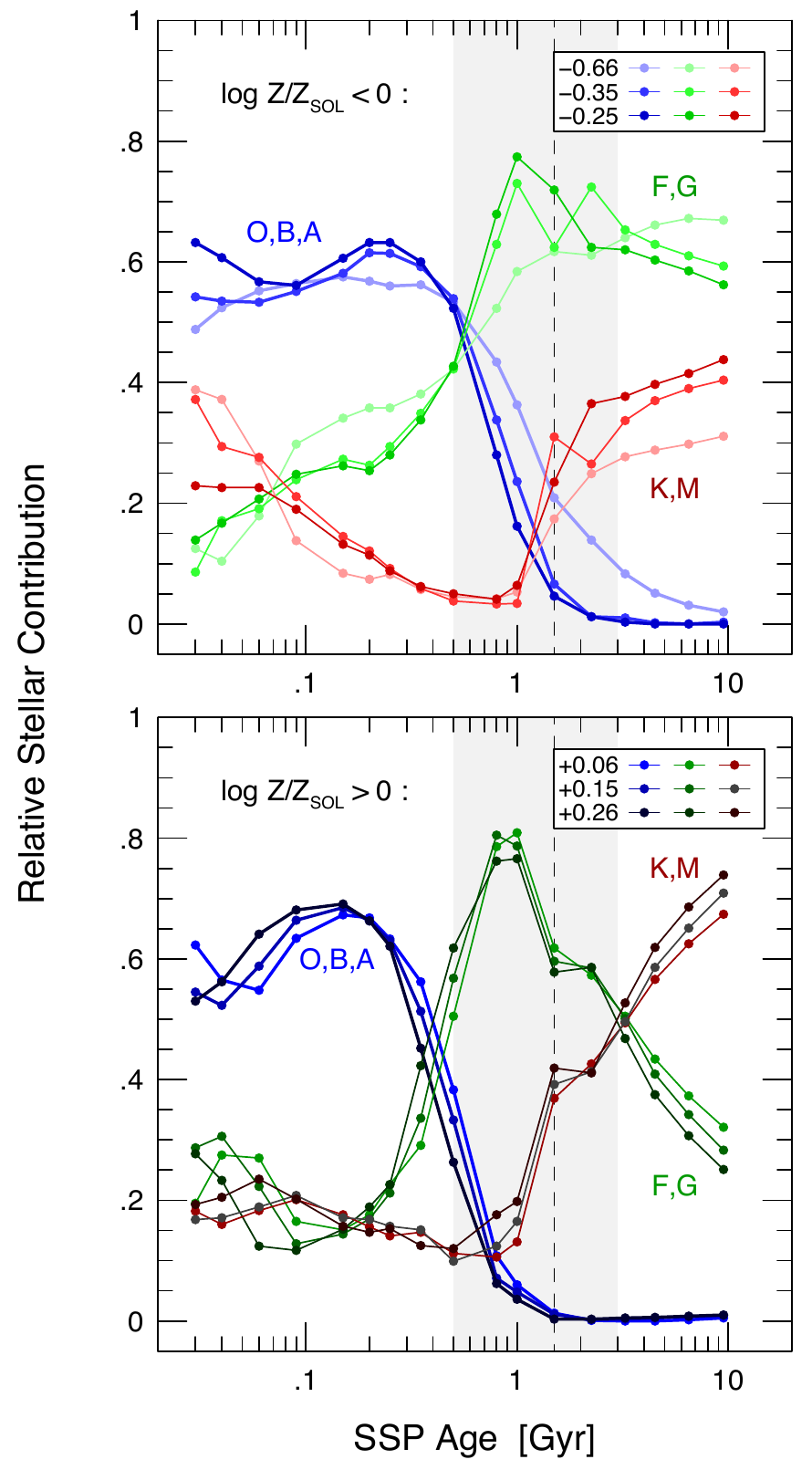}
  \caption{Relative weight of hot (O,B, and A type), intermediate (F
    and G type) and cold (K and M type) stars, regardless of
    surface-gravity, in MIUSCAT SSPs as a function of SSP age for
    different metallicity. The weight distribution (fractional
    luminosity) was derived using \pPXF\ to fit MIUSCAT SSPs with a
    subset of the Indo-US stellar library
    (Appendix~\ref{app:stellar_lib}), after the library spectra were
    flux-normalized in the mean over the fitting range (355 to 940
    nm). The hot stellar contribution begins to fall rapidly above
    0.5~Gyr, and for all but the most metal poor stellar populations
    the hot stellar contribution becomes negligible at ages above
    1.5~Gyrs. The intermediate-temperature stellar contribution peaks
    around 1~Gyr, while the cool stellar contribution becomes
    significant or dominates beyond 3~Gyr. The grey shaded region
    marks the transition between hot- and cool-stellar dominated ages,
    while the vertical dashed line represents our demarcation of
    'young' and 'old' stellar ages for the purpose of the analysis in
    this paper, as further motivated in the text.}
	\label{fig:hot_star_relative_wts}
\end{figure}

Throughout this paper we conduct full-spectrum fitting of the observed
spectra between 350 and 940 nm. The extensive wavelength coverage
yields a wide range of stellar features including the full Balmer
series dominated by hot young stars as well as strong stellar features
from Ca H\&K to the NIR triplet which are dominated by cool stars;
together these features should contain the significant information on
the the kinematics of the young and old populations.

Full-spectrum fitting uses \pPXF. This code fits an observed spectrum
with a set of spectral templates, convolved with a line-of-sight
velocity distribution (LOSVD) in pixel-space. Throughout this study,
we fit for the velocity and velocity dispersion moments of the LOSVD
of all kinematic components. The code allows multiple kinematic
components to be fit simultaneously; in our analysis we have leveraged
this to fit the observed galaxy spectrum with a single component for
the observed gas emission features and a single or multiple stellar
kinematic components.

During our full-spectrum fitting of observed galaxy spectra, we fit
for the following gas emission features; H$\alpha$, H$\beta$,
H$\gamma$, H$\delta$, [OII]$\lambda3727$, [OII]$\lambda3729$,
[OIII]$\lambda4959$, [OIII]$\lambda5007$, [OI]$\lambda6300$,
[OI]$\lambda6364$, [NII]$\lambda6548$, [NII]$\lambda6584$,
[SII]$\lambda61716$ and [SII]$\lambda6731$. The gas emission lines are
modeled as Gaussian features and are treated as a single kinematic
component independent of that measured for the stellar component(s).
While we fix the relative flux ratio between the [OIII], [OI] and
[NII] doublets, based on atomic physics, the fluxes of the gas
emission lines are free parameters during the fitting process.

When fitting two stellar components we take advantage of the
\pPXF\ feature to constrain the relative weight of two kinematic
components, referred to in the code as {\tt fraction} but shortened to
\pfrac\ in this study. This parameter is a {\it light-weighted}
quantity defined as the relative weight in the first kinematic
component to that in all components:

\begin{equation}
\label{frac_def}
        {\tt yfrac} = \frac{\sum w_i}{\sum w_i + \sum w_j}
\end{equation}
where $w_i$ are the weights of the first component while $w_j$ are the
weights of second component. In our analysis the first component will
always correspond to the young stellar population and hence the use of
the subscript 'y.' We flux normalize the template spectra for our
stellar components and hence, if the first (second) component
represents the young (old) stellar population, \pfrac\ is a {\it
  relative} indicator of star formation history. Though by default
\pPXF\ limits the user to constrain the relative weight of two
kinematic components, in the course of our analysis we removed this
limit to allow the development the algorithm in
Section~\ref{sec:stellar_algo}. This allowed us to constrain the
relative weight of more than two kinematic components simultaneously
during our full-spectrum fitting by providing a vector of relative
fractional values ({\tt frac}$_n$ for the $n^{th}$ element), i.e.:
\begin{equation}
	{\tt frac}_n = \frac{\sum w_n}{\sum_{k=1}^{c} \sum w_k}, n=[1, 2, ..., c]
\end{equation}
where $c$ is the number of kinematic components being fit, $\sum w_n$
is the sum of the weights of the $n^{th}$ component and
$\sum_{k=1}^{c} \sum w_k$ is the sum of the weights of all
components. Throughout our study, we allow the relative weight of the
gas component to be free and unconstrained by \pfrac.

The \pPXF\ code can also minimize errors caused by imperfect spectral
calibration, template mismatch, or scattered light by fitting
multiplicative and/or additive Legendre polynomials or trigonometric
series. In our analysis we simultaneously use additive and
multiplicative polynomials of order 8 when fitting MaNGA galaxy
spectra for their stellar kinematics with independent stellar
population templates.\footnote{This order is based on experience
  gained from analysis of MaNGA data (Westfall et al. submitted) as
  well as analysis of SAURON data of galaxies in the Coma cluster
  (Shetty et al. submitted).} When fitting mock spectra using the same
stellar population templates we do not use any polynomials.

\subsection{Choice of SSPs}
\label{sec:ssp_choice}

For our analysis, we adopt the SSPs of MIUSCAT
\citep{Vazdekisetal2012} as our template spectra for the stellar
components. These models were developed using empirical stellar
spectral libraries, particularly the MILES \citep{Falcon-Barroso11},
Indo-US Library of Coud\'{e} Feed Stellar spectra
\citep{Valdesetal2004}, and the CaT Stellar Library
\citep{Cenarroetal2001}. These models were selected primarily due to
their significant overlap in rest-frame wavelength coverage with
MaNGA. The 346.5-946.9 nm span for MIUSCAT almost exactly matches the
360-1030 nm span for MaNGA at the typical redshift of the MaNGA
sample. The spectral resolution (0.251~nm FWHM) is slightly higher at
all wavelengths than the MaNGA data \citep[\cf~][respectively their
  Figures 18 and 20]{Lawetal2016_MaNGA_DRP,Yan16b}, which is important
for properly convolving the data to the MaNGA instrumental resolution
for kinematic analysis, as described in Westfall et al. (submitted).
MIUSCAT SSPs used in our analysis cover an age range 0.03-14~Gyrs in
53 steps, metallicities ([M/H]) -0.66, -0.35, -0.25, 0.06, 0.15 and
0.26 (without $\alpha$-enhancement), and use the BaSTI isochrones
\citep{BaSTI_1} and a Kroupa initial mass function (IMF)
\citep{Kroupa01}. The metallicity range of these SSPs encompass the
observed range of metallicities seen in large integral-field surveys
sampling all galaxy types over a wide range of luminosity and radius
\citep[e.g.,][]{Sanchez-Blazquez14,Gonzalez-Delgado15,Zheng2017}.

\subsection{Defining Young and Old Components}
\label{sec:young_old}

For the old stellar population we use SSPs older than 1.5~Gyrs as
template spectra, and the remaining for the young stellar population
modeling. This choice is motivated in
Figure~\ref{fig:hot_star_relative_wts}. Here we conduct a
light-weighted full spectrum fit of each MIUSCAT SSP with a set of
empirical stellar spectra representative of a broad range of stellar
spectral types (see Appendix~\ref{app:stellar_lib}) and sum the
weights for the presented bins of stellar types. The plot demonstrates
that except for the lowest metallicity models the relative weight in
hot stars (O, B, and A type) falls below 10\% above ages of 1.5~Gyrs,
while the coolest stars begin to contribute substantially ($>$30\%)
after this time, namely the times-scale for the formation of the
red-giant branch.

It is also the case that the intermediate stars (F,G) peak around
1~Gyr. While the hottest of these (F0~V) will still have
  significant Balmer absorption and Main Sequence (MS) anlifetimes of
  roughly 3 Gyr, for regions with on-going or recent star-foration the
  F-star contribution to the Balmer absorption equivalent width will be
  small compared contributions from hotter, more luminous MS stars.
Nonetheless, based on this Figure, an argument can be made for an
intermediate age range, say, between 0.5 and 3 Gyr, particularly
  for modeling galaxies or regions of galaxies that have not had
  recent star-formation.  We take advantage of the multimodel
  distributions in Figure~\ref{fig:hot_star_relative_wts} to constrain
  the mix of stars used in the stellar library algorithms
  (Section~\ref{sec:stellar_algo_evol}).  However, while an additional
  time-bin would greatly aid in determining the AVR, for simplicity of
  our development in this work we distinguish only between two age
  bins for measuring differential asymmetric drift signals.

In the MW solar cylinder from \cite{Aumer09} or from MW and M33 star
clusters \citep{Beasley15} there is observed to be roughly a factor of
3 change in velocity dispersion between t=0 and 1.5 Gyr (assuming the
birth population has the same dispersion as the molecular gas --
something that is not clearly the case for the star-cluster {\bf {\it
    measurements}}), and about a factor of 2 increase between 1.5 and
10 Gyr; in M31 \citep{Dorman15} the increase in velocity dispersion is
roughly the same during these two time periods. Hence our choice of
age bins appears a sensible starting point {\it if} the three large
galaxies in the Local Group are representative of the larger
population of intermediate-to-massive spiral disks.


\begin{figure*}[ht!]
\begin{center}
  \includegraphics[width=0.95\linewidth]{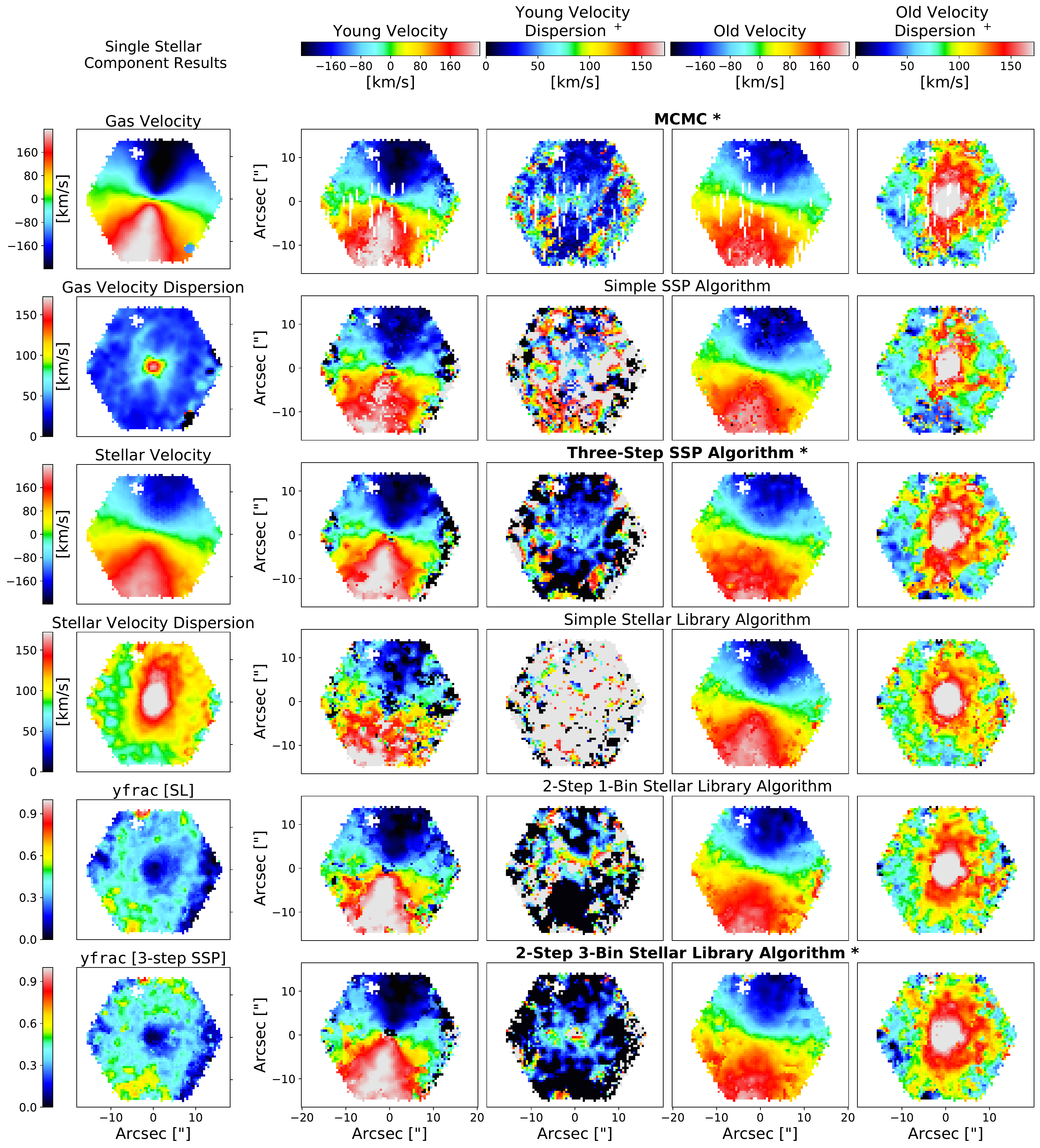}
  \caption{Derived kinematics of MaNGA galaxy MID 1-339041 (data-cube
    8138\_12704) using \pPXF\ to fit for a single stellar component
    and ionized gas component (left-hand column), and using different
    algorithms to fit for the two stellar components (young and old)
    and the ionized gas component (right-hand 4 columns).  All
    kinematics are line-of-sight (projected) measurements.  For the
    single stellar component fits the panels show (top to bottom): gas
    velocity, gas velocity dispersion, stellar velocity, stellar
    velocity dispersion, \pfrac\ used in the stellar library algorithm
    (Section~\ref{subsec:stellib_algo_summary}), and final measured
    \pfrac\ from the final iteration of the SSP algorithm
    (Section~\ref{subsec:ssp_algo_summary}). For two-component fits,
    the right-most 4 columns show (left to right) derived velocities
    and velocity dispersions of the young and old stellar
    populations. Gas velocity and velocity dispersion maps are
    identical to the fits using two stellar components and hence are
    not repeated.  Different rows show different algorithms: (top row)
    MCMC analysis from Section~\ref{sec:hypothesis} showing {\it all}
    spaxels; (second row) one-step SSP algorithm described in
    Section~\ref{sec:ssp_simple}; (third row) three-step (final)
      SSP algorithm described in Section~\ref{sec:ssp_multi} and
      Section~\ref{subsec:ssp_algo_summary}); (fourth row) 2-step
      stellar library algorithm that provides instructive initial
      kinematic values for decomposition, described in
      Section~\ref{sec:stellar_twostep}; (fifth row) final 2-step
      stellar library algorithm constrained for three bins in stellar
      temperature, described in
      Section~\ref{subsec:stellib_algo_summary}. Our two final
    algorithms are marked in bold and asterisks.}
	\label{8138_12704_bp}
\end{center}
\end{figure*}

\begin{figure*}[ht!]
\centering
\includegraphics[width=0.166\linewidth]{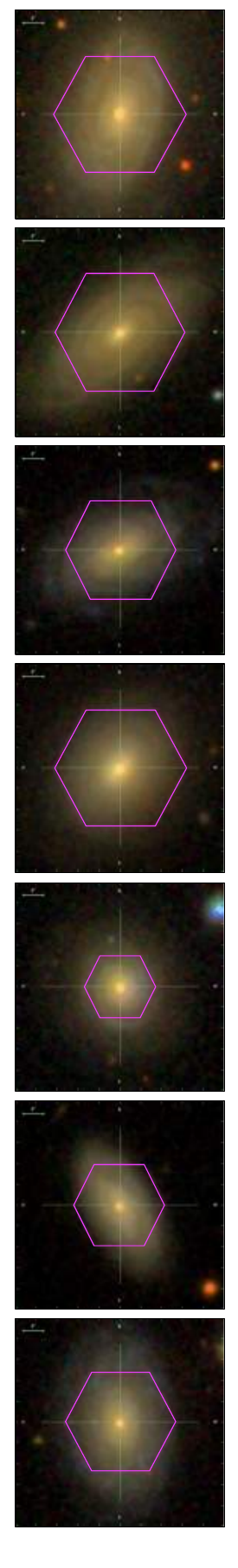}
\includegraphics[width=0.5935\linewidth]{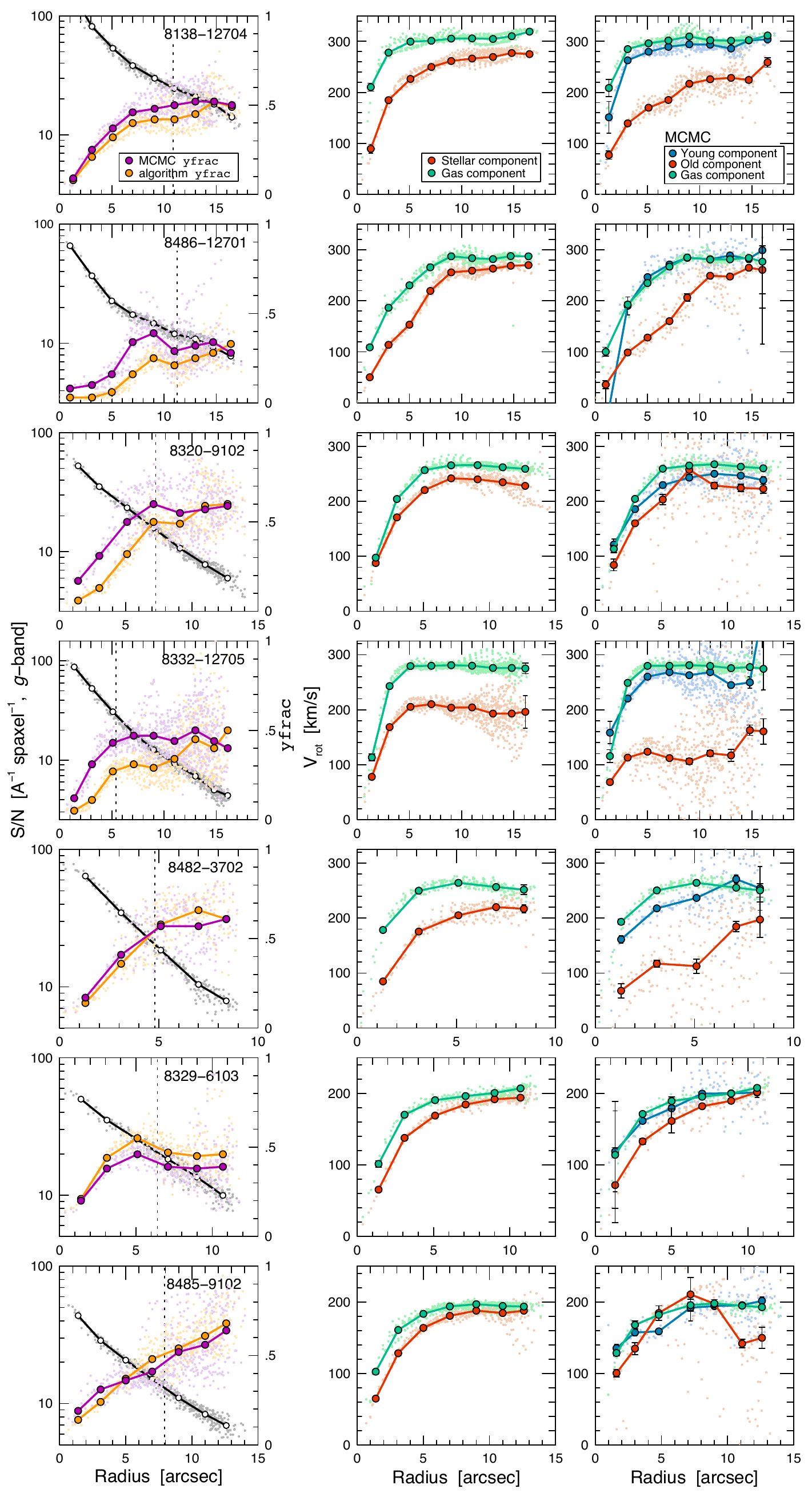}
\caption{MaNGA survey galaxies MID 1-339041, 1-209537, 1-532459,
  1-251279, 1-542358, 1-265988, 1-209199 (top to bottom, as given in
  Table~\ref{tab:sample}). All measurements are from data-cubes with
  plate-ifu designation given in the second column.  First column
  (left): False-color ({\it gri}) $50\times50$ arcsec image. The
  magenta hexagon marks the MaNGA IFU footprint. Second column: {\it
    g}-band signal-to-noise (S/N) and \pfrac\ radial profiles.  S/N
  (black) is defined per spaxel as the product of the median spectral
  flux and the median of the square-root of the inverse-variance ({\tt
    IVAR} in the datacubes). Values of \pfrac\ are shown as derived
  from the MCMC computation (purple) and from our subsequent
  algorithms (orange; refer to Section.~\ref{sec:ssp_multi} for the
  2-step SSP algorithm). Vertical dotted lines mark the half-light
  radius. Third column: Rotation curves for a {\it single} stellar
  component and a single gas component from a full-spectrum fit to
  individual spaxels via \pPXF. Fourth column: Rotation curves from
  the MCMC algorithm for {\it two} stellar population components
  (young and old, as defined in the text) and a third component for
  ionized gas. All rotation-curve measurements use {\it deprojected}
  velocities to derive the tangential speeds of the components using
  geometries defined in Table~\ref{tab:geometry}.  In the three
  right-most panels small points represent individual spaxel
  measurements within a $\pm30^{\circ}$ wedge about the major axis,
  while larger points and error bars represent medians and the
  associated robust uncertainty (mostly smaller than the points) for
  the ensemble of these individual measurements in 2 arcsec radial
  bins.\\ \\}
\label{fig:emceespaxels}
\end{figure*}

\section{A Robust Two-component AD Signal}
\label{sec:hypothesis}

\begin{deluxetable*}{lrrrrr}[ht!]
\tablewidth{0pt}
\tabletypesize{\footnotesize}
\tablecaption{Geometric Parameters}
\tablehead{
  \colhead{plate-IFU} &
  \colhead{xoff} & 
  \colhead{yoff} & 
  \colhead{PA$_{\rm kin}$} & 
  \colhead{i$_{\rm kin}$} & 
  \colhead{i$_{\rm phot}$} \\
  \colhead{} &
  \colhead{(arcsec)} &
  \colhead{(arcsec)} &
  \colhead{(deg)} &
  \colhead{(deg)} &
  \colhead{(deg)} \\
  \colhead{(1)} &
  \colhead{(2)} &
  \colhead{(3)} &
  \colhead{(4)} &
  \colhead{(5)} &
  \colhead{(6)} 
}
\startdata
 8138-12704 & +0.10$\pm$0.01 & +0.02$\pm$0.01 & 165.92$\pm$0.04 & 53.4$\pm$0.1 & 46.8 \\
 8486-12701 & -0.04$\pm$0.01 & +0.03$\pm$0.01 & 309.72$\pm$0.04 & 59.3$\pm$0.1 & 51.3 \\
 8320-9102  & +0.03$\pm$0.01 & -0.07$\pm$0.01 & 111.56$\pm$0.05 & 50.1$\pm$0.2 & 48.0 \\
 8332-12705 & +0.04$\pm$0.01 & +0.00$\pm$0.01 & 146.25$\pm$0.05 & 42.6$\pm$0.2 & 36.6 \\
 8482-3702  & -0.00$\pm$0.02 & -0.20$\pm$0.01 &  14.75$\pm$0.11 & 27.1$\pm$0.9 & 22.1 \\
 8329-6103  & +0.17$\pm$0.01 & -0.17$\pm$0.01 & 207.07$\pm$0.06 & 54.9$\pm$0.2 & 50.6 \\
 8485-9102  & -0.23$\pm$0.01 & -0.09$\pm$0.01 & 181.54$\pm$0.06 & 49.8$\pm$0.2 & 44.6
 \enddata
 \label{tab:geometry}
\tablecomments{ Column (1) gives the galaxy plate-IFU
  identifier matching Table~\ref{tab:sample}.  Columns (2) and (3)
  give the x- and y-offsets of the galaxy barycenter from the IFU
  center in arcsec from kinematic modeling described in the text.
  Column (4) gives the position angle from the same kinematic
  modeling.  Column (5) gives the inclination from the same kinematic
  modeling.  Column (6) give the photometric inclination derived from
  the b/a values in Table~\ref{tab:sample} assuming an intrinsic disk
  oblateness of 0.2.  }
\end{deluxetable*}

Outside of the Milky Way's solar neighborhood and few of the nearest
galaxies in the Local Group, we lack evidence demonstrating the
presence of an AVR in other galaxies. All existing evidence is based
on measurements of resolved stellar populations.  Before developing a
technique to measure AVR in spectra of integrated star-light as
observed by MaNGA, we demonstrate that this information is present and
extractable from such spectra. The purpose of this section is to show
unequivocally, via MCMC analysis, that a stellar population model with
at least two kinematic components can be robustly constrained by the
data.

The MCMC analysis undertakes full-spectrum fitting of the observed
galaxy spectrum using three kinematic components; a young stellar
component (age $\leq 1.5$~Gyrs), an old stellar component (age $>$
1.5~Gyrs) and a gas component -- as defined in the previous
section. The MCMC helps us explore how robust the \pPXF\ likelihood
maximization is at deriving distinct kinematics of a young and old
stellar populations. This is important given the degrees of freedom in
the fitting, the subtlety of the signal (a few 10's of \kms\ in
velocity difference expected between young and old populations), the
degeneracies in stellar population synthesis (e.g., age and
metallicity), and a range of data quality -- all of which contribute
to introducing local minima in $\chi^{2}$ space.  We use Bayesian
inference to sample the posterior probability distribution of the
kinematics, employing an implementation of Affine-Invariant
Markov-Chain Monte-Carlo (MCMC) proposed by \citet{Goodman&Weare2010}
and encoded by \citet{Foremanetal2013}. This technique minimizes the
need for a lengthy ``burn-in'' phase which, given the extensive
computation time of the full spectrum fit with the complete set of SSP
models, requires significant computing time.

Here, the MCMC uses 14 independent `walkers' to sample
$\chi^{2}$ for a model with two stellar populations and gas with
distinct kinematics. Each step of the chain fixes the kinematics
(velocity and dispersion) and the \pfrac\ of the two stellar
populations and allows \pPXF\ to optimize the SSP weights,
polynomials, and gas emission line intensities. The kinematics of the
young and old stellar components are then randomly `walked' through in
order to sample $\chi^{2}$ and determine the location and shape of the
global minima of this parameter space\footnote{In what follows we
  assume all models have equal probability so probability scaling can
  be ignored.}. If the kinematics of the young and old stellar
populations in the galaxy are similar, then we expect to see so within
the posterior distribution sampled by the MCMC. However if this is not
the case, then the peaks of the posterior distributions should not
overlap.

The MCMC parameterizes over the six kinematic parameters of the
components, i.e. the velocities and dispersions for the young and old
stellar component and gas component, and \pfrac\ (defined in
Equation~\ref{frac_def}). The initial starting values for the Markov
Chain are randomly generated using the results from an initial
full-spectrum fit of the galaxy spectrum with a single stellar
component and a gas component. We hypothesize that the kinematics of
the young stellar component is close to that derived for the gas
component, and the kinematics of the old component is close to that of
the single stellar component. Hence the initial kinematics for the two
components are randomly generated from a normal distribution centered
at the measured kinematics and with a standard deviation of
25~\kms\ for the velocities and 50~\kms\ for the velocity
dispersions. The initial starting values for \pfrac\ is also randomly
generated from a normal distribution centered at total relative weight
in the young SSP templates in the initial fit, with a standard
deviation of 0.1. For each step of the MCMC, we bound the parameters
of the fit such that the velocities of the components are within
250~\kms, while the velocity dispersions are within 200~\kms\ of the
starting values. The parameter \pfrac\ is bound to values between 0
and 1. Within these bounds, we assume no prior on the parameters.

\begin{figure*}[ht!]
  \centering
  \includegraphics[width=0.23\linewidth]{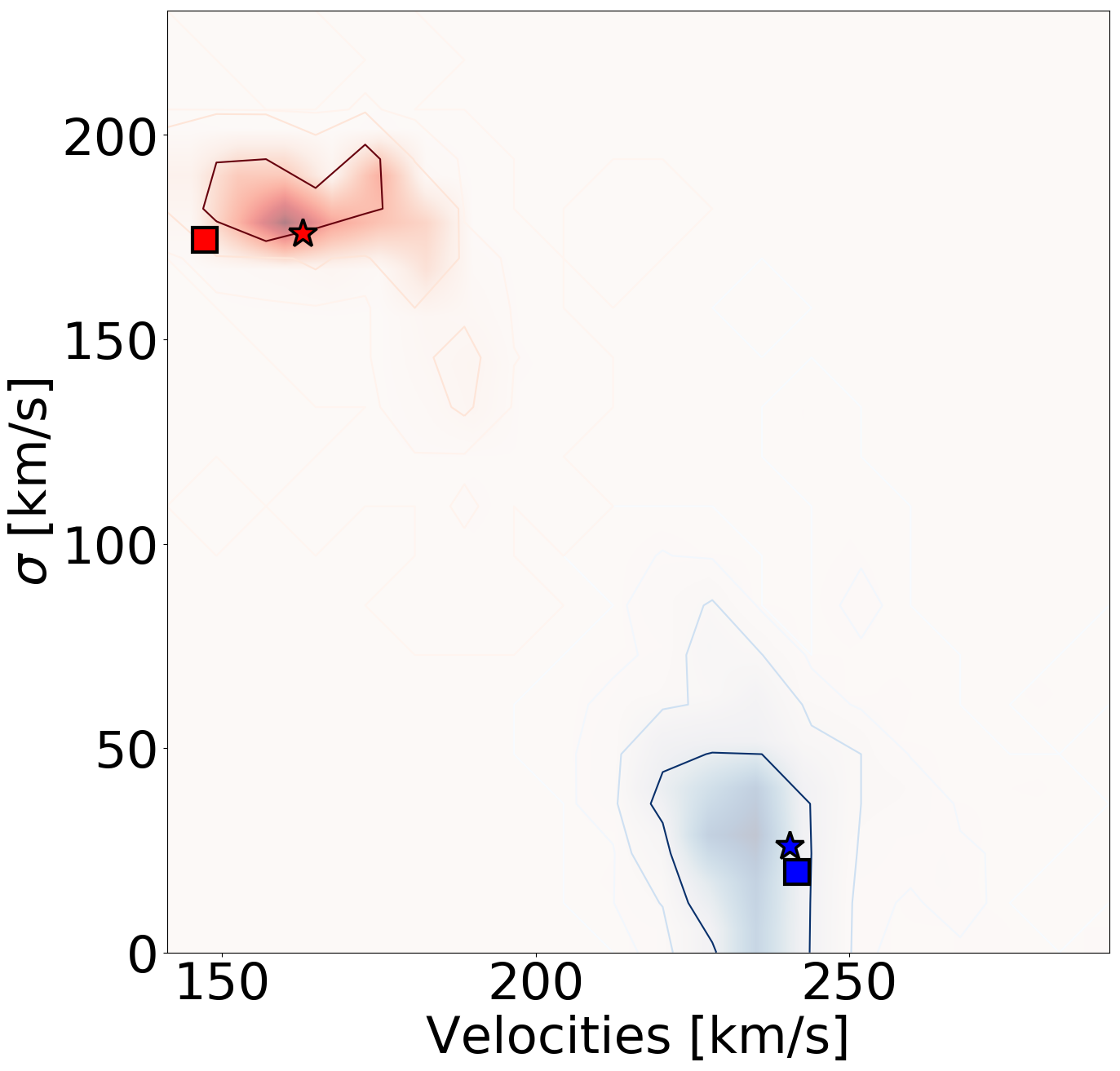}
  \includegraphics[width=0.23\linewidth]{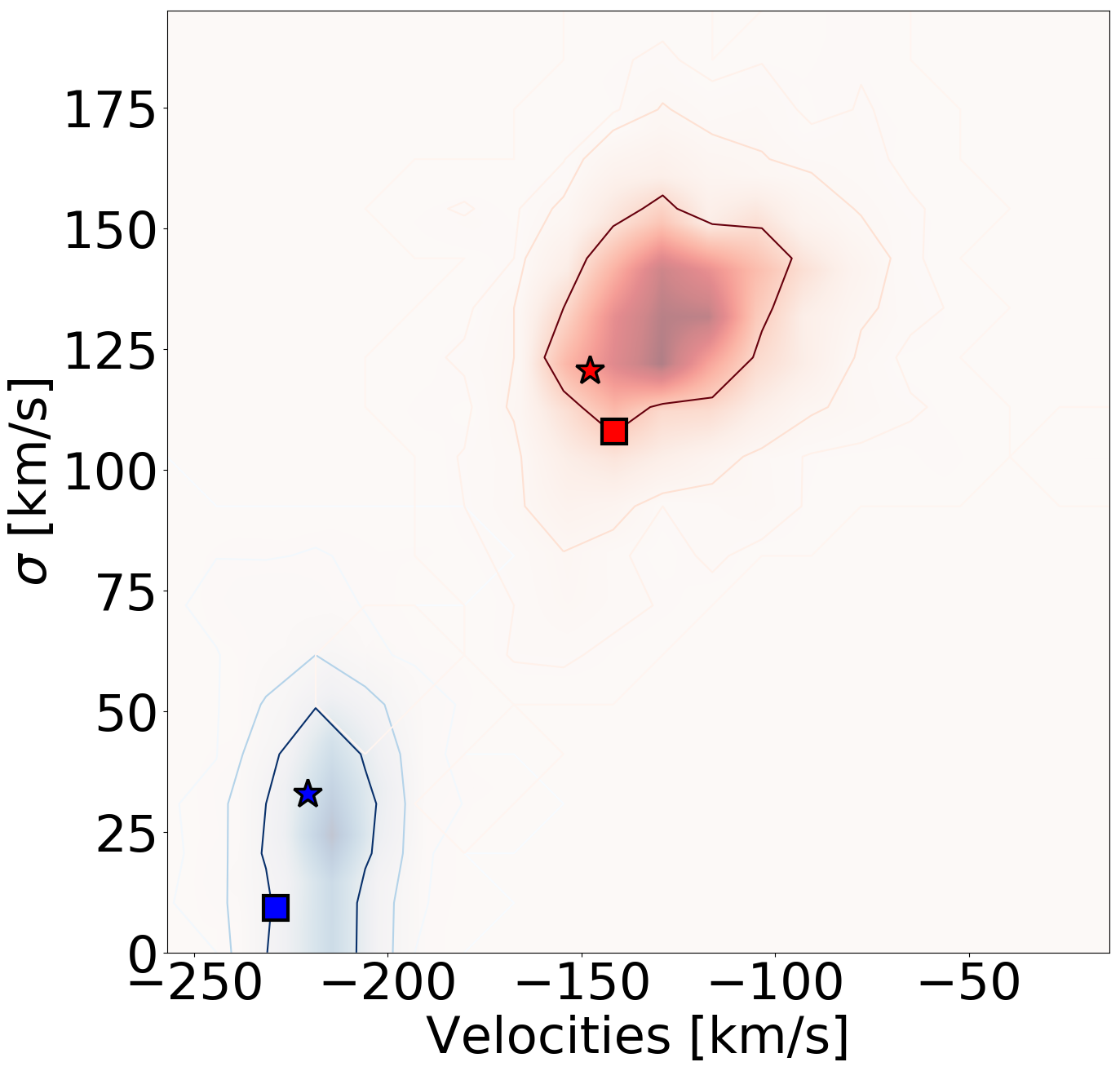}
  \includegraphics[width=0.23\linewidth]{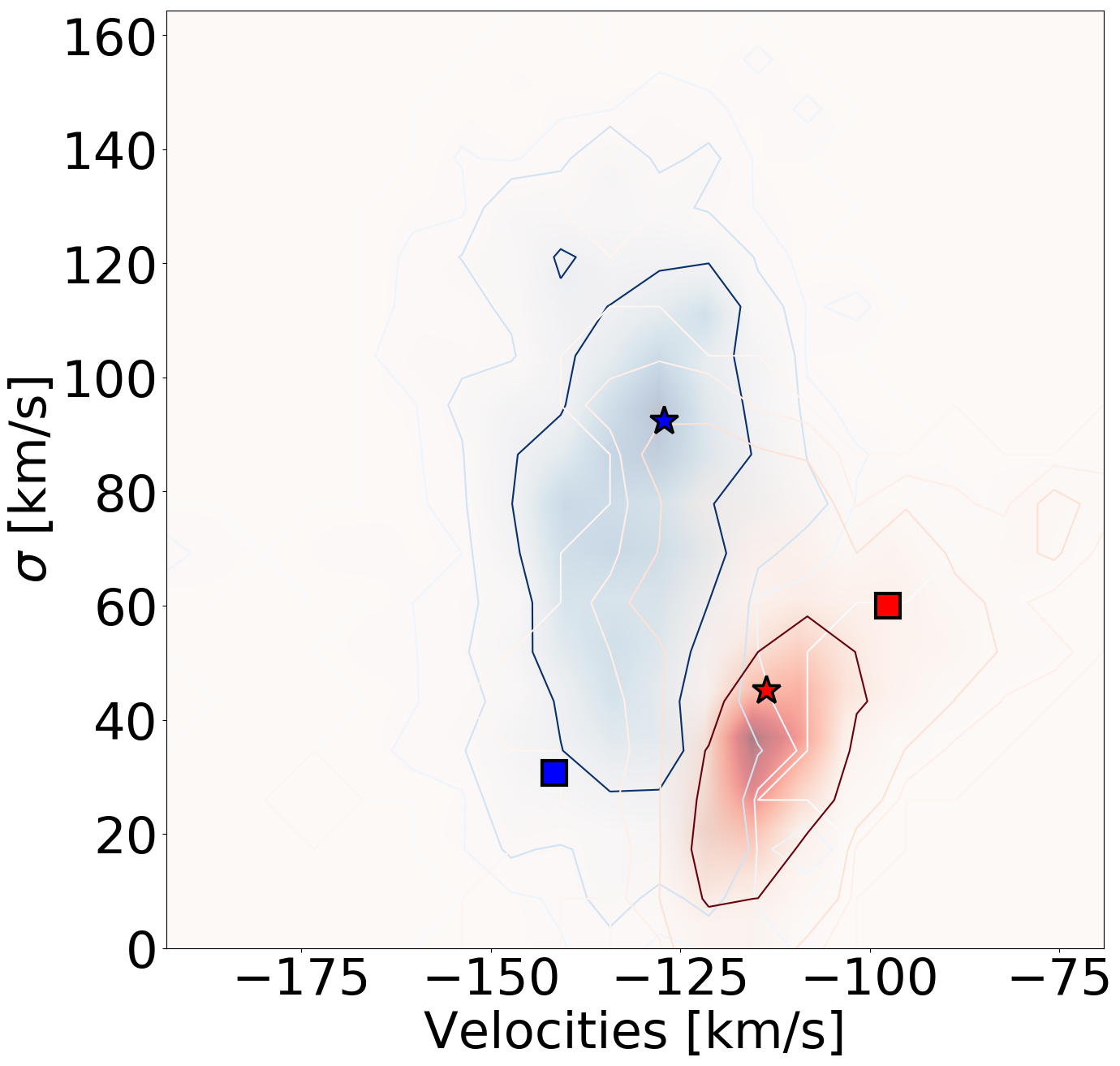}
  \includegraphics[width=0.23\linewidth]{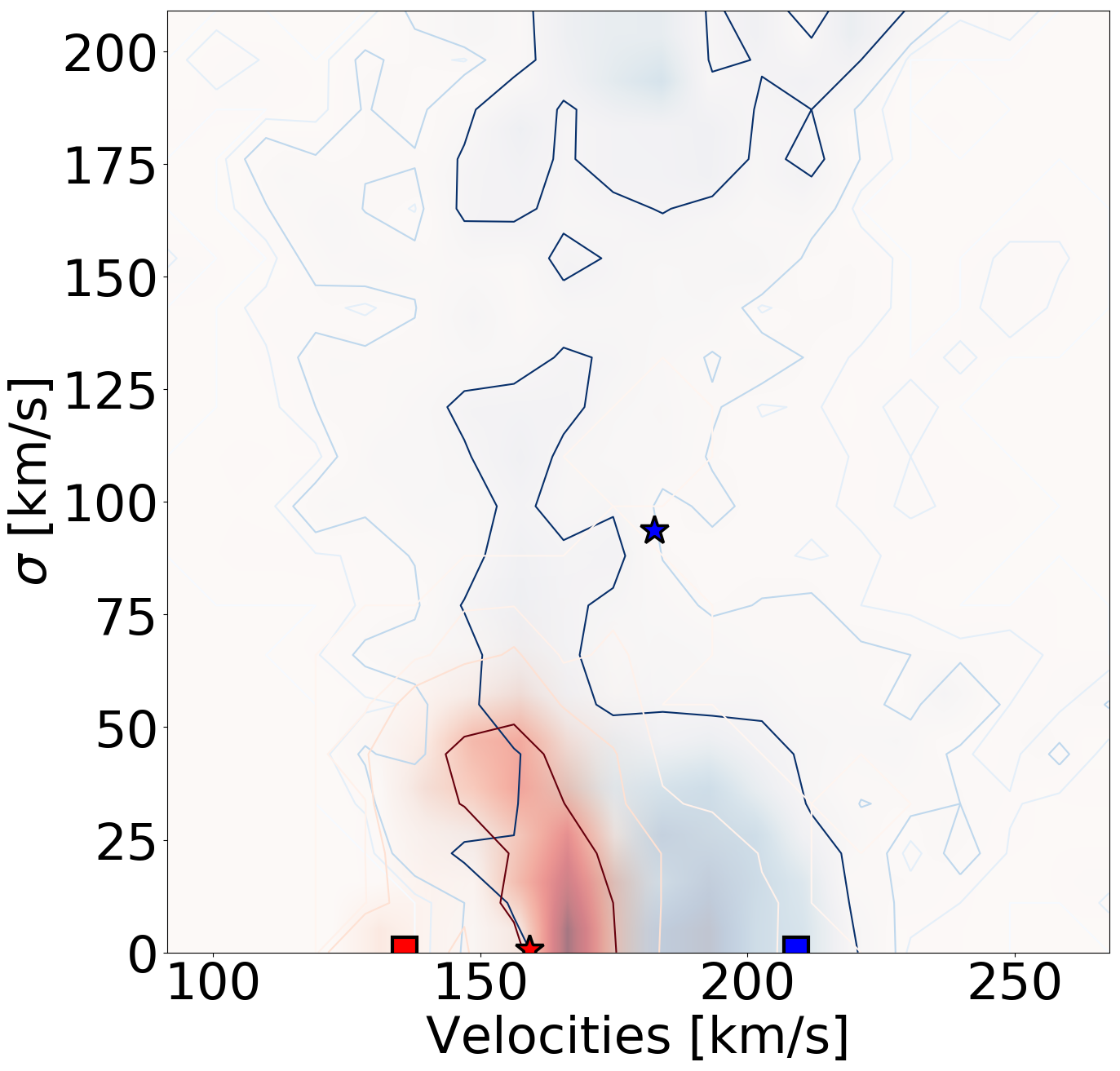}
  \caption{Results of our MCMC analysis for four spaxels taken, two
    each, from from MaNGA galaxies MID 1-339041 (8138-12704) and
    1-265988 (8329-6103).  Each panel shows the posterior distribution
    in velocity and velocity dispersion of the young (blue) and old
    (red) components resulting from iterations of the Markov
    Chain. Contours enclose 68\% 95\%, and 99.7\% of the
    distribution. From left to right, spaxels were chosen at 3.5
    arcsec and 12 arcsec radius along the major axis in 1-339041
    [spaxels (0.5, -3.5) and (-3.5, 11.5) respectively in the format
      (RA, Dec) where +RA implies East and +Dec implies North] and at
    4 and 8.5 arcsec in 1-265988 [(2, 3.5) and (-4, -7.5)] to
    illustrate (i) the flip in sign on receding and approaching sides
    between young and old velocities, (ii) the expected relation
    between asymmetric drift and velocity dispersion when the observed
    dispersion is well above the instrumental resolution ($\sigma_{\rm
      inst}\sim 65$ \kms) for the old component; and (iii) to
    demonstrate how small the signal can get with robust
    discrimination between {\it velocities} consistent with that
    expected from AVR even when velocity dispersion measurements are
    non-sensical or far below the instrumental resolution.  These
    spaxels also demonstrate the effect of S/N on the $\chi^2$ space
    of the solution; the spaxels from left to right have S/N 65, 23,
    30 and 15 respectively.  Preferred solutions identify young and
    old stellar populations with discrete kinematics at very high
    significance in all cases, with the results consistent with that
    expected from AVR. Blue and red squares represent the results from
    our final 3-step SSP algorithm described in
    Section~\ref{subsec:ssp_algo_summary}, while blue and red stars
    represent the results from our final 2-step Stellar-library
    algorithm described in Section~\ref{subsec:stellib_algo_summary}.}
	\label{fig:emcee_v_sigma}
\end{figure*}

\subsection{MCMC Results}
\label{sec:mcmc_results}

We applied the MCMC algorithm to our sample of seven galaxies on a
{\it spaxel-by-spaxel basis}. Figure~\ref{8138_12704_bp} shows the
measured, line-of-sight (projected) velocity and velocity dispersion
maps for one galaxy for young and old stellar components in the second
to fourth columns of the top row. (Maps for the remaining six galaxies
in our sample are in Appendix~\ref{sec:maps}.)  There is no smoothing
applied to these maps; the color of each data point in the image
corresponds to the kinematics of individual spaxels.  Blank spaxels in
these maps {\it within} the MaNGA IFU footprint are either regions
masked in the DRP as foreground stars or, in the case of the MCMC
only, represent spaxels where the MCMC did not converge within the
maximum run-time permitted for jobs on the computer cluster to which
we had access.\footnote{Job run-times were limited to 72 hours. Each
  job consisted of processing for several contiguous spaxels, leading
  sometimes to vertical sets of incompletion.} For comparison we also
show the derived velocity and velocity dispersion maps for gas and
stars for a single stellar component. (In later sections these serve
as the initial fitting step in multi-step alrgorithms.)  These were
measured directly from \pPXF. Note the coherency of the velocity
fields with the expected spider-diagram of isovels, as well as the
smoothness of the kinematics on scales much larger than the beam size
(2 arcsec fiber diameter). This smoothness and coherency persists for
both single and two-component stellar kinematics. Keeping in mind that
the kinematics are derived {\it independently} for each spaxel, this
coherency indicates that our two-component stellar solutions from the
MCMC algorithm are robust and measuring astrophysical kinematic
signals. While there is much more scatter in the velocity {\it
  dispersion} maps, our approach is to avoid using these and higher
kinematic moments.\footnote{NB: stellar velocity dispersion maps
  provided for the young and old components, denoted by a cross, are
  presented as qualitative metric to guide the development of the
  algorithm and are considered quantitatively unreliable much below
  the instrumental resolution until forthcoming improvements in the
  accuracy in the instrumental line-spread-function are concluded
  (D. Law, private communication).} It is reassuring nonetheless to
see that on average the younger component has lower dispersions than
the older component as anticipated, and the old component has the
qualitatively expected radial decline in value.  These trends are seen
for all galaxies in our sample.

While the asymmetric drift signal is readily apparent from
Figure~\ref{8138_12704_bp} to the practiced eye, this is more easily
seen by extracting rotation curves using some fiducial
geometry. Accordingly, Figure~\ref{fig:emceespaxels} shows the results
of our MCMC algorithm in decoupling the {\it deprojected} tangential
speeds of young and old stellar populations. In most cases and at most
radii we see that both young and old stellar components lag in their
tangential speed relative to the ionized gas (i.e., this is asymmetric
drift), with the old component lag being larger, {\it as one would
  expect from the presence of AVR.} For comparison we show the
deprojected tangential speeds for gas and stars for a single stellar
component (again, derived directly from \pPXF), from which the
asymmetric drift signal also is readily apparent.

For Figure~\ref{fig:emceespaxels} we adopt geometric parameters
derived from full, two-dimensional kinematic modeling of a monolithic
inclined-disk. We use the method described in \cite{Westfall11} and
\cite{Andersen2013}, applied simultaneously to the DAP gas and stellar
kinematics. These parameters are summarized in
Table~\ref{tab:geometry}. However, these geometric parameters are only
used to define radial and azimuthal bins and deproject the kinematics;
{\it the individual fits to each spaxel are independent of these
  geometric parameters}. Consequently the rotation curves in
Figure~\ref{fig:emceespaxels} make no assumptions about kinematic or
rotation-curve models except insofar as they share the geometries
given in Table~\ref{tab:geometry}. Figure~\ref{8138_12704_bp} and
associated Figures in Appendix~\ref{sec:maps} are completely
independent of {\it any} assumed geometric or rotation-curve model.

We focus on two fairly extreme cases: 8138-12704 and 8329-6103.  Both
galaxies have S/N above 10 \AA$^{-1}$ even at large radii, and \pfrac\ values
between 0.1 to 0.2 (in the central regions) and 0.4 to 0.55 in the
outer regions. The radial trend in \pfrac\ is qualitatively what is
to be expected for age gradients in spiral galaxies where the older
population is centrally more concentrated.

The former galaxy 8138-12704 was chosen because of its exceptionally
large asymmetric drift signal as measured for a {\it single} stellar
kinematic component with respect to the ionized gas, yet clear
presence of young and old stellar populations. At R = R$_e$, D$_{\rm
  n}$4000 = 1.42 while $V_{\rm gas}-V_{\rm stars} = 41.8$~\kms. Of a sample of
nearly 500 galaxies in MPL-5 selected to have regular kinematics and
moderate inclinations (see Section ~\ref{sec:sample}), this galaxy is
in the upper quartile in terms of deprojected gas rotation speed
($V_{\rm gas} = 304.8$~\kms). Since in general the asymmetric drift signal
scales with velocity, it is useful to also look at the fractional
difference between tangential speed of the stars and gas relative to
the gas ($1-V_{\rm stars}/V_{\rm gas}$). For 8138-12704 this fraction is
13.7\%, and it too is in the upper quartile.

In contrast 8329-6103 has an asymmetric drift signal that is in the
lowest quartile for the sample of galaxies in both an absolute and
relative sense: $V_{\rm gas}-V_{\rm stars} = 11.8$~\kms, and
$1-V_{\rm stars}/V_{\rm gas}$ is only 6\%. The galaxy has about 25\% of the
total stellar mass and 41\% of the dynamical mass at R = R$_e$ ($V_{\rm gas} =
194.4$~\kms) compared to 8138-12704, yet has a comparable D$_{\rm n}$4000 of
1.36. Indeed, our analysis shows the two galaxies have \pfrac\
values of 0.42 and 0.53 respectively. Compared to the MW
\citep[e.g.,][]{Licquia16}, 8329-6103 is almost a factor of two less
luminous.

For the more massive galaxy, 8138-12704, the differential asymmetric
drift signal is readily apparent at all radii sampled. To quantify
this excellent kinematic separation of young and old stellar
populations, Figure~\ref{fig:emcee_v_sigma} shows the bivariate
posterior probability distribution for velocity and velocity
dispersion of two spaxels selected from near the major axis. One
spaxel is on the approaching side at 0.3 R$_e$ (3.5 arcsec radius) and
the other is on the receding side at 1.1 R$_e$ (12 arcsec radius). In
both cases the peaks and 99.7\% contours are well separated with the
relative velocities and dispersions for the two populations. Moreover,
they are differentiated {\it as expected} toward higher dispersions
and lower speeds for the old population and vice versa for the younger
population. While this is consistent with our priors, the presence of
these distinct, age-dependent kinematics independently determined for
all spaxels using an unconstrained model suggests there is ample
information to extract a {\it quantitative} measure consistent with
our {\it qualitative} astrophysical expectations.

For the less massive galaxy 8329-6103, Figure~\ref{fig:emceespaxels}
shows the kinematics for the young and old populations are
differentiated at smaller radii, as expected, but at larger radii the
differentiation is less evident. This is reflected in
Figure~\ref{fig:emcee_v_sigma} as well. One spaxel is on the receding
side at 0.6 R$_e$ (4 arcsec radius) and the other is on the approaching
side at 1.3 R$_e$ (8.5 arcsec radius). For the inner spaxel, we see the
anticipated velocity difference between young and old population,
although the velocity dispersions are flipped in the opposite sense of
what we might physically expect, i.e., the faster-rotating (young
component) has a larger dispersion. However, the 67\% probability
contours are quite broad, and both components are at, or well below,
the instrumental resolution. At the larger radius the young stellar
component continues to appear to rotate more quickly, even though the
probability contours substantially overlap. The velocity dispersions
are well below the instrumental resolution in this limit.

These results are promising, indicating we should be able to derive
separate kinematics for most of the MaNGA sample that have regular
disk kinematics. Clearly Figure~\ref{fig:emceespaxels} shows some
irregularities in the tangential speeds of young and old components at
large radii, which correspond to low S/N.  The behavior reflects an
increasing frequency with decreasing signal-to-noise where the
measured kinematics of the young stellar populations are dynamically
hotter and rotate slower than the old stellar component. In
Appendix~\ref{app:mcmc_ppd} we demonstrate this behavior is due to
degeneracies in disentangling the kinematics of the young and old
stellar populations in this regime when using a global minimizer
(i.e., MCMC) with a very broad range of template age and metallicity.
Our subsequent algorithms that use a local minimizer with carefully
constructed initial conditions largely eliminate this problem.


\section{SSP-Based Algorithms to Measure AD}
\label{sec:ssp_algo}

While MCMC provides a robust means for determining the kinematic
parameters of our two-component stellar kinematic model, it is also
computationally too time-consuming for application to data-cubes of
hundreds to tens-of-thousands of galaxies in e.g., MaNGA. In this
section we explore how we can make this measurement by directly using
the likelihood maximization technique of \pPXF\ since this
reduces computation times by factors of order a thousand.  
To evaluate the performance of such efficient algorithms we
rely on two different metrics:

\begin{enumerate}

\item Quantitative comparison of algorithm recovered velocities and \pfrac\
to model or MCMC values: In the early stages of algorithm development
we used 2,500 MaNGA-like mock spectra with realistic star formation
histories.  We summarize the key results for this metric in the main
text and refer the reader to Appendix~\ref{app:mocks} for further
details. Since the mocks and fitting spectra were based on the same
SSPs, they mocks were less useful for refinning our
algorithms. Consequently we also compared the young and old
stellar-component velocities derived by our efficient algorithms to
those quantities derived from our MCMC results on real data. This
served as our summary performance metric, with results given in
Table~\ref{tab:MCMC_SSP_metric} and discussed throughout this Section and
in Section~ \ref{sec:stellar_algo} for the stellar library algorithm.

\item Qualitative comparison of velocity and velocity dispersion maps:
We used the smoothness of velocity and velocity dispersion maps (e.g.,
Figure~\ref{8138_12704_bp}) as a qualitative assessment of our
algorithms. The smoothness of the field maps are suggestive of
astrophysically real and observationally reliable results since each
spaxel is analyzed independently. This second metric, only available
for real data, is particularly useful because the quantitative metrics
don't capture systemmatic uncertainties in measured velocities.

\end{enumerate}

\begin{deluxetable*}{lcrrrrrrrrrrr}
\tablewidth{0pt}
\tabletypesize{\footnotesize}
\tablecaption{SSP Algorithm Performance Metrics Referenced to MCMC Results}
\tablehead{
  \multicolumn{1}{l}{Algorithm} &
  \colhead{S/N} &
  \multicolumn{2}{c}{$\Delta V_y$} &
  \colhead{} &
  \multicolumn{2}{c}{$\Delta V_o$} &
  \colhead{} &
  \multicolumn{2}{c}{$\delta V_{\rm Algo} / \delta V_{\rm MCMC}$} &
  \colhead{} &
  \multicolumn{2}{c}{$\Delta{\rm \pfrac}$} \\  [0.05in] \cline{3-4} \cline{6-7} \cline{9-10} \cline{12-13}
  \multicolumn{13}{c}{} \\  [-0.08in]
  \colhead{} &
  \colhead{Range} &
  \multicolumn{1}{r}{med} &
  \multicolumn{1}{r}{$\sigma_{\rm MAD}$} &
  \colhead{} &
  \multicolumn{1}{r}{med} &
  \multicolumn{1}{r}{$\sigma_{\rm MAD}$} &
  \colhead{} &
  \multicolumn{1}{r}{med} &
  \multicolumn{1}{r}{$\sigma_{\rm MAD}$} &
  \colhead{} &
  \multicolumn{1}{r}{med} &
  \multicolumn{1}{r}{$\sigma_{\rm MAD}$} \\
  \colhead{} &
  \colhead{} &
  \multicolumn{2}{c}{(\kms)} &
  \colhead{} &
  \multicolumn{2}{c}{(\kms)} &
  \colhead{} &
  \colhead{} &
  \colhead{} &
  \colhead{} &
  \colhead{} &
  \colhead{} \\
  \multicolumn{1}{l}{(1)} &
  \colhead{(2)} &
  \multicolumn{1}{r}{(3)} &
  \multicolumn{1}{r}{(4)} &
  \colhead{} &
  \multicolumn{1}{r}{(5)} &
  \multicolumn{1}{r}{(6)} &
  \colhead{} &
  \multicolumn{1}{r}{(7)} &
  \multicolumn{1}{r}{(8)} &
  \colhead{} &
  \multicolumn{1}{r}{(9)} &
  \multicolumn{1}{r}{(10)}
}
\startdata
Simple SSP      & [10,35] & -1.9 & 18.3 &&  0.7 &  9.9 && 0.84 & 0.66 && 0.040 & 0.088 \\
Simple SSP      & [35,50] & -3.3 &  7.5 &&  1.6 &  4.6 && 0.81 & 0.32 && 0.057 & 0.058 \\
Simple SSP      &   $>$50 & -0.1 &  7.1 &&  1.7 &  4.8 && 0.82 & 0.29 && 0.038 & 0.044 \\ [0.05in]
2-Step SSP      & [10,35] & -2.7 & 13.5 &&  0.4 &  9.3 && 0.82 & 0.59 && 0.018 & 0.050 \\
2-Step SSP      & [35,50] & -4.9 &  3.9 &&  1.4 &  3.8 && 0.88 & 0.18 && 0.019 & 0.022 \\
2-Step SSP      &   $>$50 & -2.1 &  3.5 &&  1.4 &  3.2 && 0.88 & 0.12 && 0.023 & 0.016 \\ [0.05in]
3-Step SSP      & [10,35] & -0.1 & 10.8 &&  0.7 &  7.6 && 0.93 & 0.47 && 0.006 & 0.038 \\
3-Step SSP      & [35,50] & -0.5 &  2.3 &&  0.8 &  2.2 && 0.96 & 0.09 && 0.005 & 0.014 \\
3-Step SSP      &   $>$50 &  0.4 &  2.1 &&  1.8 &  2.0 && 0.98 & 0.08 && 0.005 & 0.013
\enddata
\label{tab:MCMC_SSP_metric}
\tablecomments{SSP algorithms (column 1) are defined in Sections~\ref{sec:ssp_simple}
  and \ref{sec:ssp_multi}.  Three S/N bins (column 2) contain 12945,
  587, and 512 spaxels, respectively from low to high S/N. Spaxels are
  culled from the seven galaxies in our sample to be at radii greater
  than 2 arcsec in radius (to avoid low values of {\tt yfrac}) and to
  available MCMC solutions. The MCMC measurements use the Maximum
  Likelihood solutions. The metrics include the median (med) and mean
  absolute deviation ($\sigma_{\tt MAD}$) scaled by a factor of 1.4
  for $\Delta V_y \equiv V_{y,{\rm MCMC}}-V_{y,{\rm Algorithm}}$ in
  columns (3) and (4); $\Delta V_o \equiv V_{o,{\rm MCMC}}-V_{o,{\rm
  Algorithm}}$ in columns (5) and (6); the ratio $\delta V_{\rm
  Algorithm} / \delta V_{\rm MCMC}$ in columns (7) and (8), where
  $\delta V \equiv V_y - V_o$; and $\Delta {\rm \pfrac} \equiv
  {\rm \pfrac}_{\rm MCMC}-{\rm \pfrac}_{\rm Algorithm}$ in columns (9)
  and (10).}
\end{deluxetable*}

In the following sub-sections, we first present the simplest possible
SSP-based algorithm for measuring efficiently the two-component
velocities of young and old stellar populations
(Section~\ref{sec:ssp_simple}). This involves a single fitting step.
Its shortcomings motivate using better initial conditions for the
kinematics and constraints on \pfrac\ in algorithms that involve two
fitting steps (Section~\ref{sec:ssp_multi}). A third fitting step that
leaves \pfrac\ unconstrained is found to improve metric performance
further, and yield results that are within the uncertainties of the
MCMC algorithm results. This final, three-step, algorithm is
summarized in Section~\ref{subsec:ssp_algo_summary}.

\subsection{A Simple SSP Algorithm}
\label{sec:ssp_simple}

The simplest scheme for measuring the kinematics of two discrete
stellar population components consists of giving \pPXF\ the set of
SSPs described in Section \ref{sec:methods}. We again split the
assignment of stellar population components at 1.5~Gyr regardless of
metallicity. There are no constraints placed on \pfrac; \pPXF\ is
free to assign weights to the templates as needed to minimize
$\chi^2$. A single full-spectrum fit to the galaxy spectrum is made
with these two (young and old) stellar kinematic components and, in
the case of real galaxy spectra (as opposed to emission-line free
mocks) a kinematically independent gas component. The initial
velocities and velocity dispersions for the two stellar components are
identical and are set at 0 and 210 \kms respectively.

This simple algorithm performs well on our mock spectra (metric
1), yielding derived AD signals within 10\% of the expected value
67\% of the time even for AD signals as low as 5-10~\kms (see
Appendix~\ref{sec:ssp_mock_performance} and
Table~\ref{tab:SSP_realistic_mocks_metric}). While these results look
promising, the test is highly idealized, absent noise and using the
same templates in the mocks and the fitting templates. Consequently,
we turn to apply this simple algorithm to real MaNGA spectra as a more
definitive performance test. In applying the simple SSP algorithm to
real galaxy data, we fit for the gas kinematic component
simultaneously along with the stellar kinematic components and match
the LSF of the SSP templates to that of the spaxels.\footnote{We use
the {\tt PRESPECRES} datacube extension providing each spaxels'
pre-pixelized ({\tt PRE}) spectral resolution ({\tt SPECRES}).}

Application of the {\it simple} SSP algorithm to 8138-12704 (Section
\ref{sec:hypothesis}) indicates the algorithm does not appear to
reliably converge to the kinematics measured by the MCMC (metric 2). In Figure~\ref{8138_12704_bp} we observe that for both the MCMC
  results and the simple SSP algorithm the younger component appears
  to have on average faster rotation than its older counterpart, as
  expected. However, the young velocities for the simple SSP algorithm
  are systematically lower (in amplitude) than derived from MCMC,
  while the older velocities are systematically higher; both young and
  old velocity fields are less smooth as well.  Further, the $\sigma$
  maps are noisy, particularly for the young component, and lack the
  expected structure of being azimuthally smooth with a radial
  decline. These qualitative conclusions for 8138-12704 are also
  found in the remaining six galaxies in our sample
  (Appendix~\ref{sec:maps}). This is quantified in the first
  three lines of Table~\ref{tab:MCMC_SSP_metric}.

Given that the more robust MCMC technique derived a solution for the
kinematics different from our simple SSP algorithm, this suggests that
the spectral fitting code (\pPXF), which uses a Levenberg-Marquardt
technique to find the local $\chi^2$ minima, is not able to converge
to the global minima with real data.

\subsection{Multi-step SSP Algorithms}
\label{sec:ssp_multi}

To circumvent the pitfalls of a complex $\chi^2$ terrain we augment
our simple SSP algorithm by adding an additional fitting step. This
additional step allow us to improve initial conditions to position
\pPXF\ fairly close to where the global minimum is expected.

In a new first step, we conduct a full spectrum fit with only one
stellar kinematic component and one gas emission-line component, using
all SSPs (regardless of age) as templates for the single stellar
component. This fit is essentially a stellar population synthesis of
the spaxel and, accordingly, we use only multiplicative Legendre
polynomials during this fit. Expectations based on the MW AVR are that
the kinematics of the young stellar population should be close to that
of the gas. Simultaneously the kinematics of the old stellar component
should be similar to that from a single stellar component than the gas
kinematics. Results from our MCMC analysis support these
expectations. Hence, in the second fit we use the derived {\it gas}
kinematics from our first fit as the initial conditions for the {\it
young} component, and the derived {\it stellar} kinematics from our
first fit as the initial conditions for the {\it old}
component. During this second fit, we fix the kinematics of the gas to
that derived from the first fit and use additive and multiplicative
polynomials during the fit. We refer to this algorithm as
`2-Step SSP' in Table~\ref{tab:MCMC_SSP_metric}.

As seen in the Table this two-step SSP algorithm exhibits some
modest quantitative improvement in the derived kinematics compared to
the simple SSP algorithm.  This demonstrates that the irregular
$\chi^2$ space of the solution affects the reliability of the
kinematic decomposition unless reasonable initial conditions can be
placed on the kinematics. However, some observable differences
persist in the velocities of the young component. The discrepancy
suggests a lack of full convergence of the two-step SSP algorithms to
the global minima of the $\chi^2$ space identified by the MCMC
analysis. We find that although \pfrac\ determined in the first step
of the two-step SSP algorithm correlates well with that determined by
the MCMC analysis, the relation has significant scatter; better
initial conditions on \pfrac\ could also be
beneficial. Previously \citet{Katkov_2011} used a Monte-Carlo
technique to demonstrate that when conducting a spectral decomposition
of co-spatial components an incorrect estimation of the relative light
contribution of the components could bias the results.

To test if constraining the relative weights of the young and old
components further improves results we provide a data-driven
constraint on \pfrac\ by using the template weights from the {\it
first} full spectrum fit with a single stellar component. From this
weight distribution \pfrac\ is set in the {\it second} fitting step to
the sum of the weights associated with the young SSP models ($\leq$
1.5 Gyrs) over the sum of the weights of all SSPs. This constraint is
imposed independently for each spaxel.  We then add a {\it third} full
spectrum fitting step to our SSP algorithm (i) where we update the
initial conditions for the kinematics to that measured in the second
step, but (ii) we remove the constraint on \pfrac\ for the fit. The
motivation for this step is to assume that the kinematic solution
obtained in the second step is close enough to the global minima that
the final step can free the fit to converge onto the global minima
without constrains put in place on \pfrac. We refer to this
algorithm as `3-Step SSP' in Table~\ref{tab:MCMC_SSP_metric}. It is
immediately evident that across the board, all of the metrics improve
significantly, and hence we adopt this as our final SSP algorithm.
Figure~\ref{8138_12704_bp} and Appendix~\ref{sec:maps} show the
velocity and velocity dispersion fields for MCMC, simple and 3-step
SSP algorithms.

The aim of using a multi-step approach to nudge \pPXF\ toward a global
minimum using iteratively improved initial conditions appears to be on
target. In contrast, the initial two-step SSP algorithm had less
refined initial conditions on the kinematics produced nosier kinematic
maps for the two stellar components -- compared both to this
three-step algorithm and the MCMC results.

\subsection{SSP Algorithm Summary}
\label{subsec:ssp_algo_summary}

Our final SSP algorithm used to measure the two-component AD is a
three-step process as follows:

\begin{enumerate}
	
\item Use full spectral fitting with \pPXF\ with one stellar and one gas
(emission-line) kinematic component. The entire suite of SSP models
are used as templates for the single stellar component. Since each SSP
is normalized, the resulting spectral fitting weight distribution is
light-weighted. The gas emission-line components assume Gaussian line
profiles, and the kinematics for all lines are tied to a single value
for velocity and dispersion per spaxel. During this fit, the initial
conditions for the kinematic for all components are zero velocity with
respect to the NSA redshift and velocity dispersion of 210~\kms.  All
stellar kinematic fitting is done assuming a Gaussian LOSVD and
multiplicative Legendre polynomials of order eight are used to match
continuum shapes between observed and template spectra. The resulting
kinematics and weight distribution of this fit provide informed
constraints for the second fitting step.
	
\item Use full spectral fitting with \pPXF\ with two stellar
kinematic components (young and old) and one gas (emission-line)
kinematic component.  The young kinematic component is restricted to
use only SSP models with ages $\leq$ 1.5 Gyrs, while the remainder are
used as templates for the old component; \pfrac\ is constrained by the
ratio of these weights derived in the first step.  The initial
kinematics of the young component is set to the derived gas kinematics
from the first fit and the old component are set to those of the
single stellar component. The kinematics of the gas component in this
second fit is constrained to that derived by the previous fit, however
the fitting code is free to optimize the relative weights of the
different emission line features. Similarly, \pPXF\ is allowed to
optimize the relative weights of SSPs within young and old age bins,
but the sum of these weights is constrained by \pfrac\ from the first
fit.

\item Repeat the full spectral fitting in the same setup as the
previous step with the initial kinematics of the stellar components
updated to those derived in step 2, but with no constraint
on \pfrac. This full spectrum fitting provides the final measured
kinematics for the two stellar components, and the weights in the two
components provides the \pfrac\ in the spaxel.

\end{enumerate}


\section{Systematics from Template Mismatch}
\label{sec:mismatch}

\begin{figure}
  \includegraphics[width=1\linewidth]{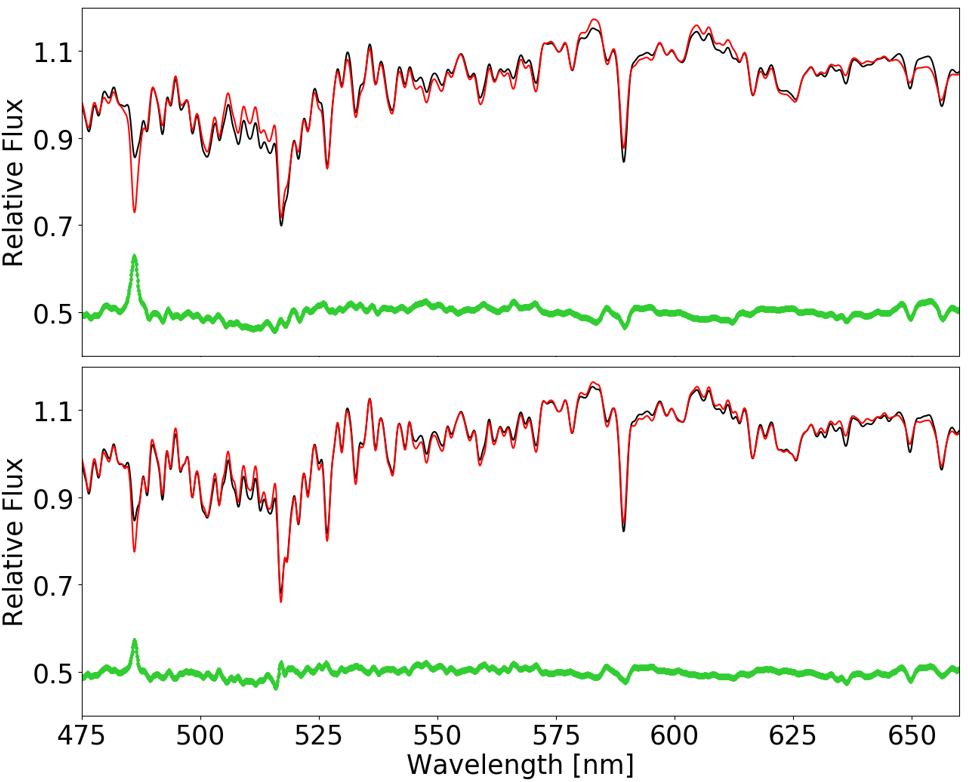}
  \caption{Demonstration of template mismatch in an extreme scenario
    in the top panel, described in text, for an old, super-metal rich
    stellar population (black curve) fit by only young and relatively
    metal poor SSPs (red curve). Residuals (green curve) illustrate
    the significant features associated with Balmer absorption. The
    bottom panel presents our more typical case of template mismatch,
    as described in the text.}
  \label{tm_demo}
\end{figure}

Template mismatch \cite[][and references therein]{Rix92,Statler95} can
lead to potential systematics in the derived kinematics caused by the
inability of the templates to accurately reproduce the true galaxy
spectrum. This may arise due to the lack of knowledge of the stellar
IMF, incomplete parameter coverage in the stellar libraries making up
the the SSPs, or systematic errors in the isochrone of the stellar
population synthesis models. Notably, the SAURON survey switched
  from using SSPs to a subset of stellar-library template
  \cite{Cappellari2007} principally to minimize template-mismatch
  errors on the derived kinematics. Previous analyses have focused on
the impact of template mismatch on systematics in velocity dispersion
and higher moments, while we are primarily interested in
velocities. It is reasonable to expect that the impact of template
mismatch on velocities will be relatively small, but since the
asymmetric drift signal is also small, it is worthy of exploration.

To create a situation where template mismatch induced systematics in
the recovered velocities we convolved the oldest and most metal rich
SSP model spectrum (14 Gyrs, 0.26 [Z/H]\textsubscript{\(\odot\)}) with
an LOSVD prescribed by a velocity of 0~\kms\ and a dispersion of
250~\kms.\footnote{Results using a 13 Gyr model are comparable.} We
then attempted to derive the kinematics of this mock by fitting the
spectra using only SSPs with ages less than 4 Gyrs and metallicities
less than solar. Since the purpose of this test is to
  demonstrate the effect template mismatch has on the measured
  kinematics, we do not include any polynomials during these
  tests. Figure~\ref{tm_demo} illustrates the quality of the best fit
(red) against the mock (black) in a feature-rich region including
H$\beta$, MgI, and NaD (recall, however, the full fit is between 360
and 745 nm). As one would expect the H$\beta$ feature at 486.2 nm is
very poorly fit while the metal features, NaD at 589.0 nm and 589.6 nm
and MgI at 517.6 nm, are significantly better fit but with notable
features being visible in their residuals. The best fit clearly does
not reproduce the mock spectra very well, so it is unsurprising the
recovered velocity from the this fit is off (in this instance) by
$\sim$18~\kms\ off the mock's model value. The bottom panel of
Figure~\ref{tm_demo} demonstrates a significantly better fit to the
same model spectrum for a more typical case of template mismatch where
we allow fitting the mock spectrum with all sub-solar metallicity,
$<$~10Gyr old templates.

\begin{figure}
  \includegraphics[width=1\linewidth]{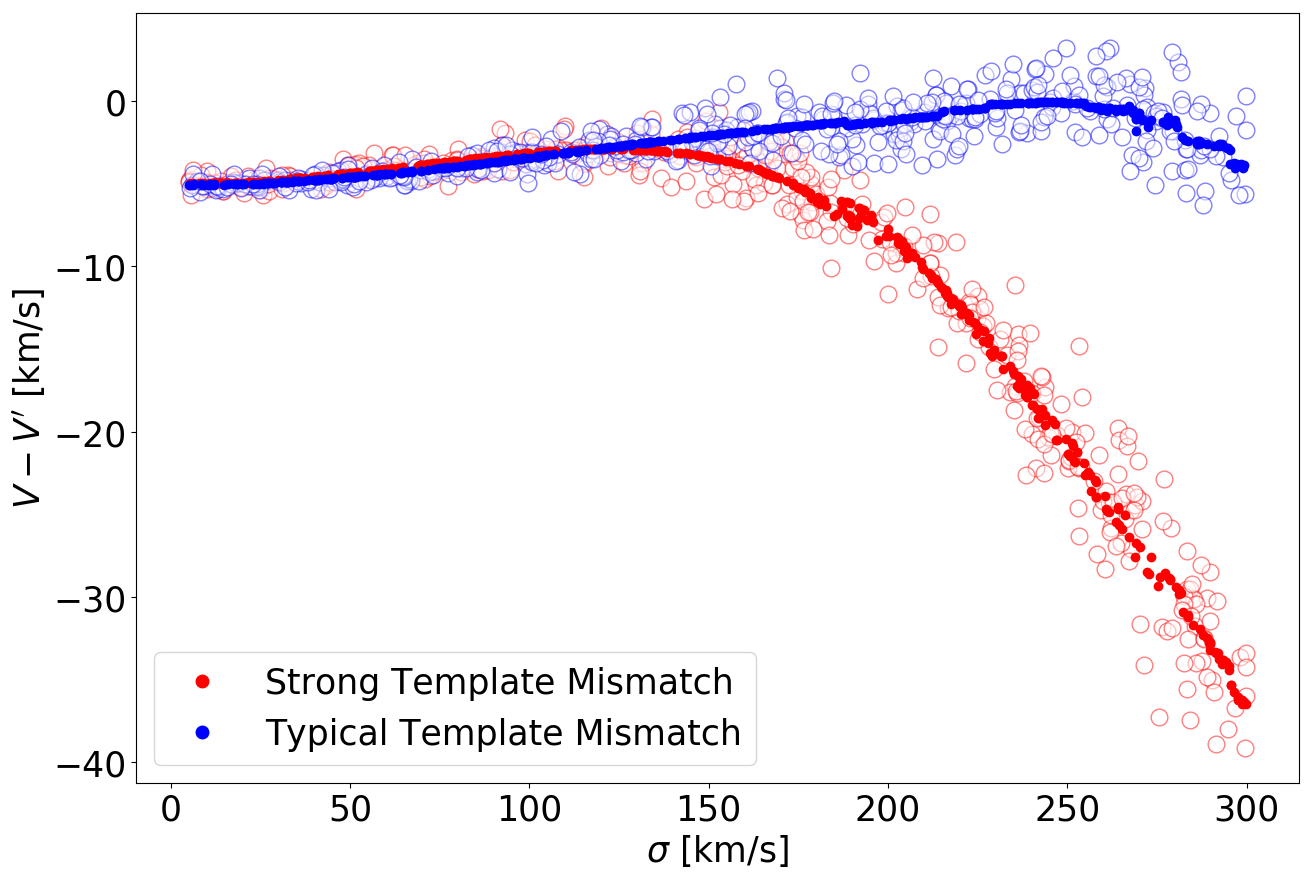}
  \caption{Effect of template mismatch on recovered velocity for our
    extreme scenario from Figure~\ref{tm_demo} (red points; 'Strong
    Template Mismatch') as a function of the mock velocity
    dispersion. A more typical template mismatch scenario, as
    described in the text, is also illustrated (blue points). Open
    circles represent the same cases with noise added such that the
    spectra have S/N of 100 per pixel where each pixel sample 70~\kms.}
	\label{tm_test}
\end{figure}

We quantified the effect on the systematic errors in recovered
velocity for this extreme scenario that induces strong template
mismatch. Results for 500 mock spectra with different LOSVDs selected
randomly with velocities between $\pm100$\kms, to remove any effect
due to pixelization, and velocity dispersions between 5 and
300~\kms\ are summarized in Figure~\ref{tm_test}.  The plot suggests
there is bias in the recovered velocities that correlates with the
mock velocity dispersion. This is likely due to the blending of
spectral features as the mock velocity dispersion increases, thereby
reducing sharp spectral features and the signal available to measure
velocity. The cause of the correlation between measured velocity
and velocity dispersion however is unclear.

recovered velocities also are biased towards
under-estimating the velocity of the mock, for reasons which are not
clear.

Were we to use this extreme case as a guide, we would note that the
upper values for the stellar dispersion in most disk galaxies should
be $<150$~\kms\ (for example 50~\kms\ would be typical of the old,
thick disk of the Milky Way at the solar circle). The effect of
template mismatch on velocity is, in this extreme case, $\sim
-5$~\kms, as reckoned from Figure~\ref{tm_test}. Since this case is
extreme, in Figure~\ref{tm_test} we also present the more typical case
for template mismatch described above. In this more typical case the
effect of template mismatch is substantially reduced relative to the
extreme case at large dispersions, but is comparable for dispersions
below 150~\kms. Examining residuals observed for fitting real galaxies
with SSP models indicates that in practice template mismatch is more
in line with our typical case. The results summarized in
Figure~\ref{tm_test}, however, shows that the resulting template
mis-match is comparable in the velocity dispersion regime of interest.

It is likely that the systematics found here are underestimates when
fitting real galaxies because in this analysis both the mocks and
fitting templates are based on the same SSPs. The differences in our
results for mock and real spectra, as illustrated in our SSP algorithm
performance, supports this conclusion. Since we are attempting to
measure a small differential velocity between different stellar
populations in integrated starlight, this potential for
template-mismatch velocity systematics cannot be ignored. It is
widely accepted that using an empirical stellar library that
efficiently samples the stellar parameter space -- rather than SSPs --
diminishes the effect of template mismatch when fitting for observed
galaxy spectra. We have therefore taken the extra step to create an
algorithm based on stellar libraries in the following
Section~\ref{sec:stellar_algo}. We use this algorithm in Section~\ref
{sec:results} to compare to our simpler, SSP-based algorithm using
real data.


\begin{figure*}[ht!]
  \centering
	\includegraphics[width=0.95\textwidth]{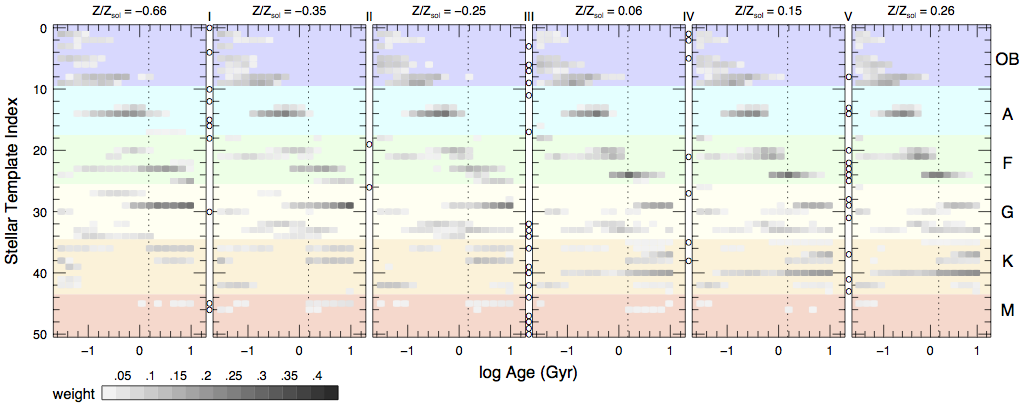}
	\caption{IndoUS-subset stellar template weights derived by
          \pPXF\ fits to MUISCAT SSPs of different age (x-axis)
          and metallicity (panels, left-to-right, as labeled). Stellar
          templates are sorted by effective temperature (decreasing
          top to bottom), with spectral types (labeled B,A,F,G,K,M at
          far right) indicated by horizontal color bands.  Luminosity
          classes (labeled I,II,III,IV,V at the top) are given by the
          open circles in the interstices between panels. The dotted
          line in each panel is at 1.5~Gyrs, the demarcation between
          young and old stellar populations. Note the expected trends
          with diminishing hot stars and increasing cool stars as a
          function of age; age-metallicity trends in the mix of F
          through K stars; and the relative dearth of the coolest (late
          K and M-type) stars even at the highest metallicity. The
          IndoUS-subset is described in Appendix~\ref{app:stellar_lib}
          and listed in Table~\ref{tab:stellar_subset}.}
	\label{Weight_Dist_SSP}
\end{figure*}

\section{Stellar Library-based Algorithms to Measure AD}
\label{sec:stellar_algo}

This section presents key steps in developing algorithms based on
stellar libraries assessed with the same metrics defined in
  Section~\ref{sec:ssp_algo}. We construct a representative set of
empirical stellar spectra (Section~\ref{sec:stellib_templates}),
incorporated first into a set of simple algorithms to disentangle
kinematics of young and old stellar components of mock spectra
(Section~\ref{sec:stellar_simple}). To guide development, we
  identify the origins of the strengths and deficiencies in these
  algorithms specific to using stellar libraries rather than SSPs.
As we found for SSP-based algorithms, initial conditions for the
kinematics are important in directing \pPXF\ away from local $\chi^2$
minima in application to real data
(Section~\ref{sec:stellar_twostep}).  In the context of stellar
library algorithms these local minima are exacerbated by the
presence of cooler stars in both young and old populations. To
  overcome this problem we implement constraints based on
expectations from stellar evolution in
Section~\ref{sec:stellar_algo_evol}. These constraints are minimally
imposed to retain the flexibility gained using stellar spectral
libraries rather than SSPs.  What emerges is a well-motivated and
robust two-step algorithm that disentangles the kinematics of young
and old stellar components and provides results qualitatively on par
with the final SSP algorithm. The final stellar library algorithm is
summarized in Section~\ref{subsec:stellib_algo_summary}.

\begin{deluxetable}{ccc}
\tablewidth{0pt}
\tablecaption{Stellar Template Sets}
\tablehead{
    \colhead{Set} &
    \colhead{Young templates} &
    \colhead{Old templates}
}
\startdata
A & OBA   & F-M \\
B & OBAFG & F-M \\
C & all   & F-M \\
D & all   & K-M
\enddata
\label{tab:template_set}
\end{deluxetable}

\subsection{Young and Old Stellar Templates}
\label{sec:stellib_templates}

We use a representative subset of 51 stars from the empirical
Indo-US Library of Coud\'{e} Feed Stellar Spectra
\citep{Valdesetal2004}. As described in Appendix~\ref{app:stellar_lib}
these are well suited for our purposes of analyzing MaNGA
  spectra. Similar to what was done for the SSP algorithms, we begin
  by defining the templates for the two age components being fit.

In the context of SSPs (Section~\ref{sec:young_old}) we define the
young-component templates to be those containing significant Balmer
features. For a stellar library, however this definition excludes
stars cooler than late-F which we know are present in young stellar
populations for any plausible present-day stellar initial mass
function. This is quantified in Figure~\ref{Weight_Dist_SSP}.
Since the younger populations have both hot and cool stars, with metal
lines contributed to the spectra by the latter, we select three ranges
that always include the hottest stars but extended to progressively
cooler limits in cases labeled A, B and C. Our aim is to include the
smallest possible range of cool stars to minimize degeneracy with the
older stellar populations, and hence simplify the fitting algorithm.

The choice of stellar templates for the old component is simpler since
hot, Main Sequence stars are short lived and hence only present in
young stellar populations, where they dominate the spectral continuum
in the blue and visible portion of the spectrum.  We therefore exclude
hot stars (hotter than F) from the templates for the old component,
effectively ignoring the possibility of extreme (very metal poor) blue
Horizontal Branch stars from contributing significantly to the
integrated light of a mixed stellar population.  In one case (D) we
restrict the older population to have only K and M type stars to
understand if there is any sensitivity to the giant-branch effective
temperature.

Table~\ref{tab:template_set} summarizes the four distinct combinations
of stellar templates we explore for the young and old stellar
components.  These template combinations place no restrictions on
luminosity class.

\subsection{Simple Stellar Library Algorithms}
\label{sec:stellar_simple}

As a starting point for the development of our algorithm, we test the
following simple algorithms to identify pitfalls to be ameliorated by
more complex techniques. These pitfalls highlight subtle but important
aspects of fitting multiple kinematic components where velocities are
offset by values comparable to the spectral resolution.
There are no constraints on \pfrac\ in any of these algorithms.

\begin{figure*}[ht!]
\centering
\includegraphics[width=0.9\textwidth]{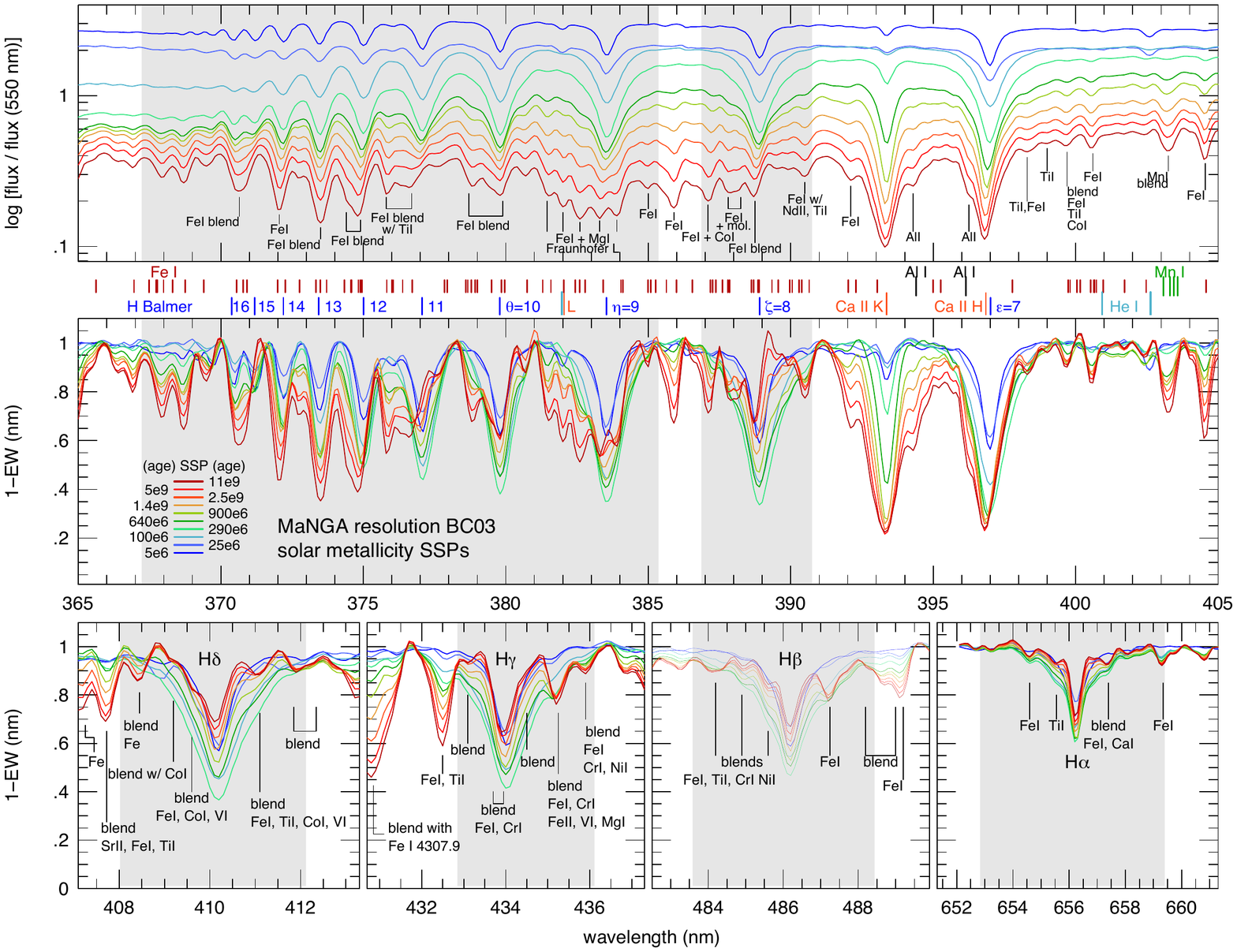}
\caption{Balmer fitting regions (see text) shown as grey shaded
  regions under solar-metallicity and solar abundance SSPs over a
  range of ages from \citet{Bruzual03}. Ages are given in the
  key. Model resolution is comparable to MaNGA data. In the bottom two
  rows spectra are normalized by their pseudo-continuum so the
  intensity has units equivalent to 1-EW, where EW is the equivalent
  width (in nm). In the top spectra are normalized at 550 nm to
  illustrate dramatic range in relative flux as a function of
  age. Balmer features and strong metal lines are marked.}
\label{fig:balmer_ad}
\end{figure*}

\begin{enumerate}

\item Full Spectrum Fitting ({\it Full}): \pPXF\ fits two kinematic
  components to the provided (mock or real) spectrum over the full
  wavelength range using the above mentioned template sets.

\item Feature Fitting ({\it Feature}): Based on the knowledge that we
  expect to see hot stars only within young stellar populations, the
  kinematics of the young component are fit with \pPXF\ using
  only wavelength regions of $\pm$1500~\kms\ about Balmer lines up to
  n=18 in the spectra while masking out the remainder, as shown in
  Figure~\ref{fig:balmer_ad}. Because of the strong contamination from
  Ca H, we exclude H$\epsilon$; and because of the G band (CN), we
  only use $\pm$500~\kms\ for H$\gamma$. Given the $\pm$1500~\kms\ velocity
  width, the region contain H$\eta$ (n=9) and above form a continuous
  band from 385.36 to 367.21 nm. During this fit, two kinematic
  components are still applied in \pPXF\: despite the dominance
  of Balmer features in these regions for young populations, there
  remains an imprint of the kinematics of the older population in
  these regions, both in the Balmer lines as well as the metal lines
  within these windows. This fit is then followed by a fit using
  \pPXF\ that {\it excludes} the Balmer regions. During this
  second fit, the velocity moments of the young component are kept
  fixed to the values derived from the first step.

\end{enumerate}

\begin{deluxetable*}{lcrrrrrrrrrrr}[ht!]
\tablewidth{0pt}
\tabletypesize{\footnotesize}
\tablecaption{SL Algorithm Performance Metrics Referenced to MCMC Results}
\tablehead{
 \colhead{Algorithm} &
 \colhead{S/N} &
 \multicolumn{2}{c}{$\Delta V_y$} &
 \colhead{} &
 \multicolumn{2}{c}{$\Delta V_o$} &
 \colhead{} &
 \multicolumn{2}{c}{$\delta V_{\rm Algo} / \delta V_{\rm mock}$} &
 \colhead{} &
 \multicolumn{2}{c}{$\Delta{\rm \pfrac}$} \\  [0.05in] \cline{3-4} \cline{6-7} \cline{9-10} \cline{12-13}
 \multicolumn{13}{c}{} \\  [-0.08in]
 \colhead{} &
 \colhead{Range} &
 \colhead{med} &
 \colhead{$\sigma_{\rm MAD}$} &
 \colhead{} &
 \colhead{med} &
 \colhead{$\sigma_{\rm MAD}$} &
 \colhead{} &
 \colhead{med} &
 \colhead{$\sigma_{\rm MAD}$} &
 \colhead{} &
 \colhead{median} &
 \colhead{$\sigma_{\rm MAD}$} \\
 \colhead{} &
 \colhead{(\kms)} &
 \multicolumn{2}{c}{(\kms)} &
 \colhead{} &
 \multicolumn{2}{c}{(\kms)} &
 \colhead{} &
 \colhead{} &
 \colhead{} &
 \colhead{} &
 \colhead{} &
 \colhead{} \\
 \colhead{(1)} &
 \colhead{(2)} &
 \multicolumn{1}{r}{(3)} &
 \multicolumn{1}{r}{(4)} &
 \colhead{} &
 \multicolumn{1}{r}{(5)} &
 \multicolumn{1}{r}{(6)} &
 \colhead{} &
 \multicolumn{1}{r}{(7)} &
 \multicolumn{1}{r}{(8)} &
 \colhead{} &
 \multicolumn{1}{r}{(9)} &
 \multicolumn{1}{r}{(10)}
}
\startdata
Simple SL       & [10,35] & 13.76 & 51.44 && 4.02 & 14.21 && 0.90 & 1.18 && 0.09 & 0.12 \\
Simple SL       & [35,50] & 21.50 & 43.47 && 8.21 & 10.88 && 0.88 & 1.38 && 0.13 & 0.07 \\
Simple SL       &   $>$50 & 69.56 & 87.29 && 5.23 &  8.82 && 1.05 & 2.36 && 0.05 & 0.05 \\
 \multicolumn{13}{c}{} \\  [-0.08in]
2-Step 1-Bin SL & [10,35] &  1.70 & 29.94 && 3.70 & 15.27 && 0.73 & 0.81 && 0.05 & 0.12 \\
2-Step 1-Bin SL & [35,50] &  1.50 & 17.25 && 6.13 &  7.86 && 1.00 & 0.40 && 0.10 & 0.07 \\
2-Step 1-Bin SL &   $>$50 &  0.90 & 24.86 && 5.46 &  6.69 && 1.09 & 0.74 && 0.04 & 0.06 \\
 \multicolumn{13}{c}{} \\  [-0.08in]
2-Step 3-Bin SL & [10,35] & -2.02 & 26.27 && 6.17 & 18.38 && 0.62 & 0.87 && 0.01 & 0.08 \\
2-Step 3-Bin SL & [35,50] &  0.66 & 11.50 && 5.44 &  6.54 && 0.98 & 0.35 && 0.05 & 0.05 \\
2-Step 3-Bin SL &   $>$50 &  1.45 & 26.45 && 7.56 &  5.54 && 1.16 & 0.70 && 0.03 & 0.04
\enddata
\label{tab:MCMC_SL_metric}
\tablecomments{SL algorithms (column 1) are defined in
  Section~\ref{sec:stellar_simple} and \ref{sec:stellar_algo_evol}.
  Columns (2) through (10) are defined identically to Table~\ref{tab:MCMC_SSP_metric}
using the same spaxels.}
\end{deluxetable*}

Based on our tests using mock spectra
  (Appendix~\ref{sec:stellar_simple_mock_performance}) and
  Table~\ref{tab:simple_mocks_metric}), we find the performance of
these two algorithms is comparable. This confirms expectations
  about where most of spectral information is stored for discriminating
  between young and old stellar populations. {\it For simplicity and
    consistency with the SSP algorithms we adopt the Full algorithm.}

Table~\ref{tab:simple_mocks_metric} also reveals important performance
differences between templates: Velocity systematics for the young
component are minimized for template set C, while velocity systematics
for the old component are minimized for template set B. Perhaps this
is unsurprising given Figures~\ref{fig:hot_star_relative_wts} and
\ref{Weight_Dist_SSP}. However it is concerning that neither template
set optimizes {\it both} young and old component systematics. A
further concern arises with template sets C and D. In these cases
there are many occurrences where \pfrac\ \textgreater 0.85, i.e., the
best-fitting solution yields too large a young-component contribution.
This is due to the degeneracy in the contribution of cool stars in the
young and old components; the \pPXF\ likelihood minimization falls
into a local minima where the stellar kinematics of the mock spectra
is almost entirely modeled by the young component. An example of
  this is given in Appendix~\ref{sec:stellar_simple_mock_performance},
  Figure~\ref{fig:high_frac_example}. We to conclude that template set
  B is the best compromise. This leaves us with performance results,
  based on mocks, that are somewhat unsatisfactory.

\subsection{Basic Two-Step Stellar Library  Algorithm}
\label{sec:stellar_twostep}

Following the success of adding additional steps to improve the
kinematic initial conditions with our SSP-based algorithms, we further
developed the {\it Full} algorithm using the template set B by adding
a step to provide initial conditions in an identical fashion as our
two-step SSP algorithm. Foreshadowing further development, we refer to
this algorithm as `2-Step 1-Bin SL', where the `1-Bin' designation
refers to the grouping of the templates as a single unit for each of
the stellar components.

An evaluation of the performance of this algorithm using mocks
(Appendix~\ref {sec:stellar_simple_mock_performance},
Table~\ref{tab:SL_realistic_mocks_metric}) indicates no improvement,
but we suspect this is due to systematics between the mocks generated
by SSPs and the fits using the stellar library. We turn instead, and
henceforth, to a comparison with the MCMC for real data. Velocity and
velocity dispersion maps (Figure~\ref{8138_12704_bp} and
Appendix~\ref{sec:maps}) show marked qualitative improvement in the
smoothness of the kinematic maps between our one-step and two-step
stellar library algorithms. Table~\ref{tab:MCMC_SL_metric} quantifies
this contrast, most pronounced for the young component velocities.
There also are performance improvements also for the old-component
velocities.

However, further inspection of the kinematic maps reveals velocities
and velocity dispersions are quite noisy for the young component of
galaxies with lower mass ($<2\times10^{10} {\rm M}_\odot$) or bluer
color ($({\rm NUV}-i)_0<3$).  (These properties are correlated in
general, and particularly so for our small sample). It is also the
case that for $\sim$33\% of the spaxels with $10<S/N<35$ both the
simple and two-step 1-bin algorithms find very high values of
\pfrac\ compared to MCMC results. This is indicative of an incorrect
allocation of cool stars to the young kinematic component in the
stellar library algorithms, and suggests further algorithm improvement
is desirable.

\subsection{Two-Step Stellar Library Algorithm with Constraints 
from Stellar Evolution}
\label{sec:stellar_algo_evol}

Since problems with our stellar library algorithms arise due to
incorrect allocation of cool stars to the young kinematic component,
we need an objective mechanism to appropriately assign weights for
these stars in young and old components. The above stellar
  library algorithms attempt to steer the allocation of weights via a
  prescriptive template set (B). Additional constraint on the relative
  weights of the young and old component (\pfrac) fails to improve
  algorithm performance with nearly all our quantitative
  metrics. Further improvement requires a better guided approach.

An astrophysically motivated mechanism to properly weight
  stellar contributions to young and old stellar populations is to
assign priors based on what we would expect from stellar evolution. A
limiting case is just to use SSPs but since the entire exercise here
is to not restrict ourselves the specific weights produced by the
mapping of theoretical isochrones to a template library, we aim to
impose a less restrictive scheme that still is informed by the
relative weights of different stellar types as we would expect from
stellar evolution. As we show, it is possible to do this by coarsely
categorizing stellar types in effective temperature while leaving
\pPXF\ the freedom to optimize template weights within these coarse
(broad) categories.

Figure~\ref{Weight_Dist_SSP} shows the derived weights for our
  stellar library from fitting SSPs with \pPXF; these results broadly
match our expectations for stellar evolution and the impact of
metallicity on, e.g., the temperature of the giant branch.  The
youngest populations, dominated by the hottest stars, indeed have most
of their weight in B and A templates, but do contain significant
weight in cooler stars. As the populations age (and the Main Sequence
burns down while the asymptotic and then red giant branches are
populated), the relative weight of the cooler stars gradually
increases and the weight of the hotter stars decrease. The oldest SSPs
are entirely dominated by the coldest stars, but the specific age
where this occurs depends on metallicity because of the temperature of
the stars on the horizontal and giant branches. Hence by constraining
the relative weights of the templates used in the kinematic fitting
such that they are at least consistent with that expected from stellar
evolution at some granularity, we could disentangle the the relative
weight of the cool stars in the two kinematic components in our
fit. Just what granularity should be adopted is one question we
answer.

To constrain the relative weights of the stellar templates in the two
kinematic components we use \pPXF\ to do a full spectrum fit of the
target spectrum with SSPs for a {\it single} stellar kinematic
component (plus a second component for gas, if the target is a real
galaxy spectrum; see Section~\ref{sec:methods}). The resulting SSP
weight distribution is translated into weights of our stellar
templates for a young and old population via a reference look-up table
containing the stellar weights derived from fitting our Indo-US
stellar subset to each SSP (Figure~\ref{Weight_Dist_SSP}). The weights
are divided between young and old populations by summing over SSP
weights younger or older than 1.5~Gyrs.

While this scheme provides reasonable estimates for individual stellar
template weights, the detailed result can be strongly effected by
noise in the galaxy spectra. Further, fixing the relative weights is
almost equivalent to using the SSP spectra as templates themselves,
which defeats the purpose of this algorithm. To relax these
constraints we divided the stellar template into temperature bins.
The purpose of this division is to constrain the relative weights only
{\it between but not within} these bins for young and old populations
during the full-spectrum fitting of the galaxy spectrum. The detailed
distribution of weights for the individual stellar templates within
each bin is a free parameter, tuned (by \pPXF) to optimize the
likelihood of the solution.

\begin{figure*}[ht!]
  \center
  \includegraphics[width=0.9\linewidth]{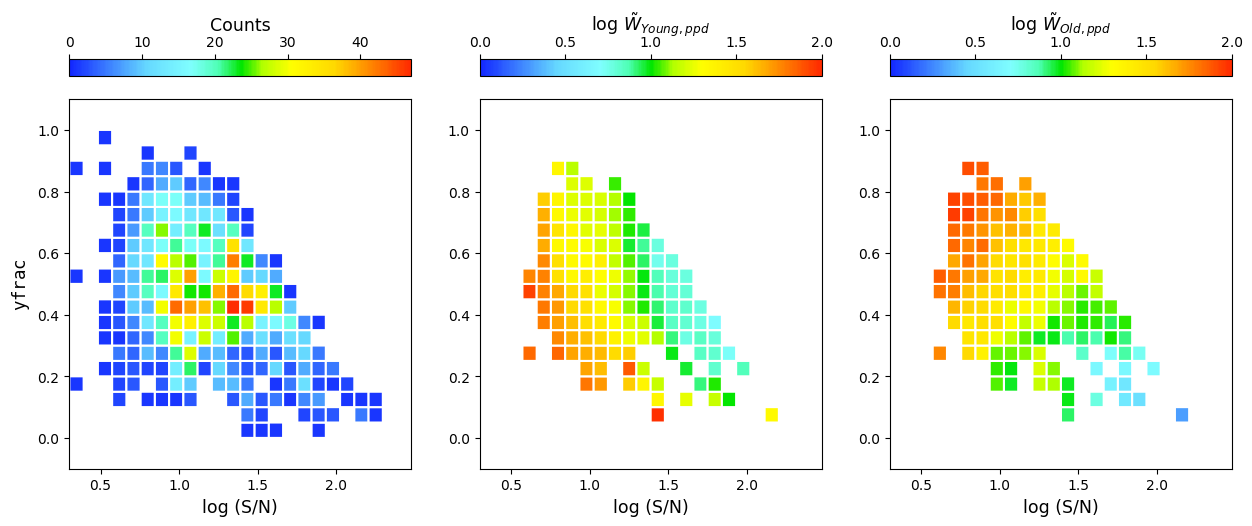}
  \caption{Visualization of random errors in two-component stellar
    velocities. The left panel shows the spaxel count for all galaxies
    in our sample, culled as described in Appendix~\ref{app:mcmc_ppd}
    for $\delta V < 0$ and $\delta \sigma < 0$ (preferentially spaxels
    at low S/N), as a function of S/N (measured per \AA) and \pfrac.
    Middle and right panels show the median value of the PPD width
    ($\tilde{w}_{\rm ppd}$) as a function of \pfrac\ and S/N for
      young and old stellar populations, respectively.}
  \label{fig:random_errors}
\end{figure*}

We divide templates into three temperature bins (each contains stars
with a range of luminosity classes and metallicity) motivated by the
fact that the young and old kinematic components both have cooler
stellar types (F-M) as common templates, but the mix of
intermediate-temperature stars (F-G) changes strongly with population
age (see Figure~\ref{fig:hot_star_relative_wts}), while only the young
component has the hottest stellar types (O-A).  The three bins are:
Hot (which contains templates of stellar types O, B and A with $T_{\rm
  eff}$ roughly above 9000$^\circ$ K), Intermediate (F, G with $T_{\rm
  eff}$ roughly between 5000$^\circ$ and 9000$^\circ$ K) and Cold (K
and M roughly below 5000$^\circ$ K).  Since the unrealistic
distribution of weights for the cool stellar types is causing the
catastrophic failure in \pfrac when deriving the kinematics of
the young and old kinematic components for our simple and 2-step
  1-bin algorithms, this three-bin division fix the
hot~:~intermediate~:~cool contributions in each kinematic component to
reasonable values consistent with both the observed spectra and rough
expectations from stellar evolution. Consequently we can dispense
  with the restrictions of the template sets A-D from
  Section~\ref{sec:stellib_templates}. We refer to this as our 
  2-step 3-bin stellar library algorithm.

To implement constraints on the relative weights of the bins we
utilize features in \pPXF\ that (i) allow the relative weight to be
fixed between multiple kinematic components, and (ii) grants the user
the ability to tie together the kinematics of different
components. Effectively we fit a set of template bins with tied
kinematics for each stellar component.  By fixing the relative weights
of the bins of the young and old components we can avoid unrealistic
template weights while giving freedom for template
optimization. Nominally \pPXF\ limits users to fixing the relative
weight between only two kinematic components. For this study we
modified the \pPXF\ code to disable this limit, enabling us to split
the templates for each component into as many bins as needed.

We find the addition of these multiple constraints on \pfrac\ yield a
significant qualitative improvements in the smoothness of the velocity
and velocity dispersion maps for the lowest-mass and bluest galaxies
in Appendix~\ref{sec:maps}. As seen in these Figures, this stellar
library algorithm provides comparable {\it qualitative} performance to
the SSP algorithm, and based on this we adopt this three-bin
approach. Further comparison with the SSP algorithm is presented in
Section~\ref{sec:results}.

\subsection{Stellar Library Algorithm Summary}
\label{subsec:stellib_algo_summary}

The final algorithm (SL hereafter) can be summarized as a two-step
process. The first step is identical to what is described in
Section~\ref{subsec:ssp_algo_summary}; it is a full-spectrum fit for a
single stellar kinematic component using SSPs. This step is used to
constrain the relative weights of empirical stellar spectra via a
decomposition of the SSPs into relative weights for the empirical
stellar spectra.  Henceforth SSPs are not used. A second and final,
two-stellar-component fit (also full-spectrum) uses the empirical
stellar spectra with weights broadly constrained in three
effective-temperature classes (Hot, Intermediate, and Cool).  The
second step uses a modified version of \pPXF\ with fixed fractional
contributions to the total fit for two stellar {\it kinematic}
components, each with three different sub-populations. By definition,
the three sub-populations for each kinematic component share the same
kinematics. In detail:

\begin{enumerate}

\item The relative weights for the Hot, Intermediate and Cool stellar
  bins for each (young and old) kinematic components are derived from
  the decomposition of the SSPs into individual stellar templates, as
  illustrated in Figure~\ref{fig:hot_star_relative_wts}. The stellar
  template weights for the SSPs younger or older than 1.5~Gyr,
  respectively, are assigned to the young and old stellar components
  via the look-up table generated from the decomposition of each SSP
  (Figure~\ref{Weight_Dist_SSP}).

\item The initial kinematics of the young component are set to the
  derived gas kinematics from the initial fit, and the old component
  are set to those of the single stellar component.

\item The kinematics of the gas component in this second fit is
  constrained to that derived by in the first fit. The fitting code is
  free to optimize the relative fluxes of the different emission
  lines.

\item The fitting code is allowed to optimize the relative weights of
  the stellar templates {\it within} each temperature bin for each age
  component (young and old) but the {\it relative} sum of these
  weights is constrained by six \pfrac\ values defined above. This
  flexibility minimizes the effect of template mismatch.

\end{enumerate}


\section{Results}
\label{sec:results}

In this section we explore the random and systematic uncertainties in
the our methods to disentangle the tangential velocities of young and
old stellar populations in spiral galaxies observed by MaNGA.
Section~\ref{sec:measured_kin} presents the measured velocities
for our seven test galaxies.  Section~\ref{sec:eran} and
Section~\ref{sec:esys}, respectively, explore the behavior of random
and systematic uncertainties in our measurements.  Finally,
Section~\ref{sec:summ_measure} puts these uncertainties into the
astrophysical context of a two-component asymmetric drift signal.

\subsection{Measured Kinematics}
\label{sec:measured_kin}

The velocity and velocity dispersion maps in Figure~\ref{8138_12704_bp}
and Appendix~\ref{sec:maps} show that our different algorithms display
qualitatively similar kinematics. We focus further analysis on
kinematics from spaxels within a $\pm30^{\circ}$ wedge of each
galaxies' major axis, defined using the geometry in
Table~\ref{tab:geometry}.  Tangential velocities for each galaxy and
tracer are corrected for their own systemic velocity determined by
minimizing the difference in the approaching and receding components,
excluding data at radii less than 2 arcsec. We find these systemic
velocity values are consistent (at the $\sim$1~\kms\ level, on average)
with values estimated from full two-dimensional kinematic modeling of
a monolithic inclined-disk using the method described in
\citep{Westfall11,Andersen2013}. There are variations between systemic
velocities between tracers that are of order 3 to 5~\kms. These
differences are consistent with random errors, and they are negligible
in our overall error budget.

\subsection{Random Errors}
\label{sec:eran}

Our two algorithms (SSP, SL) both use \pPXF\ to minimize $\chi^2$.
Because \pPXF\ does not map the shape of the minima in $\chi^2$ space,
these algorithms do not provide a direct estimate of the measurement
uncertainty. In contrast the Markov-Chain Monte Carlo analysis we
conducted for our sample galaxies do provide this information.  From
the MCMC chains we have marginalized the posterior probability
distribution (PPD) for the kinematics (velocities, velocity
dispersions) and relative light contribution of the young and old
stellar populations of the galaxies.  As mentioned in
Appendix~\ref{app:mcmc_ppd}, the shapes of the PPD for individual
spaxels can sometimes be multi-modal, particularly at low S/N. We
quantify the PPD width using a standard deviation (rather than, e.g.,
the median absolute deviation) specifically to include the impact of
multi-modality.

We find that the uncertainty in the measured velocities of the two
components correlates with the S/N of the spectra and
\pfrac. Figure~\ref{fig:random_errors} illustrates these trends by
plotting the median width of the PPDs ($\tilde{w}_{\rm ppd}$) as a
function of \pfrac\ and S/N. The middle-panel shows that for the
lowest values of \pfrac\ the uncertainty in the measured velocity of
young stellar component is the lowest for a given S/N. This is
expected given that lower value of \pfrac\ corresponds to lower
contribution of the young stellar component to observed galaxy
spectrum. Likewise, the right-panel illustrates that at a given S/N
the reliability of the measured velocity of the old stellar component
is higher when \pfrac\ is lower, the relative contribution of the old
stellar component is higher. Both panels illustrate the expected
correlation between $\tilde{w}_{\rm ppd}$ and S/N. Since the
  systematic differences (seen in the next section) and the scatter
  between methods is smaller than the random errors exhibited in
  Figure~\ref{fig:random_errors}, these values are well suited for
  characterizing the random errors in appplication of either the SSP
  or SL algorithms.

For reference, at a S/N (\AA$^{-1}$) = 10 and \pfrac\ = 0.4, the
random errors in the young and old velocities are roughly 30~\kms,
while at S/N (\AA$^{-1}$) = 30 the errors are between 5 and
10~\kms\ for young and old components respectively. On-line
  tables of quantities plotted in the middle and left panels of
  Figure~\ref{fig:random_errors} are available.

\begin{figure*}
  \center
  \includegraphics[width=0.9\linewidth]{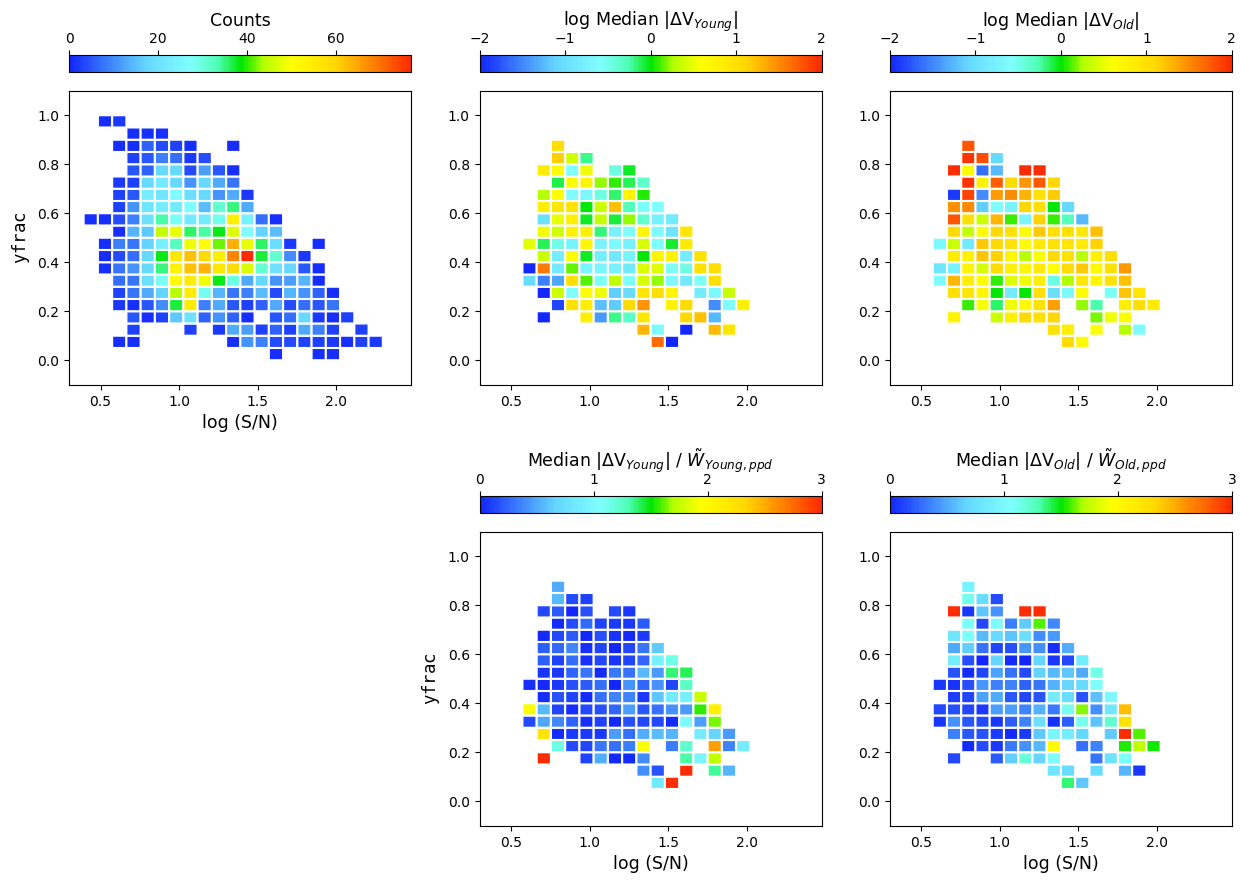}
  \caption{Visualization of systematics in two-component stellar
    velocities. Left panel shows the spaxel count for all galaxies in
    our sample culled such that $\delta V > 0$ for both of our two
    algorithms (SSP and SL) as a function of S/N (measured per
    \AA) and \pfrac\ (using the same \pfrac\ values as in
    Figure~\ref{fig:random_errors}).  Middle and right panels show
    spaxel-by-spaxel differences in young- and old-component stellar
    velocities from our two algorithms (SSP and SL), with the
    same culling and binning. Color-coding is for the signed-log of
    the absolute value of the median difference within each bin.  The
    bottom row shows the ratio of the absolute value of the median
    difference to the expected random error in each bin, taken as
    $\tilde{w}_{\rm ppd}$ from Figure~\ref{fig:random_errors}.}
  \label{fig:systematics}
\end{figure*}

\subsection{SL vs SSP: Systematics}
\label{sec:esys}

Figure~\ref{fig:systematics} compares the median difference (spaxel by
spaxel) between the derived velocities from our two algorithms binned
in S/N and \pfrac. This shows that the systematic differences are
small in absolute value, but nonetheless non-zero. The SL algorithm's
(Section~\ref{subsec:stellib_algo_summary}) measurement of the velocity of
the young component is consistent with that of the SSP algorithm
(Section~\ref{subsec:ssp_algo_summary}) across S/N and \pfrac\ to within
$\pm$8\kms, with an overall median difference of -5~\kms\ such that
the SL algorithm tends to find the young component rotating faster.
For the old component, the median difference is 7~\kms\ such that the
SL algorithm tends to find the old component rotating slower.These
systematic differences may reflect template mismatch or systematics in
the final \pfrac\ values of the two algorithms.  We will further
explore the cause of these systematics in future work, but it suffices
to conclude here that the systematics are small and have little
correlation with S/N or \pfrac. As shown in the bottom panels of
Figure~\ref{fig:systematics}, random errors almost always dominate
over systematic differences between algorithms {\it at the spaxel
  level}.

\subsection{Summary Two-component Asymmetric Drift Measurements}
\label{sec:summ_measure}

As a summary measurement of asymmetric drift for two independent
stellar components differentiated by age, we take the mean velocity
from our two algorithms (SSP and SL), spaxel by spaxel, as our
measure. For each spaxel we take $\tilde{w}_{\rm ppd}$ as the random
error, and half the difference between the velocities from the two
algorithms as the systematic error. When averaging the kinematics
derived from, e.g., a radial bin of spaxels, we assume the random
errors diminish with the usual quadrature averaging, while the
systematic errors do not. Systematic errors in the binned data are
taken as the median of the systematic error for individual spaxels in
the bin.

\begin{figure*}[ht!]
  \center
  \includegraphics[width=0.9\linewidth]{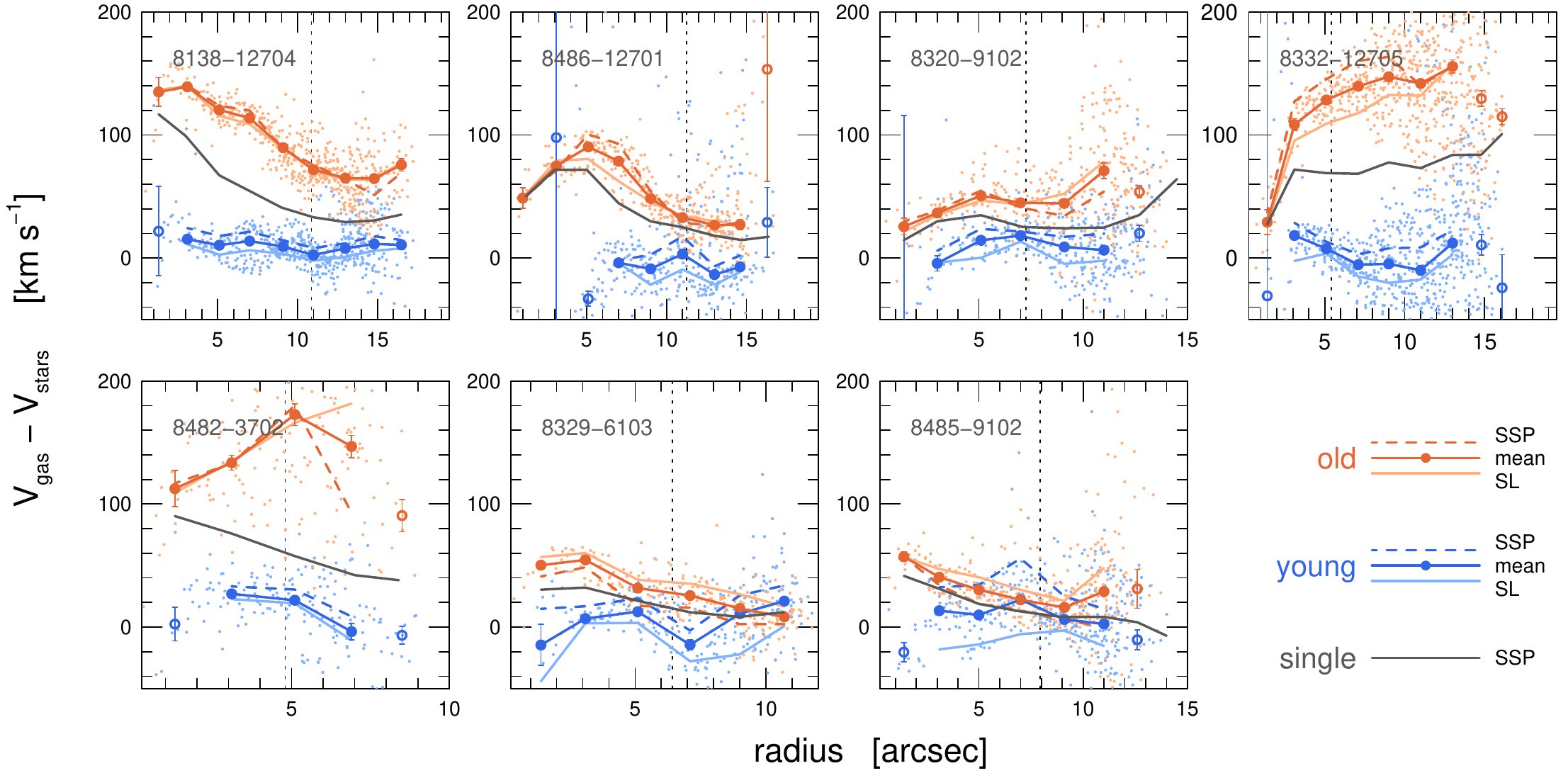}
  \caption{Asymmetric drift measurements for young and old stellar
    components of our galaxy sample as a function of radius, measured
    within $\pm30^\circ$ of the kinematic major axis, and corrected
    for projection in azimuth and inclination. Vertical dotted lines
    mark half-light radii. Small points represent individual spaxel
    measurements (blue and red for young and old components,
    respectively) while solid lines connecting large points represent
    the mean of these spaxel values in 2 arcsec radial bins. As given
    in the key, light and dashed lines, respectively, show the results
    for the SL and SSP algorithms alone, as an estimate of systematic
    error. For comparison the single-stellar-component asymmetric
    drift signal is shown as a black curve. Error-bars on large points
    represent the random error in the mean. Open circles represent
    young velocities where \pfrac$<$0.15 (inner radii) or young and
    old velocities where individual spaxel measurements have S/N$<$10
    (outer radii).}
  \label{fig:final_AD}
\end{figure*}

Figure~\ref{fig:final_AD} shows the implementation of this summary
scheme to a measure of asymmetric drift for young and old components
relative to the ionized gas velocity for the seven galaxies in our
sample. Line ratios (e.g., [OIII]$\lambda$5007/H$\beta$ versus
[NII]/H$\alpha$) are consistent with HII-like regions for most
galaxies at radii outside several arcsec; hence ionized gas velocities
should be close to the circular speed of the potential.  Results from
Section~\ref{sec:eran} indicate that young-component velocities are
unreliable for \pfrac$<$0.15, and both young- and old-component
velocities are unreliable when individual spaxels have S/N$<$10. These
radial points have been indicated with open symbols. Referring to
Figure~\ref{fig:emceespaxels}, we see that most of the observed radial
range for the galaxies in our sample are above these limits.

This final figure shows that we can indeed measure a distinct
asymmetric drift for two stellar components for this sample of
galaxies. Systematic errors dominate random errors in radial bins, but
in most cases and at most radii the errors are substantially smaller
than the difference between the asymmetric drift of young and old
components. While the difference between ionized gas and young
  stellar component is small, it is non-zero ($6\pm2$~\kms\ in the
  mean, with a median value of 8~\kms). In all cases the asymmetric
drift signals for young and old components straddle what is measured
for a single stellar component, as one would expect. The asymmetric
drift signal for the young component is generally small, with value of
order 20~\kms\ or less, also as one might expect. The net effect of
isolating the two population components by age is to elevate the
asymmetric drift signal for the old component relative to the
single-component asymmetric drift signal.

\subsection{Discussion}
\label{sec:discuss}

The results from Figure~\ref{fig:final_AD}, combined with population
age information will be exploited in future work to measure the AVR in
MaNGA galaxies. However, it is already interesting to remark on the
large range of AD signals observed in our sample of seven galaxies.
There is also some indication of a trend with mass: The differential
asymmetry drift signal for older populations is largest in the more
massive galaxies, but this trend is far less regular (noting the small
and large AD signals for 83320-9102 and 8482-3702, respectively)
compared to the standard mass--rotation-speed scaling seen in
Figure~\ref{fig:emceespaxels}. While the addition of age estimates
may transform the AD signal into a smoother trend in AVR with mass,
the rather uniform $g-r$ colors of our sample suggests this
expectation may be illusory. Given that there are at least three
likely astrophysical paths to generating an AVR (see
Section~\ref{sec:intro}), perhaps it is unsurprising to see a wide
range in AD over such a modest range of stellar mass. Application of
our methods to larger samples will provide statistics on the
distribution of AD as a function of galaxy mass.

It is worth noting the limitations of the present methods and scope of
analysis. We have selected galaxies with regular kinematics and with
modest values of \pfrac. One may well expect that interpretation of AD
measurements will be problematic in cases where the kinematics are
less regular or there are significant bar or oval distortions. It is
imperative, therefore, to have an assessment of galaxy kinematic
regularity, and to limit analysis to galaxies or radial zones within
galaxies where the gas is on near-circular orbits and low-order
distortions (e.g., m=1,2 modes) are weak or absent for both stars and
gas.

We have also seen that our algorithms are limited in precision once
the continuum S/N falls much below $\sim 20$ \AA$^{-1}$. However,
there is nothing preventing the methods described here to be applied
in a more sophisticated analysis that fits multiple spaxels
simultaneously in a parameterized model to increase the effective S/N;
even a relatively simple tilted-ring approach would likely improve
upon the S/N limitations of our current spaxel-by-spaxel scheme, e.g.,
at larger radii.

Nonetheless, the kinematics of the young component are not well
measured for \pfrac\ $< 0.2$. This means that the AVR within the
centers of galaxies with prominent, old bulge or pseudo-bulge
populations, while typically observed at sufficient S/N, are not
easily accessible given our current methods.  While we have not probed
\pfrac\ $>0.8$ with our modest sample, we can expect that at least in
the outskirts of later-type galaxies this condition will exist. We can
also expect that measuring the kinematics for the old component will
be difficult in this regime. Hence there is likely to be a sweet
spot for our methods at intermediate radius, where both young and old
disk components are comparably present. Such a sweet-spot is likely
well-matched to a comparison with the MW solar neighborhood.

Finally, while we cannot offer a definitive recommendation between our
SSP and SL algorithms, our preference is for the SL algorithm. Despite
its greater complexity, the SL algorithm in principle should suffer
less from template-mismatch systematics. Possibly supporting this
statement is the observation that the SL algorithms'
velocity-dispersion maps for the young stellar component look smoother
and more realistic (in amplitude) than for the SSP algorithm
(Figure~\ref{8138_12704_bp} and Appendix~\ref{sec:maps}).

\subsubsection{Gas Kinematics}
\label{sec:gas}

It has become increasingly apparent that diffuse ionized gas (DIG)
contributes significantly to the projected area of many early-type
galaxies as well as the centers of some intermediate-type galaxies
\citep[e.g.,][]{Belfiore2016}. Recent studies have shown that in some
cases the ionized gas of even rotationally supported systems lags
behind that of the molecular gas \citep{Levy2018}. In general this is
not the case for intermediate and late-type galaxies where the
emission from ionized gas is dominated from HII regions; in these
systems the agreement between H$\alpha$ and HI velocities is excellent
\citep[e.g.,][]{Martinsson2016}.  Unsurprisingly, the discrepancy
tends to occur in more massive systems or early-type dwarfs where the
specific star-formation is depressed. As recently shown by
\cite{denBrok2020} when the ionized gas is DIG like not HII-like (as
reckoned by e.g., line ratios of [OI], [NII] or [SII] to H$\alpha$.)
In these cases the ionized gas cannot be used as an accurate surrogate
for the potential's circular speed, and hence also cannot be used to
estimate the absolute amplitude of asymmetric drift in the
stars. However the differential asymmetric drift between young and old
can be measured regardless. 

For the galaxy sample in this paper, as we noted in the previous
Section, we find the ionized gas is predominately HII-like at most
radii outside of the central regions. We note that application of our
technique broadly should consider the limitations of using ionized gas
as a tracer of circular speed; the situations where the ionized gas is
predominantly DIG-like can be readily determined with the same
spectroscopy used to measure velocities.


\section{Summary and Conclusions}
\label{sec:conclude}

The study of the stellar age--velocity-dispersion relationship (AVR)
in galaxies has been limited to date to a handful of nearby galaxies
where individual stars are spatially resolved. This limitation has
been a significant obstacle in understanding the processes of galaxy
evolution that give rise to the phenomenon. With the advent of large
IFS surveys, such as MaNGA, we have an opportunity to
spectroscopically resolve the spatial profiles of galaxy populations
on an unprecedented scale, but not to spatially resolve the individual
stars therein, nor necessarily to spectroscopically resolve their
velocity dispersions at radii of interest. Consequently no tool has
existed to probe the AVR in these galaxies despite the recent wealth
of IFS data.

In this paper we have developed two algorithms to efficiently measure
the asymmetric drift (which depends only on an accurate measure of
tangential speed) in modern IFS data for spatially-unresolved stellar
populations. Using these algorithms we have presented the first
measurements of asymmetric drift in two stellar age components for
seven galaxies from the MaNGA survey. Since asymmetric drift can be
used as a proxy for velocity dispersion, combined with age information
for multiple components, the measurement of AVR in external galaxies
is now possible in unresolved star-light. Future efforts to
  increase the number of age components will further enhance our
  ability to broadly constrain AVR.

In Section~\ref{sec:hypothesis} we tested the hypothesis via a MCMC
analysis that the observed MaNGA galaxy spectra contain sufficient
information between 360 to 940 nm to constrain velocities for a young
and old stellar component, as well as a gas component.  We demonstrate
not only that the two stellar components have distinct kinematics but
also that the best-fitting kinematics for our galaxy sample have
velocity and velocity dispersion maps that are reasonably smooth and
have the expected trends with radius and azimuth. The measured
tangential velocities for the two stellar components are also
qualitatively consistent with that observed for nearby galaxies with
resolved stellar populations, i.e., the young stellar component has a
higher measured tangential velocity than its older counterpart. The
radial profiles of the derived fraction of young to old stellar
components (\pfrac), which for the purpose of our analysis was
split at an age of 1.5 Gyr, rise with radius, also in agreement with
expectations. These results demonstrate, respectively, that the random
and systematic errors in measured velocities for young and old stellar
populations are small. We show these errors are sufficiently small to
independently measure the asymmetric drift for at least two
age-components in spiral galaxies using MaNGA IFU data. This opens the
possibility for the whole-sale exploration of thin and thick-disk
properties in today's galaxy population.

Since our MCMC analysis is computationally prohibitive for application
to large data-sets, in Section~\ref{sec:ssp_algo} we developed an
algorithm using SSP models to disentangle the kinematics of the young
and old component as {\it accurately} as the MCMC method at a fraction
of the computational cost. Due to the astrophysical degeneracies
between age and metallicity on the appearance of the integrated-light
spectra, our more efficient SSP algorithm -- that uses a local
minimizer based on \pPXF\ -- is more robust at lower S/N than the
global minimization inherent to the MCMC analysis.

While the results of our SSP-based algorithm are promising, because
the asymmetric drift signal is small (of order 10's of \kms), there
remains the perennial but rarely addressed concern in studies of
galaxy kinematics of what systematic effects template-mismatch could
have on our conclusions. Rather than dismiss the concern out of hand,
we present an estimate of what we might expect for systematics on
tangential velocity due to template mismatch
(Section~\ref{sec:mismatch}). The potential effect is large enough to
warrant attention. We therefore developed an independent algorithm in
Section~\ref{sec:stellar_algo} based on empirical stellar spectra
rather than SSPs; the former are known to minimize the effect of
template mismatch. This algorithm is also computationally efficiency
and robust at low S/N.

Leveraging the results from the analysis of our seven galaxy sample in
Section~\ref{sec:results}, we use (i) the posterior probability
distribution from our MCMC analysis to estimate random errors on our
derived two-age component stellar tangential velocities; and (ii) and
the difference between our SSP and stellar-library algorithms to
estimate our systematic errors. This analysis indicates that reliable
measurements require S/N (per \AA) above 10.  While random errors
dominate at the spaxel level, systematic errors (which are of order 5
to 20 \kms deprojected) dominate in radial bins containing 10's of
spaxels. 

We find the measured asymmetric drift for the young and old components
are clearly distinct for 6 of our 7 galaxies with rotation speeds
above 200 \kms. The asymmetric drift for the young stellar component
is always close to zero, as expected, but in some cases is consistent
with non-zero values (10 to 20 \kms). The old stellar component has
significantly (roughly a factor of 2) larger asymmetric drift value
than what would be inferred from an asymmetric drift measurement
adopting only a single stellar component. There is a wide range in
old-component asymmetric drift signals from 20 to 150 \kms\ (a factor
of $\sim$7) at large radii in disks with rotation speeds varying only
from 200 to 350 \kms\ (a factor $<$2). This indicates there is much to
learn about the statistical properties and potentially a wide range of
evolutionary histories of disk populations with MaNGA and other IFS
surveys.

\acknowledgements{This research was directly supported by the
  U.S. National Science Foundation (NSF) AST-1517006. We thank an
  anonynous referee for comments which led to improved clarity of the
  paper. Funding for the Sloan Digital Sky Survey IV has been provided
  by the Alfred P. Sloan Foundation, the U.S. Department of Energy
  Office of Science, and the Participating Institutions. SDSS-IV
  acknowledges support and resources from the Center for
  High-Performance Computing at the University of Utah. The SDSS web
  site is www.sdss.org.

SDSS-IV is managed by the Astrophysical Research Consortium for the
Participating Institutions of the SDSS Collaboration including the
Brazilian Participation Group, the Carnegie Institution for Science,
Carnegie Mellon University, the Chilean Participation Group, the
French Participation Group, Harvard-Smithsonian Center for
Astrophysics, Instituto de Astrof\'isica de Canarias, The Johns
Hopkins University, Kavli Institute for the Physics and Mathematics of
the Universe (IPMU) / University of Tokyo, Lawrence Berkeley National
Laboratory, Leibniz Institut f\"ur Astrophysik Potsdam (AIP),
Max-Planck-Institut f\"ur Astronomie (MPIA Heidelberg),
Max-Planck-Institut f\"ur Astrophysik (MPA Garching),
Max-Planck-Institut f\"ur Extraterrestrische Physik (MPE), National
Astronomical Observatories of China, New Mexico State University, New
York University, University of Notre Dame, Observat\'ario Nacional /
MCTI, The Ohio State University, Pennsylvania State University,
Shanghai Astronomical Observatory, United Kingdom Participation Group,
Universidad Nacional Aut\'onoma de M\'exico, University of Arizona,
University of Colorado Boulder, University of Oxford, University of
Portsmouth, University of Utah, University of Virginia, University of
Washington, University of Wisconsin, Vanderbilt University, and Yale
University.

}

\begin{appendices}
\setcounter{table}{0}
\renewcommand{\thetable}{\Alph{section}\arabic{table}}
\setcounter{figure}{0}
\renewcommand{\thefigure}{\Alph{section}\arabic{figure}}
\setcounter{equation}{0}
\renewcommand{\theequation}{\Alph{section}\arabic{equation}}

\section{Stellar Template Library}
\label{app:stellar_lib}


\begin{deluxetable}{crllrll}
\tabletypesize{\footnotesize}
\tablecaption{IndoUS Stellar Template Subset}
\tablehead{
    \colhead{} &
    \colhead{} &
    \multicolumn{2}{c}{Spectral type} &
    \colhead{} &
    \colhead{} &
    \colhead{} \\ \cline{3-4}
    \multicolumn{7}{c}{} \\ [-0.1in]
    \colhead{Index} &
    \colhead{HD} &
    \colhead{SIMBAD} &
    \colhead{Pickles} &
    \colhead{$T_{\rm eff} ({\rm K})$} &
    \colhead{$\log g$} &
    \colhead{[Fe/H]}
}
\startdata
   0  &  30614  & O9.5Iae  & A0I   & 29647 & 3.05 & $+0.3$ \\
   1  &  34816  & B0.5IV   & B2IV  & 29890 & 4.22 & $-0.24$ \\
   2  &  180163 & B2.5IV   & B2IV  & 17360 & 3.38 & $-0.01$ \\
   3  &  207330 & B3III    & B3III & 19470 & 3.49 & $-0.1$ \\
   4  &  51309  & B3Ib/II  & B3I   & 17390 & 2.7  & $-0.17$ \\
   5  &  41692  & B5IV     & B6IV  & 14400 & 3.12 & $-0.42$ \\
   6  &  155763 & B6III    & B5III & 12900 & 3.9  & $-0.95$ \\
   7  &  35497  & B7III    & B9III & 13622 & 3.8  & $-0.1$ \\
   8  &  34797  & B8/B9IV: & B8V   & 14000 & 4.5  & $-0.6$ \\
   9  &  175640 & B9III    & B9III & 12100 & 4.0  & $-0.55$ \\
  10  &  105262 & B9       & A0I   &  8542 & 1.5  & $-1.37$ \\
  11  &  18296  & B9p...   & B9III & 11200 & 3.0  & $-0.12$ \\
  12  &  87737  & A0Ib     & A0I   &  9700 & 2.0  & $-0.05$ \\
  13  &  183324 & A0V      & A0V   &  9260 & 4.22 & $-1.5$ \\
  14  &  198001 & A1V      & A2V   &  9470 & 3.64 & $+0.07$ \\
  15  &  14489  & A2Ia     & A2I   &  9000 & 1.4  & $-0.26$ \\
  16  &  223385 & A3Iae    & A2I   &  9333 & 1.0  & $\;\;\:0.0$ \\
  17  &  34578  & A5II     & A5III &  8300 & 1.85 & $+0.16$ \\
  18  &  36673  & F0Ib     & F0I   &  7400 & 1.1  & $+0.04$ \\
  19  &  25291  & F0II     & F0II  &  7600 & 1.5  & $+0.11$ \\
  20  &  90277  & F0V      & F0V   &  7412 & 3.46 & $+0.19$ \\
  21  &  33276  & F2IV     & F02IV &  7099 & 3.3  & $+0.29$ \\
  22  &  184266 & F2V      & F2V   &  5713 & 2.64 & $-1.85$ \\
  23  &  168151 & F5V      & F5V   &  6587 & 4.09 & $-0.31$ \\
  24  &  134083 & F5V      & F5V   &  6632 & 4.5  & $+0.1$ \\
  25  &  108954 & F9V      & F8V   &  6060 & 4.35 & $-0.11$ \\
  26  &  92125  & G2.5IIa  & G5II  &  5600 & 2.1  & $+0.38$ \\
  27  &  126868 & G2IV     & G2IV  &  5521 & 3.3  & $-0.06$ \\
  28  &  106210 & G3V      & G2V   &  5337 & 4.0  & $-0.54$ \\
  29  &  117176 & G5V      & G5V   &  5480 & 3.83 & $-0.11$ \\
  30  &  47731  & G5Ib     & G5I   &  4990 & 1.0  & $-0.16$ \\
  31  &  131156 & G8V      & G8V   &  5500 & 4.6  & $-0.15$ \\
  32  &  106714 & G8III    & G8III &  4897 & 2.34 & $-0.23$ \\
  33  &  107383 & G8III    & G8III &  4690 & 2.91 & $-0.39$ \\
  34  &  104985 & G9III    & G8III &  4658 & 2.2  & $-0.31$ \\
  35  &  191026 & K0IV     & K0IV  &  5150 & 3.49 & $-0.1$ \\
  36  &  124897 & K1.5III  & K1III &  4300 & 1.5  & $-0.49$ \\
  37  &  149661 & K2V      & K2V   &  5362 & 4.56 & $+0.01$ \\
  38  &  121146 & K2IV     & K3IV  &  4400 & 1.85 & $-0.13$ \\
  39  &  105043 & K2III    & K2III &  4374 & 2.67 & $+0.02$ \\
  40  &  175545 & K2III    & K2III &  4429 & 2.94 & $+0.29$ \\
  41  &  110281 & K5       & K5V   &  3950 & 0.2  & $-1.56$ \\
  42  &  113996 & K5III    & K5III &  3970 & 1.69 & $-0.26$ \\
  43  &  237903 & K7V      & K7V   &  4070 & 4.7  & $\;\;\:0.0$  \\
  44  &  102212 & M1III    & M1III &  3761 & 1.5  & $+0.06$ \\
  45  &  39801  & M1       & M2I   &  3540 & 0.0  & $+0.05$ \\
  46  &  36389  & M2Iab:   & M2I   &  3706 & 0.7  & $+0.11$ \\
  47  &  112300 & M3III    & M3III &  3700 & 1.3  & $-0.16$ \\
  48  &  44478  & M3III    & M3III &  3450 & 1.0  & $\;\;\:0.0$ \\
  49  &  148783 & M6III    & M6III &  3250 & 0.2  & $-0.01$ \\
  50  &  126327 & M7.5     & M8III &  3000 & 0.0  & $-0.58$
\enddata
\label{tab:stellar_subset}
\end{deluxetable}

The Indo-US Library of Coud\'{e} Feed Stellar Spectra
\citep{Valdesetal2004} is an empirical stellar spectral library of
1,273 stars with broad coverage of stellar types and with spectra
spanning from 346.5--946.9 nm with a resolution (FWHM) of $\sim$0.135
nm sampled at 0.044 nm pixel$^{-1}$. The wavelength range is important
because it includes strong metal features in the red, such as the Ca
Triplet, which we anticipate will constrain the kinematics of old
populations in galaxy spectra that are dominated in the blue by
younger stellar populations. The spectral resolution is roughly twice
the spectral resolution as MaNGA data. Using a library with higher
spectral resolution to that of MaNGA allows for a simple accounting of
the difference in the line spread function between the galaxy spectra
and templates. For this reason we have not used the MaStar library
\citep{Yan19} which otherwise is superior in stellar parameter
coverage and calibration. While the diversity of stellar types is
important to minimize effects of template mismatch we believe the
Indo-US Library is sufficient for our purposes. From this library we
select a subset of 51 stars which represent and span the stellar
parameters in $T_{\rm eff}$ and $\log g$ of the complete library.

The sub-sample of 51 stars from the Indo-US Coud\'{e} Feed Stellar
Library \citep{Valdesetal2004} covers stellar spectral types O to M.
Table ~\ref{tab:stellar_subset} gives the internal index used in this
analysis and corresponding HD number; classification and spectral
types and atmospheric parameters are repeated from Table 3 of
\citet[][references for these parameters are therein]{Valdesetal2004}.
We have included both the Simbad database\footnote{SIMBAD is operated
  at CDS, Strasbourg, France.} and \cite{Pickles98} spectral types to
point out inconsistencies between different spectral types and
temperatures, although luminosity classes and surface-gravities appear
to be more consistent.  Of particular note is HD 30614, which is
plainly an O-star given the relative strength of He to H lines, but
also suffers from significant redenning given its continuum shape; the
SIMBAD spectral type is preferred, as corroborated by its estimated
surface temperature. This illustrates some of the perils of mapping
stellar libraries for population synthesis and also why polynomials
are often required in, e.g., \pPXF, to both fit the relative
line-strengths and continuum shape. A number of other stars are
clearly mis-typed according to their surface-temperature, but given
the coarse binning into hot, intermediate and cool stars for this
study, this is not of significance for our purposes.

As Table ~\ref{tab:stellar_subset} shows, we have selected a wide
range of luminosity classes for all spectral types, although Main
Sequence stars dominate at intermediate temperatures (late-A through
mid-G), and giants dominating for the cooler stars (late-G and later).
Our thinking in this selection was as follows.  All of the
earliest-types are massive, short-lived stars. At the other extreme in
temperature, we expect the light to be dominated by giants, with the
exception of very early ages where red super-giants may be
significant, depending on metallicity. There is, in principle, an
uncomfortable intermediate temperature (types A-F) where significant
contributions can come either from the Main Sequence or a blue
horizontal branch from old, metal-poor populations. The selected
subset should offer sufficient flexibility to cover these various
conditions.

In the \pPXF\ decompositions of the MIUSCAT SSPs we find five stars are
given zero or near-zero maximum weight (\#4,10,12,15,44,50). These are
either hot super-giants (3) or cool giants (2). In contrast there are
11 stars (\#8,9,14,21,23,24,25,29,33,37,40) that either contribute
more than 20\% of the weight in a single SSP or an average weight
above 5\% across all SSPs (the mean values for the full sample is 11\%
and 2\% respectively). These contain stellar types between late-B and
early K, with mostly main-sequence B,A,F,G, stars and late-G and
early-K giants. The remainder of the stars appear significant in
aggregate, but we have made no attempt to define a minimal subset, and
certainly would not want to do so based on SSPs.  The purpose here is
to indicate that this library, while not unique, should be capable of
representing of a broad range of stellar populations, and the stars
selected by \pPXF\ to represent SSPs seem sensible.


\section{MCMC Posterior Probability Distribution: Bi-modality}
\label{app:mcmc_ppd}

\begin{figure*}
  \centering
  \includegraphics[width=0.95\linewidth]{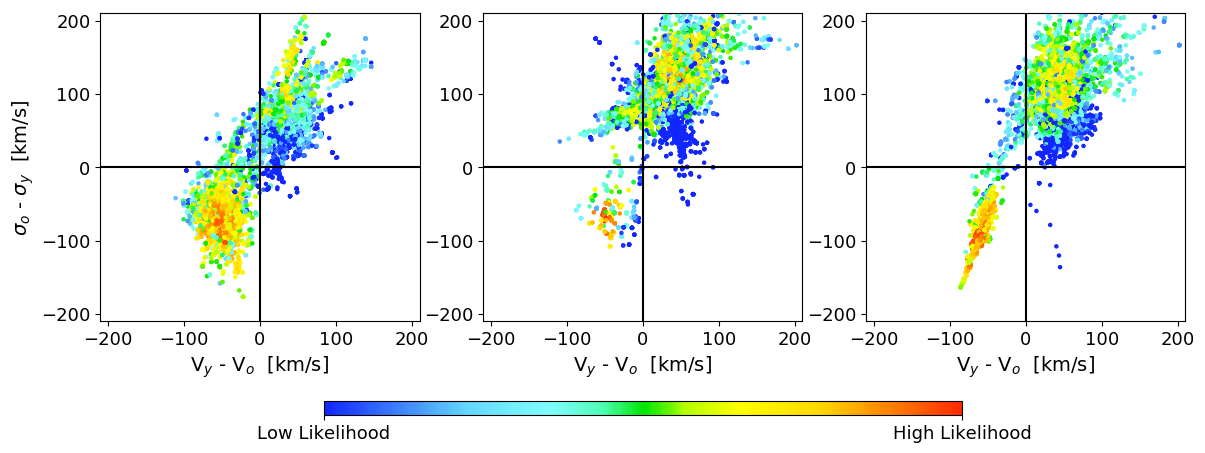}
  \caption{Locations of steps of the MCMC walkers in $\delta\sigma$
    versus $\delta V$ for some example spaxels with global maximum
    likelihoods in the bottom-left quadrant of the $\delta
    V$-$\delta\sigma$ plane. Color-coding is the likelihood of the
    model parameters in each step.}
  \label{fig:flipped_spx}
\end{figure*}

\begin{figure*}[ht!]
  \centering
  \includegraphics[width=0.95\linewidth]{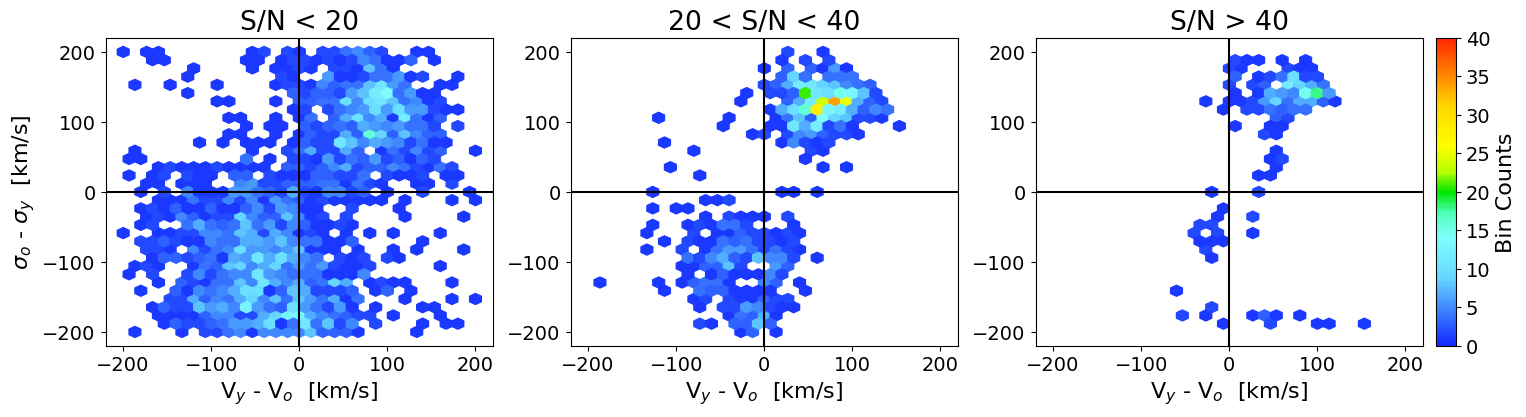}
  \caption{Distribution in $\delta V$ versus $\delta\sigma$ for
    spaxels within $\pm30^{\circ}$ of the major axis for our galaxy
    sample.  Kinematics are the Maximum Likelihood solution measured
    by the MCMC analysis presented in
    Section~\ref{sec:hypothesis}. The density of spaxels within a
    hexagonal bin is given by color. Panels right to left are for
    spaxels with S/N $<$ 20, 20 $<$ S/N $<$ 40, and S/N $>$ 40.}
  \label{fig:MCMC_dVdS}
\end{figure*}

We take advantage of the MCMC posterior probability distribution (PPD)
to understand the complexity of the likelihood space of the
two-component kinematic model tested in Section~\ref{sec:hypothesis}.
We use this understanding to assess the reliability of the MCMC
results.

For a normally distributed PPD, the best-fitting parameter and its
uncertainty can be quantified by the mean and standard deviation of
the PPD. However, such a quantification becomes problematic when the
PPD is not unimodal; the first moment may not correspond to a
parameter value at either a local or global minima in $\chi^2$. We
find there are cases where the PPD displays a bimodal distribution in
the difference between young and old stellar population velocities
($\delta V$) and the difference between old and young velocity
dispersions ($\delta \sigma$). Figure~\ref{fig:flipped_spx} presents
examples of this behavior within the steps of the MCMC walkers. These
steps are color-coded by the relative likelihood of the solution at
that step. In each case the steps outline two very different
combinations of parameters with near-equal high likelihood. We have
therefore adopted the best-fitting parameters from the MCMC analysis
as those which have the {\it maximum} likelihood.  In these particular
cases the maximum likelihood occurs where $\delta V < 0$ and $\delta
\sigma < 0 $, i.e., where the old population is rotating more quickly
and is dynamically colder than the young population. We refer to this
as a `flipped' solution.

Figure~\ref{fig:MCMC_dVdS} illustrates $\delta V$ and $\delta\sigma$
from the maximum likelihood solution of the MCMC for all seven of our
galaxies as a function of S/N. At low S/N a portion of the spaxels the
maximum likelihood solution from the MCMC is in the bottom-left
quadrant (with negative $\delta V$ and $\delta\sigma$); the kinematics
of young and old components are flipped in a significant fraction of
low S/N spaxels. The {\it spatial} distribution of the flipped spaxels
appears random, and there are no significant correlations observed
with other parameters, e.g., \pfrac. These facts suggest that the
presence of these flipped results aren't astrophysical, but rather
they demonstrate the sensitivity of the spectral-fitting maximum
likelihood solution.  This sensitivity is due to the presence of
multiple (principally two), discrete local minima, as seen in
Figure~\ref{fig:flipped_spx}. Based on a bootstrap analysis we find
these minima become shallower at lower S/N. Since we begin the walkers
in the upper-right portion of the parameter space
(Section~\ref{sec:hypothesis}), at lower S/N the walkers will more
easily find a maximum likelihood solution in a flipped portion of
parameter space.

The presence of these two local minima with discrete kinematics is not
fully understood. One possible cause may be the degeneracy of the
spectral features in the components of the young and old stellar
population.  Upon inspection of the derived young and old spectra in
these flipped cases, we find they retain their overall spectral energy
distribution: The young spectrum is bluer than the old spectrum. While
there are more subtle differences in the strengths of the spectral
features when comparing the spectrum of each stellar component, these
differences have not yet pointed to a solution for eliminating the
degeneracy in the kinematic solutions.

Results from previous studies and theoretical expectations of the AVR
of galaxies (see Section~\ref{sec:intro}) strongly suggest that the
older stellar populations of the galaxy disk tend to a lower
tangential velocity and be dynamically hotter than their young
counterparts, i.e. $\delta V > 0$ and $\delta \sigma> 0$. Placing this
prior upon our MCMC results and noting the near-equal $\chi^2$ of the
two minima, we believe that we have a strong case to ignore the
spaxels with flipped kinematics. For our non-MCMC algorithms, where
we use local minimization, we rely successfully on initial conditions
to largely eliminate these flipped cases and converge onto solution
consistent with previous studies and theoretical expectations. Hence
for quantitative comparison between our algorithms and the MCMC
results we ignore spaxels where the measured maximum likelihood
solution has negative $\delta V$ and $\delta\sigma$.

\setcounter{table}{0}
\setcounter{figure}{0}

\section{Mock Spectra}
\label{app:mocks}

Mock spectra were generated to assess the effectiveness of different
techniques to measure the kinematics of multiple stellar-population
components. These consist of `Simple' and `Realistic' mocks, as
  follows. The `Simple' mocks were useful in the early of our
  stellar-library algorithm development, while the `Realistic' mocks
  provided an early check on our SSP algorithms.

`Simple mocks' are composite stellar populations made up of two SSPs
with differing kinematics. For each mock one SSP is selected at random
from a subset of SSPs with young ages ($\leq$ 1.5Gyr), and a second
from a subset with ages greater than 1.5 Gyrs (see
Section~\ref{sec:young_old}).  Each SSP in a mock is shifted in
velocity and convolved with a Gaussian to mimic a LOSVD for each
population. The velocity and velocity dispersion of the young
population is determined at random within the bounds of 100-350
\kms\ and 5-50 \kms\ respectively. The velocity dispersion of the old
population is randomly assigned between the velocity dispersion of the
young population and 100 \kms. The velocity of the old population is
then determined such that the quadrature sum of the velocity and
velocity dispersion of the two populations are equal. The convolved
spectra are then added together after being randomly normalized
randomly between the wavelength range 360-900A~nm, allowing our simple
mock spectra to crudely approximate a range star formation histories.

`Realistic' mocks aim to simulate what we expect to observe in MaNGA
spectra of external galaxies in terms of more realistic star-formation
histories (SFH) but retaining a simple prescription for the
age-velocity relation (AVR).{\bf\footnote{We also constructed mocks
    with a range of realistic AVRs, but we found this added complexity
    made the interpretation of recovered values more ambigous, and
    hence of less use for algorithm development.}} Our mocks consist
of a set of 2,500 spectra using the MIUSCAT SSP library described in
Section \ref{sec:ssp_choice} with randomly generated parameters for
their SFH, chemical evolution and AVR relationship.

We adopt the parametric star-formation model recommended by
\citet{Simhaetal2014} based on their analysis of smooth-particle
hydrodynamical simulations. This model accurately recovers the
physical properties of their simulated data in terms of mass-to-light
ratio, population ages, and specific star formation rates. The
star-formation model is a combination of one function describing an
initial burst with time-scale $\tau$ initiated at $t_i$ and a
normalization factor $A$, followed by a linear function with slope
$\Gamma$ after a transition-time $t_{\rm trans}$:

\begin{figure}[ht!]
\begin{center}
  \includegraphics[width=1\linewidth]{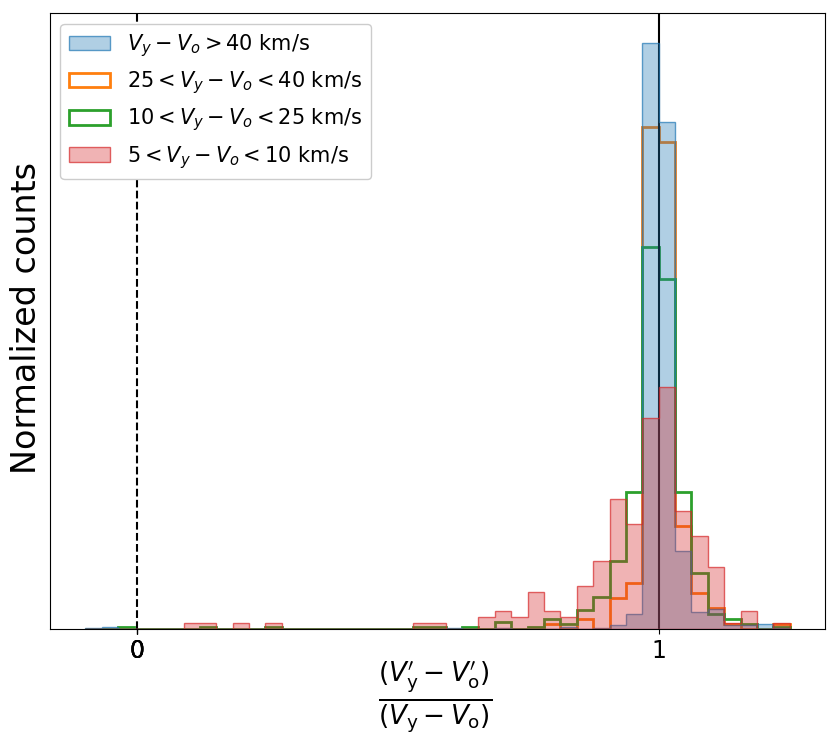}
  \caption{Performance of our simple, 1-step SSP algorithm to
    disentangle the kinematics of young and old stellar populations in
    our mock spectra.  The abscissa is a performance metric computed
    as the ratio of the recovered difference in the velocities of the
    two components to the model difference in the mocks. To illustrate
    the performance of the algorithm for different signal levels
    (i.e., the model difference between young and old stellar
    component velocities), histograms are shown for mocks in four
    ranges of velocity difference, as given in the legend.}
  \label{fig:soph_mock_results}
\end{center}
\end{figure}

\begin{deluxetable*}{lcrrrrrrrrrrrrrrr}[ht!]
\tablewidth{0pt}
\tabletypesize{\footnotesize}
\tablecaption{Simple SSP Algorithm Performance Metrics Based on Realistic Mock Spectra}
\tablehead{
 \colhead{Algorithm} &
 \colhead{$\delta V_{\rm mock}$} &
 \multicolumn{2}{c}{$\Delta V_y$} &
 \colhead{} &
 \multicolumn{2}{c}{$\Delta V_o$} &
 \colhead{} &
 \multicolumn{2}{c}{$\delta V_{\rm Algo} / \delta V_{\rm mock}$} &
 \colhead{} &
 \multicolumn{2}{c}{$\Delta{\rm \pfrac}$} &
 \colhead{} &
 \multicolumn{3}{c}{Failure Fraction} \\  [0.05in] \cline{3-4} \cline{6-7} \cline{9-10} \cline{12-13} \cline{15-17}
 \multicolumn{17}{c}{} \\  [-0.08in]
 \colhead{} &
 \colhead{Range} &
 \colhead{med} &
 \colhead{$\sigma_{\rm MAD}$} &
 \colhead{} &
 \colhead{med} &
 \colhead{$\sigma_{\rm MAD}$} &
 \colhead{} &
 \colhead{med} &
 \colhead{$\sigma_{\rm MAD}$} &
 \colhead{} &
 \colhead{median} &
 \colhead{$\sigma_{\rm MAD}$} &
 \colhead{} &
 \colhead{Cat} &
 \colhead{Flip} &
 \colhead{High} \\
 \colhead{} &
 \colhead{(\kms)} &
 \multicolumn{2}{c}{(\kms)} &
 \colhead{} &
 \multicolumn{2}{c}{(\kms)} &
 \colhead{} &
 \colhead{} &
 \colhead{} &
 \colhead{} &
 \colhead{} &
 \colhead{} &
 \multicolumn{4}{c}{} \\
 \colhead{(1)} &
 \colhead{(2)} &
 \multicolumn{1}{r}{(3)} &
 \multicolumn{1}{r}{(4)} &
 \colhead{} &
 \multicolumn{1}{r}{(5)} &
 \multicolumn{1}{r}{(6)} &
 \colhead{} &
 \multicolumn{1}{r}{(7)} &
 \multicolumn{1}{r}{(8)} &
 \colhead{} &
 \multicolumn{1}{r}{(9)} &
 \multicolumn{1}{r}{(10)} &
 \colhead{} &
 \multicolumn{1}{r}{(11)} &
 \multicolumn{1}{r}{(12)} &
 \multicolumn{1}{r}{(13)}
}
\startdata
Simple SSP      & [10,35] &  0.12 & 0.20 &&  0.08 &  0.13 &&  1.00 & 0.02 && -0.03 & 0.01 && 0.00 & 0.00 & 0.02 \\
Simple SSP      & [35,50] &  0.22 & 0.16 &&  0.16 &  0.24 &&  1.00 & 0.01 && -0.03 & 0.01 && 0.00 & 0.01 & 0.06 \\
Simple SSP      &   $>$50 &  0.23 & 0.15 &&  0.15 &  0.28 &&  1.00 & 0.01 && -0.03 & 0.01 && 0.00 & 0.02 & 0.03
\enddata
\label{tab:SSP_realistic_mocks_metric}
\tablecomments{The Simple SSP algorithm (column 1) is described in
  Section~\ref{sec:ssp_simple}.  Three bins in $\delta V \equiv V_y -
  V_o$ for the mocks ($\delta V_{\rm mock}$) given in column (2)
  contain 678, 263, and 996 mocks, respecitvely from low to high
  $\delta V_{\rm mock}$. Columns (3) to (10) are defined identically
  to the corresponding columns in Table~\ref{tab:MCMC_SL_metric}
  except they are referenced to the `Realistic' mock model values
  instead of the MCMC results. Columns (11) to (13) are defined in the
  text.}
\end{deluxetable*}

\begin{equation}
{\rm SFR}(t) =\begin{cases}
{A(t-t_i)e^{-(t-t_i)/\tau}} & \mbox{for } (t \leq t_{\rm trans})\\
{{\rm SFR}(t_{\rm trans}) + {\Gamma(t-t_{\rm trans})}}
& \mbox{for } (t > t_{\rm trans}) 
\end{cases}
\end{equation}

The time of the initial burst ($t_i$) and the transition time ($t_{\rm
  trans}$) are randomly selected from normal distributions centered at
$1\pm0.2$~Gyr and $10.7\pm2$~Gyr, respectively; the timescale of the
initial burst ($\tau$) is randomly selected from a uniform
distribution within the bounds 1 and 10~Gyr; $\Gamma$ is randomly
selected from an uniform distribution between -0.5 and 0.5. 

We implement a simple model for chemical evolution. The oldest
  stellar population in the mocks begin with a metallicity of -0.66
  dex, i.e., the lowest metallicity of the MIUSCAT SSP models
  (Section~\ref{sec:ssp_choice}). At every stellar age increment the
  metallicity of the SSP model to be added to the mock (and those that
  will follow) may increase by one step in metallicity index with a
  probability of 0.15, up to the maximum metallicity of 0.26 dex.

The AVR model changes the stellar kinematics as a step function: the
velocity and velocity dispersion of the young stellar population, ages
1.5~Gyrs and younger, are derived randomly from a uniform distribution
within the range of 100 to 250 \kms for velocity and 10 to 30 \kms for
velocity dispersion. The velocity dispersion of the old stellar
population, with ages more than 1.5~Gyrs, is also randomly drawn from
a uniform distribution between 50 and 100 \kms. Based on these
already drawn kinematics, the velocity for the older populations is
derived such that the quadrature sum of the velocity and velocity
dispersion of the spectra remains constant, i.e.  $V_{y}^2 +
\sigma_{y}^{2} = V_{o}^2 + \sigma_{o}^2$. Hence within the two
  categories of ages, the SSP models are assigned identical
  kinematics. The purpose of this AVR model is to test the impact of
providing realistic SFH in the context of clearly defined, albeit
simple, model kinematics. The {\it range} of kinematics of the
  mocks are similar to those seen in real galaxies.

\subsection{Performance of the Simple SSP Algorithm on Mock Spectra}
\label{sec:ssp_mock_performance}

`Realistic' mock spectra were used to test the abilities of the simple
(1-step) SSP algorithm to disentangle the velocities of young and old
stellar populations. We define metrics similar to those given
  in Tables~\ref{tab:MCMC_SSP_metric} and \ref{tab:MCMC_SL_metric},
  presented here in Table~\ref{tab:SSP_realistic_mocks_metric}; these
  metrics reference the mock spectra model values rather than the MCMC
  derived values from real data. We also tabulate failure fractions
defined as follows: `Cat' are instances where either one of the
measured velocities is more than 100 \kms\ from the model velocity.
`Flip' are instances where $V_y-V_o<0$. `High' are instances where
\pfrac\ $> 0.85$.  High \pfrac\ values are indicative of an incorrect
allocation of cool stars to the young kinematic component.

\begin{deluxetable*}{lccrrrrrrrrrrrrrrr}[ht!]
\tablewidth{0pt}
\tabletypesize{\footnotesize}
\tablecaption{Simple Stellar Library Algorithm Performance Metrics Based on Simple Mock Spectra}
\tablehead{
 \colhead{Algorithm} &
 \colhead{TPL} &
 \colhead{$\delta V_{\rm mock}$} &
 \multicolumn{2}{c}{$\Delta V_y$} &
 \colhead{} &
 \multicolumn{2}{c}{$\Delta V_o$} &
 \colhead{} &
 \multicolumn{2}{c}{$\delta V_{\rm Algo} / \delta V_{\rm mock}$} &
 \colhead{} &
 \multicolumn{2}{c}{$\Delta{\rm \pfrac}$} &
 \colhead{} &
 \multicolumn{3}{c}{Failure Fraction} \\  [0.05in] \cline{4-5} \cline{7-8} \cline{10-11} \cline{13-14} \cline{16-18}
 \multicolumn{18}{c}{} \\  [-0.08in]
 \colhead{} &
 \colhead{set} &
 \colhead{Range} &
 \colhead{med} &
 \colhead{$\sigma_{\rm MAD}$} &
 \colhead{} &
 \colhead{med} &
 \colhead{$\sigma_{\rm MAD}$} &
 \colhead{} &
 \colhead{med} &
 \colhead{$\sigma_{\rm MAD}$} &
 \colhead{} &
 \colhead{median} &
 \colhead{$\sigma_{\rm MAD}$} &
 \colhead{} &
 \colhead{Cat} &
 \colhead{Flip} &
 \colhead{High} \\
 \colhead{} &
 \colhead{} &
 \colhead{(\kms)} &
 \multicolumn{2}{c}{(\kms)} &
 \colhead{} &
 \multicolumn{2}{c}{(\kms)} &
 \colhead{} &
 \colhead{} &
 \colhead{} &
 \colhead{} &
 \colhead{} &
 \colhead{} &
 \multicolumn{4}{c}{} \\
 \colhead{(1)} &
 \colhead{(2)} &
 \colhead{(3)} &
 \multicolumn{1}{r}{(4)} &
 \multicolumn{1}{r}{(5)} &
 \colhead{} &
 \multicolumn{1}{r}{(6)} &
 \multicolumn{1}{r}{(7)} &
 \colhead{} &
 \multicolumn{1}{r}{(8)} &
 \multicolumn{1}{r}{(9)} &
 \colhead{} &
 \multicolumn{1}{r}{(10)} &
 \multicolumn{1}{r}{(11)} &
 \colhead{} &
 \multicolumn{1}{r}{(12)} &
 \multicolumn{1}{r}{(13)} &
 \multicolumn{1}{r}{(14)}
}
\startdata
 Feature & A & [10,35] & 26.8 & 14.8 &&  -6.8 &   5.5 && -0.92 &  1.07 &&  0.24 & 0.13 && 0.09 & 0.82 & 0.00 \\
 Feature & A & [35,50] & 21.5 & 13.4 && -19.9 &  11.2 && -0.05 &  0.58 &&  0.20 & 0.12 && 0.10 & 0.52 & 0.00 \\
 Feature & A &   $>$50 & 20.8 & 10.5 && -44.1 &  24.4 &&  0.10 &  0.42 &&  0.21 & 0.13 && 0.09 & 0.43 & 0.00 \\
 \multicolumn{18}{c}{} \\  [-0.08in]
 Feature & B & [10,35] & 18.9 & 10.4 &&  -7.3 &   7.0 && -0.78 &  1.06 &&  0.10 & 0.11 && 0.02 & 0.74 & 0.00 \\
 Feature & B & [35,50] & 14.4 &  6.0 &&  -8.6 &  11.1 &&  0.47 &  0.40 &&  0.01 & 0.08 && 0.02 & 0.35 & 0.00 \\
 Feature & B &   $>$50 & 13.2 &  4.6 &&  -7.1 &  13.2 &&  0.72 &  0.22 && -0.03 & 0.05 && 0.05 & 0.23 & 0.01 \\
 \multicolumn{18}{c}{} \\  [-0.08in]
 Feature & C & [10,35] & 14.0 &  5.1 &&  44.0 &  66.2 &&  3.01 &  4.17 && -0.20 & 0.19 && 0.43 & 0.29 & 0.45 \\
 Feature & C & [35,50] & 13.4 &  5.5 &&  15.8 &  27.2 &&  1.21 &  0.81 && -0.13 & 0.06 && 0.26 & 0.15 & 0.22 \\
 Feature & C &   $>$50 & 13.6 &  4.7 &&  18.3 &  16.6 &&  1.10 &  0.26 && -0.11 & 0.04 && 0.22 & 0.08 & 0.21 \\
 \multicolumn{18}{c}{} \\  [-0.08in]
 Feature & D & [10,35] & 15.7 &  4.5 &&  49.7 & 104.0 &&  2.72 &  5.64 && -0.32 & 0.21 && 0.46 & 0.28 & 0.65 \\
 Feature & D & [35,50] & 18.0 &  6.7 &&  34.7 &  37.3 &&  1.44 &  0.96 && -0.28 & 0.14 && 0.31 & 0.12 & 0.54 \\
 Feature & D &   $>$50 & 17.9 &  6.5 &&  57.6 &  45.2 &&  1.49 &  0.55 && -0.21 & 0.11 && 0.38 & 0.06 & 0.47 \\
 \multicolumn{18}{c}{} \\  [-0.08in]
 Full    & A & [10,35] & 17.4 &  4.4 &&  -5.9 &   5.1 && -0.26 &  0.36 &&  0.24 & 0.12 && 0.04 & 0.69 & 0.00 \\
 Full    & A & [35,50] & 17.1 &  4.8 && -20.2 &  10.5 &&  0.12 &  0.32 &&  0.21 & 0.12 && 0.04 & 0.40 & 0.00 \\
 Full    & A &   $>$50 & 15.8 &  5.6 && -43.8 &  23.8 &&  0.23 &  0.32 &&  0.21 & 0.13 && 0.05 & 0.32 & 0.00 \\
 \multicolumn{18}{c}{} \\  [-0.08in]
 Full    & B & [10,35] & 15.8 &  4.7 &&  -6.7 &   6.0 && -0.33 &  0.50 &&  0.03 & 0.09 && 0.03 & 0.69 & 0.02 \\
 Full    & B & [35,50] & 15.1 &  5.8 &&  -8.7 &  11.4 &&  0.39 &  0.46 &&  0.00 & 0.07 && 0.06 & 0.36 & 0.01 \\
 Full    & B &   $>$50 & 11.4 &  5.0 &&  -4.7 &  13.1 &&  0.79 &  0.24 && -0.02 & 0.04 && 0.08 & 0.23 & 0.03 \\
 \multicolumn{18}{c}{} \\  [-0.08in]
 Full    & C & [10,35] & 12.1 &  4.0 && 160.5 & 177.1 &&  9.11 & 10.04 && -0.30 & 0.22 && 0.57 & 0.21 & 0.60 \\
 Full    & C & [35,50] & 12.2 &  6.4 &&  27.7 &  69.3 &&  1.58 &  2.20 && -0.16 & 0.10 && 0.44 & 0.16 & 0.42 \\
 Full    & C &   $>$50 & 10.7 &  4.6 &&  26.7 &  27.0 &&  1.27 &  0.42 && -0.12 & 0.04 && 0.36 & 0.10 & 0.31 \\
 \multicolumn{18}{c}{} \\  [-0.08in]
 Full    & D & [10,35] & 13.4 &  3.6 && 265.2 & 375.5 && 18.06 & 24.56 && -0.36 & 0.22 && 0.66 & 0.18 & 0.71 \\
 Full    & D & [35,50] & 16.2 &  6.9 && 158.5 & 164.1 &&  4.13 &  3.77 && -0.38 & 0.22 && 0.59 & 0.17 & 0.69 \\
 Full    & D &   $>$50 & 16.1 &  7.6 && 109.5 &  99.4 &&  2.18 &  1.17 && -0.26 & 0.15 && 0.55 & 0.10 & 0.60
\enddata
\label{tab:simple_mocks_metric}
\tablecomments{Simple stellar library algorithms (column 1) are
  described in Section~\ref{sec:stellar_simple}; the Full algorithm
  using template set B is identified as `Simple SL' in
  Tables~\ref{tab:MCMC_SL_metric} and \ref{tab:SL_realistic_mocks_metric}.
  Column definitions and mock number are the same as in
  Table~\ref{tab:SSP_realistic_mocks_metric}, with the addition of the
  template set in column (2) and the referencing of `Simple' rather
  than `Realistic' mocks.}
\end{deluxetable*}

Figure~\ref{fig:soph_mock_results} presents the performance for on of
the metrics, $\delta V_{\rm Algorithm} / \delta V_{\rm mock}$ where
$\delta V \equiv V_y - V_o$, the `mock' values are the input
velocities used to generate the models, and the `algorithm' values are
the recovered values.  The histograms are divided into four groups
based on the strength of the signal in the mock spectra: $\delta V >
40$~\kms, $25 <\ \delta V < 40$~\kms, $10 < \delta V < 25$~\kms and $5
< \delta V < 10$~\kms. For this metric, a value of 1 suggests a
successful disentanglement of the velocities of the two components,
while other values can give clues to the cause of failure in the
measurement. As examples: A value of -1 indicates instances when the
derived velocities of the two components are flipped (the velocity of
the young is measured by the old component and vice versa).  Values
between -1 and 1 are likely caused by an over estimation of the
velocity dispersion of one component. Very large positive or negative
values suggest the observed spectrum has been reproduced by only one
component and the other has been used only to minimize the $\chi^2$ of
the solution.

For all mock $\delta V$ groups, the distribution of the performance
metric peaks at 1, as it should, consistent with the result seen in
Section~\ref{sec:hypothesis} where we demonstrate that
MaNGA/MaNGA-like spectra of galaxies can contain sufficient
information on the kinematics of co-spatial stellar populations for
disentanglement. The distribution width increases at smaller model
velocity differences, $\delta V$, because the uncertainty in the
measured velocity difference is $\sim$constant in \kms. Even for AD
signals as low as 5-10~\kms, 67\% of the derived AD signals are within
10\% of the expected value.  Performance evaluation of this algorithm
against real data is presented in Section\ref{sec:ssp_simple}.

\begin{deluxetable*}{lcrrrrrrrrrrrrrrr}[ht!]
\tablewidth{0pt}
\tabletypesize{\footnotesize}
\tablecaption{Stelar Library Algorithm Performance Metrics Based on Realistic Mock Spectra}
\tablehead{
 \colhead{Algorithm} &
 \colhead{$\delta V_{\rm mock}$} &
 \multicolumn{2}{c}{$\Delta V_y$} &
 \colhead{} &
 \multicolumn{2}{c}{$\Delta V_o$} &
 \colhead{} &
 \multicolumn{2}{c}{$\delta V_{\rm Algo} / \delta V_{\rm mock}$} &
 \colhead{} &
 \multicolumn{2}{c}{$\Delta{\rm \pfrac}$} &
 \colhead{} &
 \multicolumn{3}{c}{Failure Fraction} \\  [0.05in] \cline{3-4} \cline{6-7} \cline{9-10} \cline{12-13} \cline{15-17}
 \multicolumn{17}{c}{} \\  [-0.08in]
 \colhead{} &
 \colhead{range} &
 \colhead{med} &
 \colhead{$\sigma_{\rm MAD}$} &
 \colhead{} &
 \colhead{med} &
 \colhead{$\sigma_{\rm MAD}$} &
 \colhead{} &
 \colhead{med} &
 \colhead{$\sigma_{\rm MAD}$} &
 \colhead{} &
 \colhead{median} &
 \colhead{$\sigma_{\rm MAD}$} &
 \colhead{} &
 \colhead{Cat} &
 \colhead{Flip} &
 \colhead{High} \\
 \colhead{} &
 \colhead{(\kms)} &
 \multicolumn{2}{c}{(\kms)} &
 \colhead{} &
 \multicolumn{2}{c}{(\kms)} &
 \colhead{} &
 \colhead{} &
 \colhead{} &
 \colhead{} &
 \colhead{} &
 \colhead{} &
 \multicolumn{4}{c}{} \\
 \colhead{(1)} &
 \colhead{(2)} &
 \multicolumn{1}{r}{(3)} &
 \multicolumn{1}{r}{(4)} &
 \colhead{} &
 \multicolumn{1}{r}{(5)} &
 \multicolumn{1}{r}{(6)} &
 \colhead{} &
 \multicolumn{1}{r}{(7)} &
 \multicolumn{1}{r}{(8)} &
 \colhead{} &
 \multicolumn{1}{r}{(9)} &
 \multicolumn{1}{r}{(10)} &
 \colhead{} &
 \multicolumn{1}{r}{(11)} &
 \multicolumn{1}{r}{(12)} &
 \multicolumn{1}{r}{(13)}
}
\startdata
Simple SL       & [10,35] & 10.66 & 2.26 && -4.68 &  5.19 && -0.14 & 0.43 &&  0.02 & 0.07 && 0.00 & 0.57 & 0.00 \\
Simple SL       & [35,50] &  6.20 & 3.52 && -4.34 &  7.61 &&  0.74 & 0.25 &&  0.01 & 0.04 && 0.00 & 0.26 & 0.00 \\
Simple SL       &   $>$50 &  4.07 & 2.63 && -0.86 &  8.30 &&  0.91 & 0.15 && -0.02 & 0.02 && 0.04 & 0.22 & 0.00 \\
 \multicolumn{17}{c}{} \\  [-0.08in]
2-Step 1-Bin SL & [10,35] & 12.06 & 3.43 && -4.52 &  4.86 && -0.18 & 0.41 &&  0.10 & 0.11 && 0.00 & 0.60 & 0.00 \\
2-Step 1-Bin SL & [35,50] &  6.39 & 3.88 && -3.54 &  7.40 &&  0.75 & 0.25 &&  0.01 & 0.05 && 0.00 & 0.30 & 0.00 \\
2-Step 1-Bin SL &   $>$50 &  3.62 & 2.35 &&  1.14 &  5.46 &&  0.96 & 0.09 && -0.02 & 0.02 && 0.02 & 0.17 & 0.00 \\
 \multicolumn{17}{c}{} \\  [-0.08in]
2-Step 3-Bin SL & [10,35] &  8.46 & 8.87 && -4.06 &  4.75 &&  0.41 & 0.54 && -0.04 & 0.03 && 0.02 & 0.36 & 0.04 \\
2-Step 3-Bin SL & [35,50] &  8.93 & 8.73 && -17.1 & 15.54 &&  0.41 & 0.39 && -0.05 & 0.03 && 0.02 & 0.26 & 0.07 \\
2-Step 3-Bin SL &   $>$50 &  8.11 & 7.80 && -25.8 & 27.23 &&  0.54 & 0.36 && -0.07 & 0.04 && 0.06 & 0.19 & 0.05
\enddata
\label{tab:SL_realistic_mocks_metric}
\tablecomments{Algorithms (column 1) are described in
  Sections~\ref{sec:stellar_simple} and
  \ref{sec:stellar_algo_evol}. The remaining columns are defined
  identically as in Table~\ref{tab:SSP_realistic_mocks_metric}.}
\end{deluxetable*}

\subsection{Performance of Stellar Library Algorithms on Mock Spectra}
\label{sec:stellar_simple_mock_performance}

Table~\ref{tab:simple_mocks_metric} presents metrics using our
`simple' mocks for all simple stellar library algorithms described and
discussed in Section~\ref{sec:stellar_simple}. The adopted best simple
algorithm corresponds to `Full' with template set B.

\begin{figure*}[ht!]
\begin{center}
\includegraphics[width=0.9\linewidth]{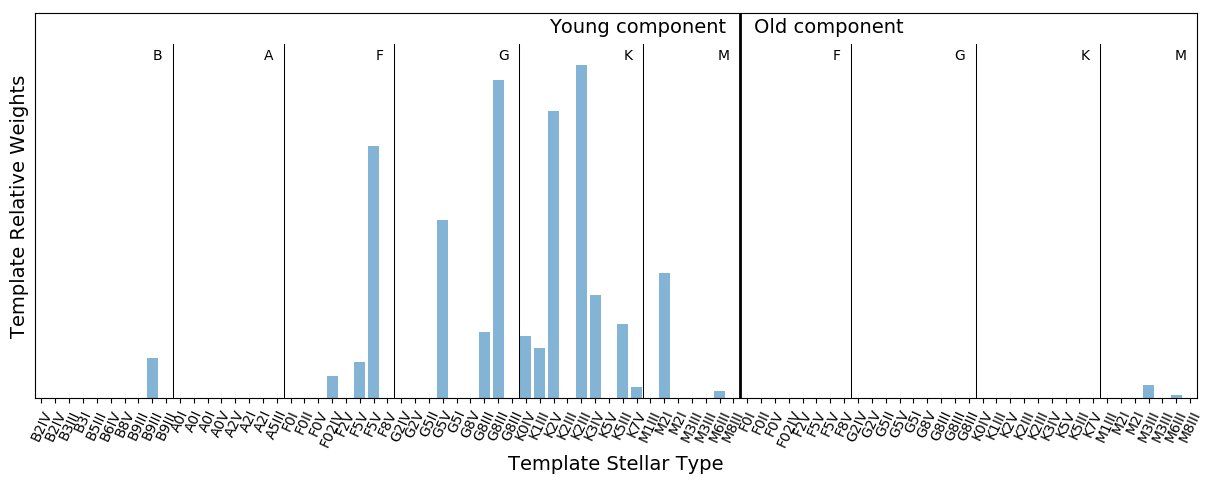}
\includegraphics[width=0.9\linewidth]{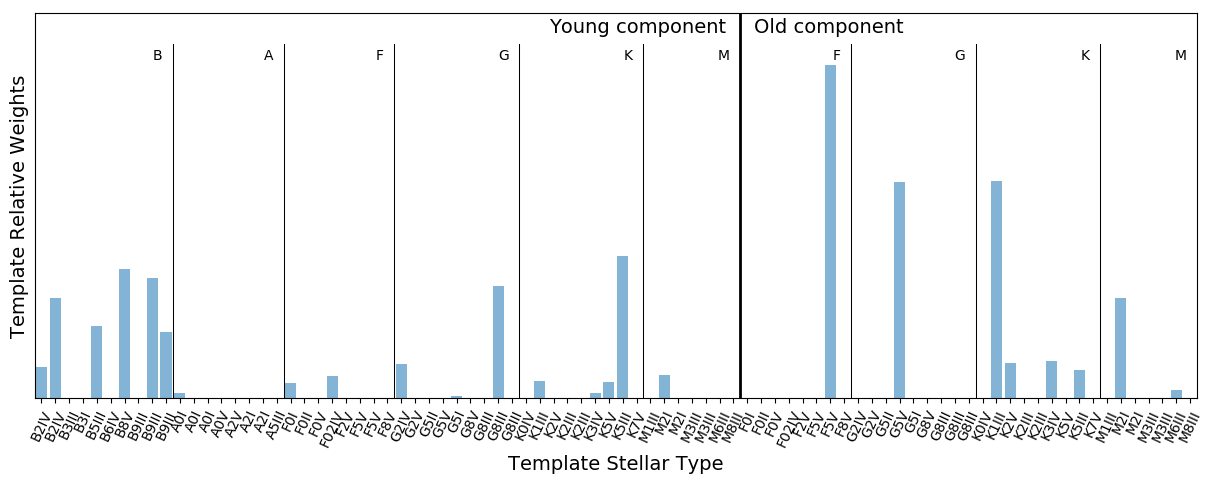}

\caption{Examples of a failed (top) and successful (bottom) fit to
  realistic mock spectra using our simple stellar library algorithm
  with template set C. The failed fit recovers a high value of
  \pfrac\ ($>$0.85) and incorrectly places too much weight in cool
  stars for the young stellar component.  The successful fit yields a
  stellar temperature distribution much more representative of
  expectations for young and old stellar populations
  (Figure~\ref{Weight_Dist_SSP}).}
\label{fig:high_frac_example}
\end{center}
\end{figure*}

Figure ~\ref{fig:high_frac_example} shows two instances of a failed
and successful fit of the simple stellar library algorithm using
template set C, where the failed case is characterized by a high
\pfrac\ value $>0.85$.

Table~\ref{tab:SL_realistic_mocks_metric} presents metrics using
`realistic' mocks for the best simple stellar library algorithm (Full
with template set B); the two-step stellar library algorithm (2-Step
1-Bin S); and our final two-step stellar library algorithm that
constrains the relative weight of young and old stellar components in
three stellar temperature bins (2-Step 3-Bin).

\setcounter{figure}{0}

\section{Maps}
\label{sec:maps}

\begin{figure*}
\center
\includegraphics[width=\linewidth]{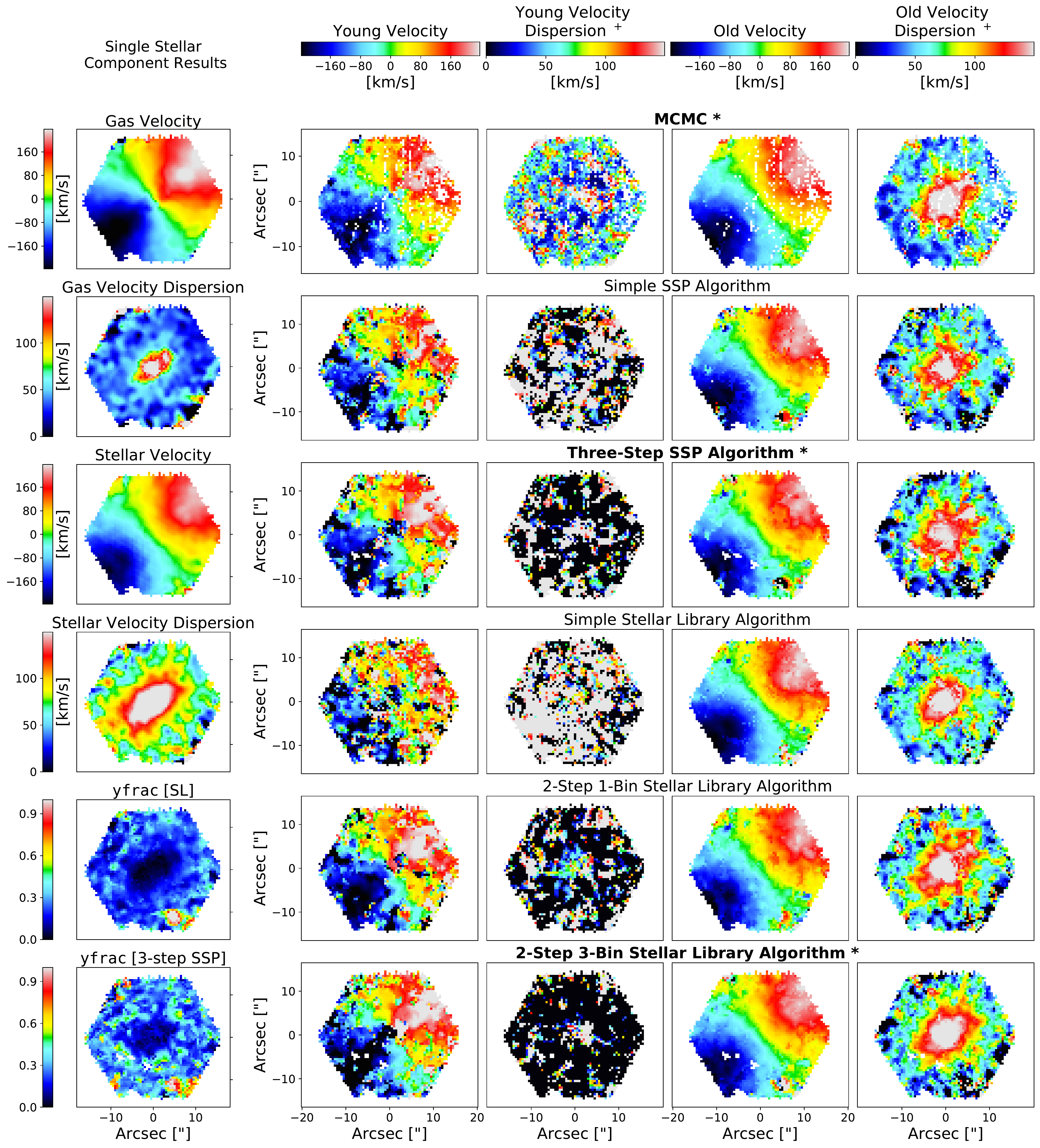}
\caption{Derived kinematics of the young and old stellar populations
  of MaNGA galaxy MID 1-209537 (8486-12701), formatted as
  Figure~\ref{8138_12704_bp}.}
\label{8486_12701_bp} 
\end{figure*}

\begin{figure*}
\center
\includegraphics[width=\linewidth]{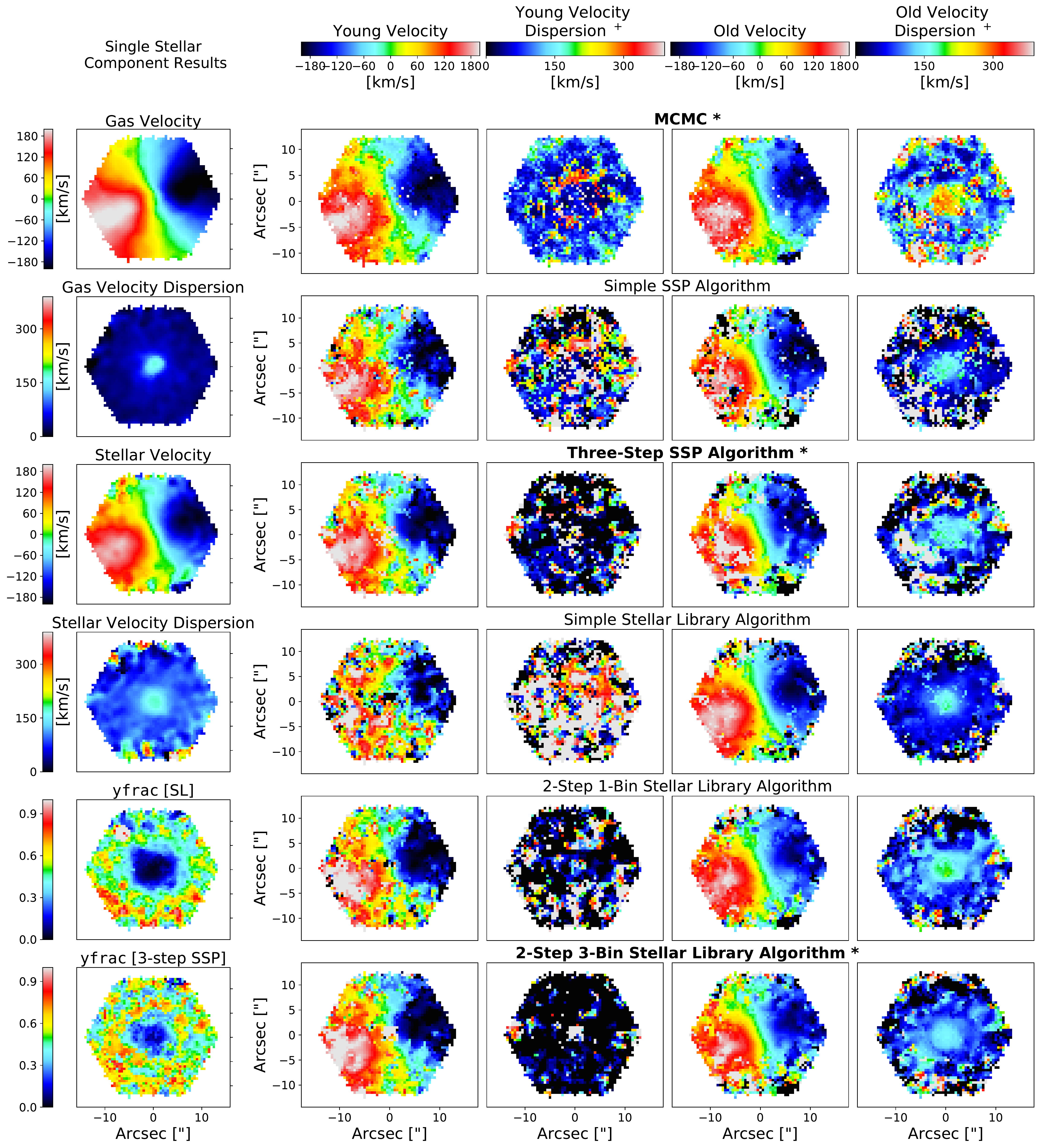}
\caption{Derived kinematics of the young and old stellar populations
  of MaNGA galaxy MID 1-532459 (8320-9102), formatted as
  Figure~\ref{8138_12704_bp}.}
\label{8320_9102_bp} 
\end{figure*}

\begin{figure*}
\center
\includegraphics[width=\linewidth]{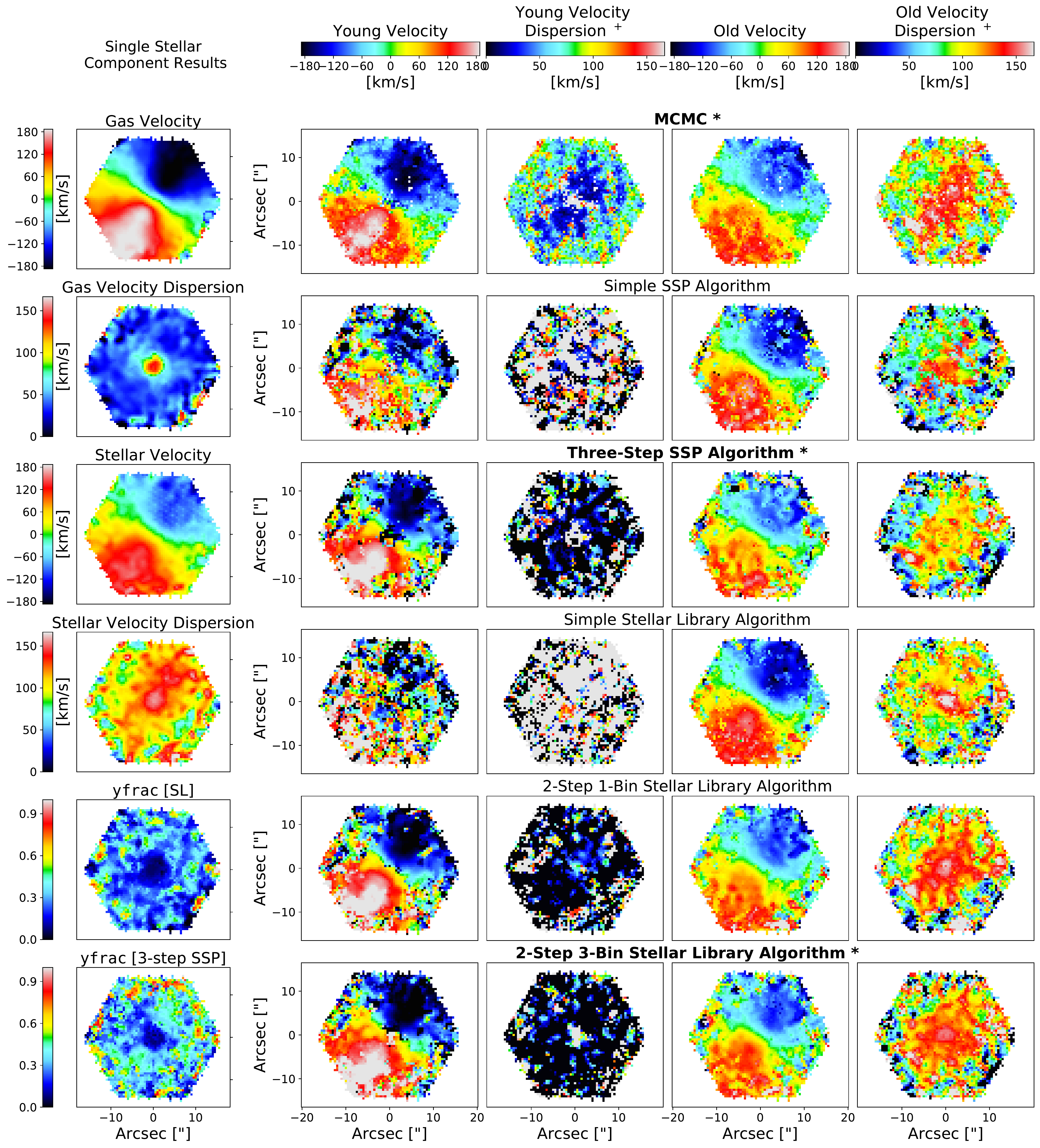}
\caption{Derived kinematics of the young and old stellar populations
  of MaNGA galaxy MID 1-251279 (8332-12705), formatted as
  Figure~\ref{8138_12704_bp}.}
\label{8332_12705_bp} 
\end{figure*}

\begin{figure*}
\center
\includegraphics[width=\linewidth]{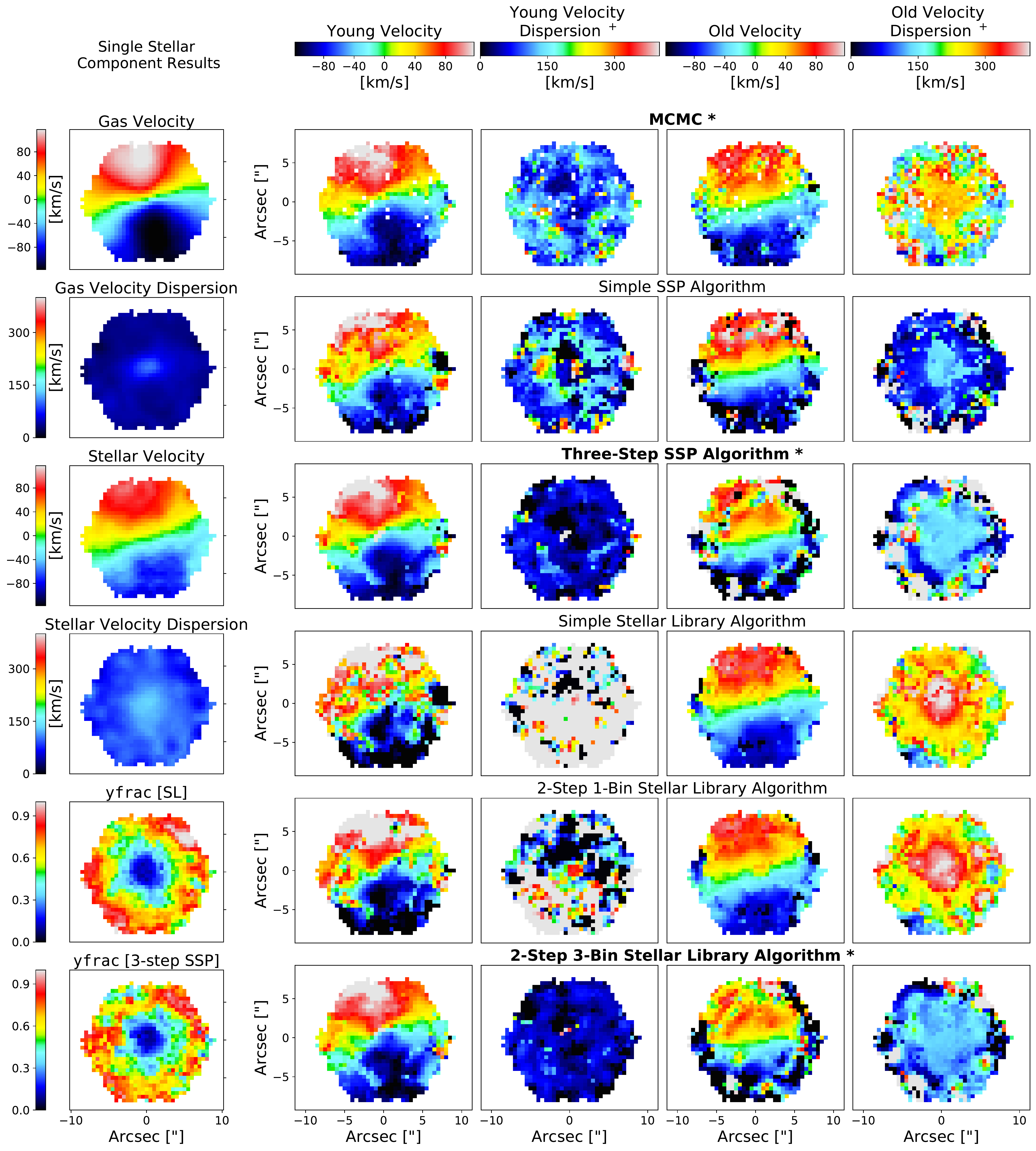}
\caption{Derived kinematics of the young and old stellar populations
  of MaNGA galaxy MID 1-542358 (8482-3702), formatted as
  Figure~\ref{8138_12704_bp}.}
\label{8482_3702_bp} 
\end{figure*}

\begin{figure*}
\center
\includegraphics[width=\linewidth]{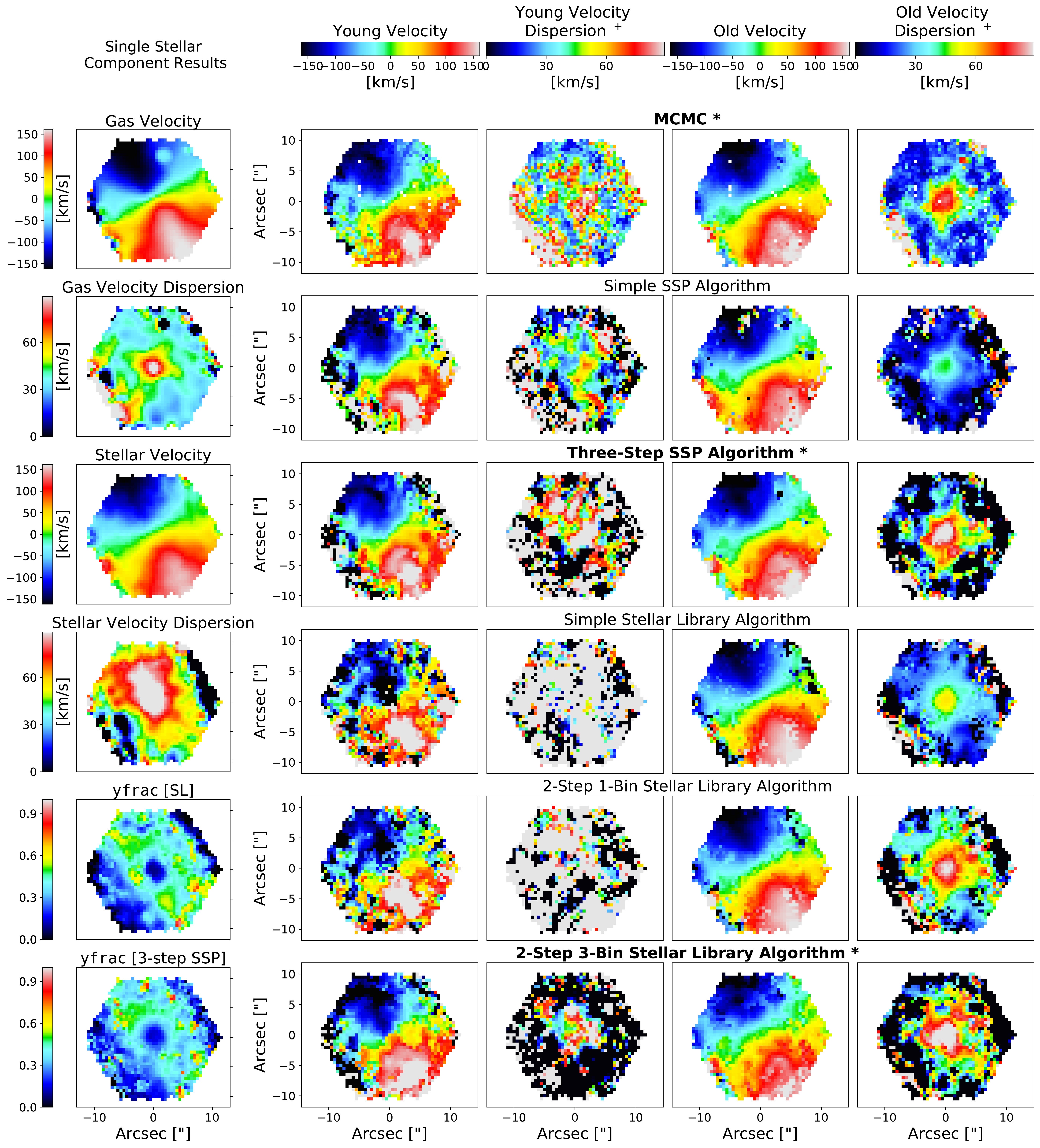}
\caption{Derived kinematics of the young and old stellar populations
  of MaNGA galaxy MID 1-265988 (8329-6103), formatted as
  Figure~\ref{8138_12704_bp}.}
\label{8329_6103_bp} 
\end{figure*}

\begin{figure*}
\center
\includegraphics[width=\linewidth]{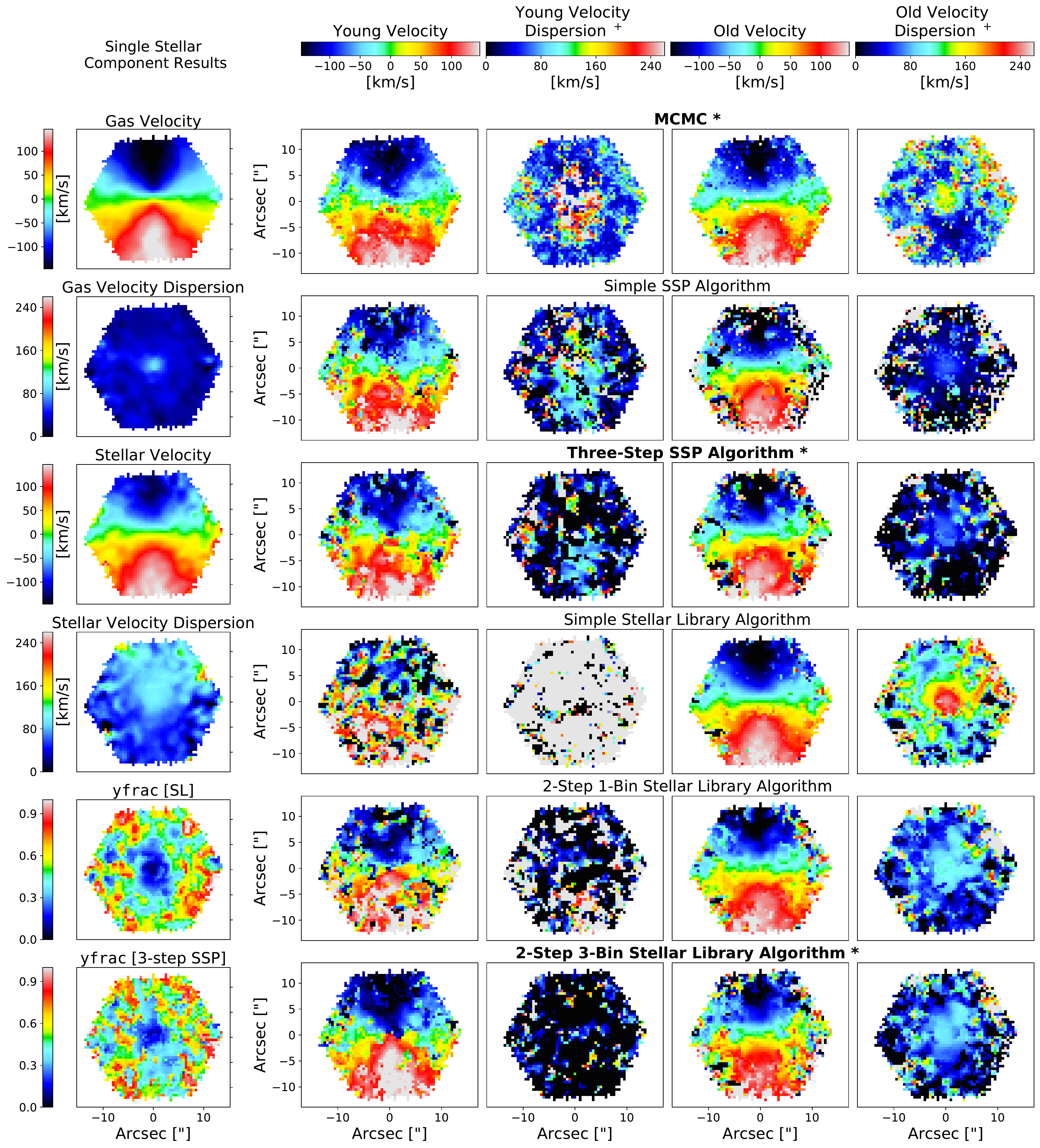}
\caption{Derived kinematics of the young and old stellar populations
  of MaNGA galaxy MID 1-209199 (8485-9102), formatted as
  Figure~\ref{8138_12704_bp}.}
\label{8485_9102_bp} 
\end{figure*}

Figures~\ref{8486_12701_bp}-\ref{8485_9102_bp} contain maps of the
following information for remaining six galaxies in our sample,
formatted as Figure~\ref{8138_12704_bp}. Maps are displayed at
  the pixel level with no interpolation between pixels.

First (left) column: Kinematics (velocity and dispersion) of the
ionized gas and single-component stellar population as well as the
decomposition of the single-component stellar population into young
and old, as presented by \pfrac\ are presented in the following
Figures. These are derived as the initial step of all algorithms
developed in this paper.

Second column: Stellar velocity for the young stellar component for
the five methods outlining the development in this paper. Top to
bottom these are the simple SSP algorithm, and then two-step SSP
algorithms with inferred initial conditions, and then constraints on
\pfrac; the two- and three-bin stellar library algorithms (with
similar inferred initial conditions and constraints on \pfrac.

Third column: Stellar velocity dispersion for the young stellar
component, ordered as the previous column.

Fourth column: Stellar velocity for the old stellar component, ordered
as the previous column.

Fifth column: Stellar velocity dispersion for the old stellar
component, ordered as the previous column.

\end{appendices}

\clearpage

\bibliographystyle{apj}
\bibliography{ad}

\begin{thebibliography}{}
\expandafter\ifx\csname natexlab\endcsname\relax\def\natexlab#1{#1}\fi

\bibitem[{{Abadi} {et~al.}(2003){Abadi}, {Navarro}, {Steinmetz}, \&
  {Eke}}]{Abadietal2003b}
{Abadi}, M.~G., {Navarro}, J.~F., {Steinmetz}, M., \& {Eke}, V.~R. 2003, \apj,
  597, 21

\bibitem[{{Abolfathi} {et~al.}(2018){Abolfathi}, {Aguado}, {Aguilar}, {Allende
  Prieto}, {Almeida}, {Ananna}, {Anders}, {Anderson}, {Andrews}, \&
  {Anguiano}}]{SDSS_DR14}
{Abolfathi}, B., {Aguado}, D.~S., {Aguilar}, G., {et~al.} 2018, \apjs, 235, 42

\bibitem[{{Aguado} {et~al.}(2019){Aguado}, {Ahumada}, {Almeida}, {Anderson},
  {Andrews}, {Anguiano}, {Aquino Ort{\'\i}z}, {Arag{\'o}n-Salamanca},
  {Argudo-Fern{\'a}ndez}, \& {Aubert}}]{SDSS_DR15}
{Aguado}, D.~S., {Ahumada}, R., {Almeida}, A., {et~al.} 2019, \apjs, 240, 23

\bibitem[{{Andersen} \& {Bershady}(2013)}]{Andersen2013}
{Andersen}, D.~R., \& {Bershady}, M.~A. 2013, \apj, 768, 41

\bibitem[{{Aumer} \& {Binney}(2009)}]{Aumer09}
{Aumer}, M., \& {Binney}, J.~J. 2009, \mnras, 397, 1286

\bibitem[{{Beasley} {et~al.}(2015){Beasley}, {San Roman}, {Gallart},
  {Sarajedini}, \& {Aparicio}}]{Beasley15}
{Beasley}, M.~A., {San Roman}, I., {Gallart}, C., {Sarajedini}, A., \&
  {Aparicio}, A. 2015, \mnras, 451, 3400

\bibitem[{{Belfiore} {et~al.}(2016){Belfiore}, {Maiolino}, {Maraston},
  {Emsellem}, {Bershady}, {Masters}, {Yan}, {Bizyaev}, {Boquien}, {Brownstein},
  {Bundy}, {Drory}, {Heckman}, {Law}, {Roman-Lopes}, {Pan}, {Stanghellini},
  {Thomas}, {Weijmans}, \& {Westfall}}]{Belfiore2016}
{Belfiore}, F., {Maiolino}, R., {Maraston}, C., {et~al.} 2016, \mnras, 461,
  3111

\bibitem[{{Benson} {et~al.}(2004){Benson}, {Lacey}, {Frenk}, {Baugh}, \&
  {Cole}}]{Benson2004}
{Benson}, A.~J., {Lacey}, C.~G., {Frenk}, C.~S., {Baugh}, C.~M., \& {Cole}, S.
  2004, \mnras, 351, 1215

\bibitem[{{Bird} {et~al.}(2013){Bird}, {Kazantzidis}, {Weinberg}, {Guedes},
  {Callegari}, {Mayer}, \& {Madau}}]{Bird13}
{Bird}, J.~C., {Kazantzidis}, S., {Weinberg}, D.~H., {et~al.} 2013, \apj, 773,
  43

\bibitem[{{Blanton} {et~al.}(2011){Blanton}, {Kazin}, {Muna}, {Weaver}, \&
  {Price-Whelan}}]{Blanton11}
{Blanton}, M.~R., {Kazin}, E., {Muna}, D., {Weaver}, B.~A., \& {Price-Whelan},
  A. 2011, \aj, 142, 31

\bibitem[{{Blanton} {et~al.}(2017){Blanton}, {Bershady}, {Abolfathi},
  {Albareti}, {Allende Prieto}, {Almeida}, {Alonso-Garc{\'{\i}}a}, {Anders},
  {Anderson}, {Andrews}, \& et~al.}]{Blanton17}
{Blanton}, M.~R., {Bershady}, M.~A., {Abolfathi}, B., {et~al.} 2017, \aj, 154,
  28

\bibitem[{{Bournaud} {et~al.}(2009){Bournaud}, {Elmegreen}, \&
  {Martig}}]{Bournaud09}
{Bournaud}, F., {Elmegreen}, B.~G., \& {Martig}, M. 2009, \apjl, 707, L1

\bibitem[{{Brook} {et~al.}(2004){Brook}, {Kawata}, {Gibson}, \&
  {Freeman}}]{Brook04}
{Brook}, C.~B., {Kawata}, D., {Gibson}, B.~K., \& {Freeman}, K.~C. 2004, \apj,
  612, 894

\bibitem[{{Bruzual} \& {Charlot}(2003)}]{Bruzual03}
{Bruzual}, G., \& {Charlot}, S. 2003, \mnras, 344, 1000

\bibitem[{{Bundy} {et~al.}(2015){Bundy}, {Bershady}, {Law}, {Yan}, {Drory},
  {MacDonald}, {Wake}, {Cherinka}, {S{\'a}nchez-Gallego}, {Weijmans}, {Thomas},
  {Tremonti}, {Masters}, {Coccato}, {Diamond-Stanic}, {Arag{\'o}n-Salamanca},
  {Avila-Reese}, {Badenes}, {Falc{\'o}n-Barroso}, {Belfiore}, {Bizyaev},
  {Blanc}, {Bland-Hawthorn}, {Blanton}, {Brownstein}, {Byler}, {Cappellari},
  {Conroy}, {Dutton}, {Emsellem}, {Etherington}, {Frinchaboy}, {Fu}, {Gunn},
  {Harding}, {Johnston}, {Kauffmann}, {Kinemuchi}, {Klaene}, {Knapen},
  {Leauthaud}, {Li}, {Lin}, {Maiolino}, {Malanushenko}, {Malanushenko}, {Mao},
  {Maraston}, {McDermid}, {Merrifield}, {Nichol}, {Oravetz}, {Pan}, {Parejko},
  {Sanchez}, {Schlegel}, {Simmons}, {Steele}, {Steinmetz}, {Thanjavur},
  {Thompson}, {Tinker}, {van den Bosch}, {Westfall}, {Wilkinson}, {Wright},
  {Xiao}, \& {Zhang}}]{Bundy15}
{Bundy}, K., {Bershady}, M.~A., {Law}, D.~R., {et~al.} 2015, \apj, 798, 7

\bibitem[{{Cappellari}(2017)}]{Cappellari2017}
{Cappellari}, M. 2017, \mnras, 466, 798

\bibitem[{{Cappellari} {et~al.}(2007){Cappellari}, {Emsellem}, {Bacon},
  {Bureau}, {Davies}, {de Zeeuw}, {Falc{\'o}n-Barroso}, {Krajnovi{\'c}},
  {Kuntschner}, {McDermid}, {Peletier}, {Sarzi}, {van den Bosch}, \& {van de
  Ven}}]{Cappellari2007}
{Cappellari}, M., {Emsellem}, E., {Bacon}, R., {et~al.} 2007, \mnras, 379, 418

\bibitem[{{Carlberg} {et~al.}(1985){Carlberg}, {Dawson}, {Hsu}, \&
  {Vandenberg}}]{Carlberg85}
{Carlberg}, R.~G., {Dawson}, P.~C., {Hsu}, T., \& {Vandenberg}, D.~A. 1985,
  \apj, 294, 674

\bibitem[{{Cenarro} {et~al.}(2001){Cenarro}, {Cardiel}, {Gorgas}, {Peletier},
  {Vazdekis}, \& {Prada}}]{Cenarroetal2001}
{Cenarro}, A.~J., {Cardiel}, N., {Gorgas}, J., {et~al.} 2001, \mnras, 326, 959

\bibitem[{{Coccato} {et~al.}(2011){Coccato}, {Morelli}, {Corsini}, {Buson},
  {Pizzella}, {Vergani}, \& {Bertola}}]{Coccatoetal2011}
{Coccato}, L., {Morelli}, L., {Corsini}, E.~M., {et~al.} 2011, \mnras, 412,
  L113

\bibitem[{{Collins} {et~al.}(2011){Collins}, {Chapman}, {Ibata}, {Irwin},
  {Rich}, {Ferguson}, {Lewis}, {Tanvir}, \& {Koch}}]{Collins2011}
{Collins}, M.~L.~M., {Chapman}, S.~C., {Ibata}, R.~A., {et~al.} 2011, \mnras,
  413, 1548

\bibitem[{{Croom} {et~al.}(2012){Croom}, {Lawrence}, {Bland-Hawthorn},
  {Bryant}, {Fogarty}, {Richards}, {Goodwin}, {Farrell}, {Miziarski}, {Heald},
  {Jones}, {Lee}, {Colless}, {Brough}, {Hopkins}, {Bauer}, {Birchall}, {Ellis},
  {Horton}, {Leon-Saval}, {Lewis}, {L{\'o}pez-S{\'a}nchez}, {Min}, {Trinh}, \&
  {Trowland}}]{Croom12}
{Croom}, S.~M., {Lawrence}, J.~S., {Bland-Hawthorn}, J., {et~al.} 2012, \mnras,
  421, 872

\bibitem[{{Dalcanton} \& {Stilp}(2010)}]{Dalcanton2010}
{Dalcanton}, J.~J., \& {Stilp}, A.~M. 2010, \apj, 721, 547

\bibitem[{{Davis} {et~al.}(2013){Davis}, {Alatalo}, {Bureau}, {Cappellari},
  {Scott}, {Young}, {Blitz}, {Crocker}, {Bayet}, {Bois}, {Bournaud}, {Davies},
  {de Zeeuw}, {Duc}, {Emsellem}, {Khochfar}, {Krajnovi{\'c}}, {Kuntschner},
  {Lablanche}, {McDermid}, {Morganti}, {Naab}, {Oosterloo}, {Sarzi}, {Serra},
  \& {Weijmans}}]{Davis2013}
{Davis}, T.~A., {Alatalo}, K., {Bureau}, M., {et~al.} 2013, \mnras, 429, 534

\bibitem[{{De Bruyne} {et~al.}(2004){De Bruyne}, {De Rijcke}, {Dejonghe}, \&
  {Zeilinger}}]{DeBruyne04}
{De Bruyne}, V., {De Rijcke}, S., {Dejonghe}, H., \& {Zeilinger}, W.~W. 2004,
  \mnras, 349, 461

\bibitem[{{den Brok} {et~al.}(2020){den Brok}, {Carollo}, {Erroz-Ferrer},
  {Fagioli}, {Brinchmann}, {Emsellem}, {Krajnovi{\'c}}, {Marino}, {Onodera},
  {Tacchella}, {Weilbacher}, \& {Woo}}]{denBrok2020}
{den Brok}, M., {Carollo}, C.~M., {Erroz-Ferrer}, S., {et~al.} 2020, \mnras,
  491, 4089

\bibitem[{{Dorman} {et~al.}(2015){Dorman}, {Guhathakurta}, {Seth}, {Weisz},
  {Bell}, {Dalcanton}, {Gilbert}, {Hamren}, {Lewis}, {Skillman}, {Toloba}, \&
  {Williams}}]{Dorman15}
{Dorman}, C.~E., {Guhathakurta}, P., {Seth}, A.~C., {et~al.} 2015, \apj, 803,
  24

\bibitem[{{Drory} {et~al.}(2015){Drory}, {MacDonald}, {Bershady}, {Bundy},
  {Gunn}, {Law}, {Smith}, {Stoll}, {Tremonti}, {Wake}, {Yan}, {Weijmans},
  {Byler}, {Cherinka}, {Cope}, {Eigenbrot}, {Harding}, {Holder}, {Huehnerhoff},
  {Jaehnig}, {Jansen}, {Klaene}, {Paat}, {Percival}, \& {Sayres}}]{Drory15}
{Drory}, N., {MacDonald}, N., {Bershady}, M.~A., {et~al.} 2015, \aj, 149, 77

\bibitem[{{Falc{\'o}n-Barroso} {et~al.}(2011){Falc{\'o}n-Barroso},
  {S{\'a}nchez-Bl{\'a}zquez}, {Vazdekis}, {Ricciardelli}, {Cardiel}, {Cenarro},
  {Gorgas}, \& {Peletier}}]{Falcon-Barroso11}
{Falc{\'o}n-Barroso}, J., {S{\'a}nchez-Bl{\'a}zquez}, P., {Vazdekis}, A.,
  {et~al.} 2011, \aap, 532, A95

\bibitem[{{Few} {et~al.}(2012){Few}, {Gibson}, {Courty}, {Michel-Dansac},
  {Brook}, \& {Stinson}}]{Few2012}
{Few}, C.~G., {Gibson}, B.~K., {Courty}, S., {et~al.} 2012, \aap, 547, A63

\bibitem[{{Forbes} {et~al.}(2012){Forbes}, {Krumholz}, \&
  {Burkert}}]{Forbesetal2012}
{Forbes}, J., {Krumholz}, M., \& {Burkert}, A. 2012, \apj, 754, 48

\bibitem[{{Foreman-Mackey} {et~al.}(2013){Foreman-Mackey}, {Hogg}, {Lang}, \&
  {Goodman}}]{Foremanetal2013}
{Foreman-Mackey}, D., {Hogg}, D.~W., {Lang}, D., \& {Goodman}, J. 2013, \pasp,
  125, 306

\bibitem[{{F{\"o}rster Schreiber} {et~al.}(2009){F{\"o}rster Schreiber},
  {Genzel}, {Bouch{\'e}}, {Cresci}, {Davies}, {Buschkamp}, {Shapiro},
  {Tacconi}, {Hicks}, {Genel}, {Shapley}, {Erb}, {Steidel}, {Lutz},
  {Eisenhauer}, {Gillessen}, {Sternberg}, {Renzini}, {Cimatti}, {Daddi},
  {Kurk}, {Lilly}, {Kong}, {Lehnert}, {Nesvadba}, {Verma}, {McCracken},
  {Arimoto}, {Mignoli}, \& {Onodera}}]{Forster-Schreiber09}
{F{\"o}rster Schreiber}, N.~M., {Genzel}, R., {Bouch{\'e}}, N., {et~al.} 2009,
  \apj, 706, 1364

\bibitem[{{Gonz{\'a}lez Delgado} {et~al.}(2015){Gonz{\'a}lez Delgado},
  {Garc{\'{\i}}a-Benito}, {P{\'e}rez}, {Cid Fernandes}, {de Amorim},
  {Cortijo-Ferrero}, {Lacerda}, {L{\'o}pez Fern{\'a}ndez}, {Vale-Asari},
  {S{\'a}nchez}, {Moll{\'a}}, {Ruiz-Lara}, {S{\'a}nchez-Bl{\'a}zquez},
  {Walcher}, {Alves}, {Aguerri}, {Bekerait{\'e}}, {Bland-Hawthorn}, {Galbany},
  {Gallazzi}, {Husemann}, {Iglesias-P{\'a}ramo}, {Kalinova},
  {L{\'o}pez-S{\'a}nchez}, {Marino}, {M{\'a}rquez}, {Masegosa}, {Mast},
  {M{\'e}ndez-Abreu}, {Mendoza}, {del Olmo}, {P{\'e}rez}, {Quirrenbach}, \&
  {Zibetti}}]{Gonzalez-Delgado15}
{Gonz{\'a}lez Delgado}, R.~M., {Garc{\'{\i}}a-Benito}, R., {P{\'e}rez}, E.,
  {et~al.} 2015, \aap, 581, A103

\bibitem[{{Goodman} \& {Weare}(2010)}]{Goodman&Weare2010}
{Goodman}, J., \& {Weare}, J. 2010, Communications in Applied Mathematics and
  Computational Science, Vol.~5, No.~1, p.~65-80, 2010, 5, 65

\bibitem[{{Gunn} {et~al.}(2006){Gunn}, {Siegmund}, {Mannery}, {Knapp}, {York},
  {Boroski}, {Kent}, {Lupton}, {Rockosi}, {Evans}, {Waddell}, {Anderson},
  {Annis}, {Barentine}, {Bartoszek}, {Bastian}, {Bracker}, \& et~al.}]{Gunn06}
{Gunn}, J.~E., {Siegmund}, W.~A., {Mannery}, E.~J., {et~al.} 2006, \aj, 131,
  2332

\bibitem[{{Helmi} {et~al.}(2012){Helmi}, {Sales}, {Starkenburg}, {Starkenburg},
  {Vera-Ciro}, {De Lucia}, \& {Li}}]{Helmi2012}
{Helmi}, A., {Sales}, L.~V., {Starkenburg}, E., {et~al.} 2012, \apjl, 758, L5

\bibitem[{{Holmberg} {et~al.}(2007){Holmberg}, {Nordstr{\"o}m}, \&
  {Andersen}}]{Holmbergetal2007}
{Holmberg}, J., {Nordstr{\"o}m}, B., \& {Andersen}, J. 2007, \aap, 475, 519

\bibitem[{{House} {et~al.}(2011){House}, {Brook}, {Gibson},
  {S{\'a}nchez-Bl{\'a}zquez}, {Courty}, {Few}, {Governato}, {Kawata},
  {Ro{\v{s}}kar}, {Steinmetz}, {Stinson}, \& {Teyssier}}]{House2011}
{House}, E.~L., {Brook}, C.~B., {Gibson}, B.~K., {et~al.} 2011, \mnras, 415,
  2652

\bibitem[{{Huang} \& {Carlberg}(1997)}]{HuangCarlberg1997}
{Huang}, S., \& {Carlberg}, R.~G. 1997, \apj, 480, 503

\bibitem[{{Johnston} {et~al.}(2013){Johnston}, {Merrifield},
  {Arag{\'o}n-Salamanca}, \& {Cappellari}}]{Johnstonetal2013}
{Johnston}, E.~J., {Merrifield}, M.~R., {Arag{\'o}n-Salamanca}, A., \&
  {Cappellari}, M. 2013, \mnras, 428, 1296

\bibitem[{{Katkov} \& {Chilingarian}(2012)}]{Katkov_2011}
{Katkov}, I.~Y., \& {Chilingarian}, I.~V. 2012, in IAU Symposium, Vol. 284, The
  Spectral Energy Distribution of Galaxies - SED 2011, ed. R.~J. {Tuffs} \&
  C.~C. {Popescu}, 69--71

\bibitem[{{Kokubo} \& {Ida}(1992)}]{KokuboIda1992}
{Kokubo}, E., \& {Ida}, S. 1992, \pasj, 44, 601

\bibitem[{{Kroupa}(2001)}]{Kroupa01}
{Kroupa}, P. 2001, \mnras, 322, 231

\bibitem[{{Law} {et~al.}(2007){Law}, {Steidel}, {Erb}, {Larkin}, {Pettini},
  {Shapley}, \& {Wright}}]{Law07}
{Law}, D.~R., {Steidel}, C.~C., {Erb}, D.~K., {et~al.} 2007, \apj, 669, 929

\bibitem[{{Law} {et~al.}(2015){Law}, {Yan}, {Bershady}, {Bundy}, {Cherinka},
  {Drory}, {MacDonald}, {S{\'a}nchez-Gallego}, {Wake}, {Weijmans}, {Blanton},
  {Klaene}, {Moran}, {Sanchez}, \& {Zhang}}]{Law15}
{Law}, D.~R., {Yan}, R., {Bershady}, M.~A., {et~al.} 2015, \aj, 150, 19

\bibitem[{{Law} {et~al.}(2016){Law}, {Cherinka}, {Yan}, {Andrews}, {Bershady},
  {Bizyaev}, {Blanc}, {Blanton}, {Bolton}, {Brownstein}, {Bundy}, {Chen},
  {Drory}, {D'Souza}, {Fu}, {Jones}, {Kauffmann}, {MacDonald}, {Masters},
  {Newman}, {Parejko}, {S{\'a}nchez-Gallego}, {S{\'a}nchez}, {Schlegel},
  {Thomas}, {Wake}, {Weijmans}, {Westfall}, \& {Zhang}}]{Lawetal2016_MaNGA_DRP}
{Law}, D.~R., {Cherinka}, B., {Yan}, R., {et~al.} 2016, \aj, 152, 83

\bibitem[{{Leaman} {et~al.}(2017){Leaman}, {Mendel}, {Wisnioski}, {Brooks},
  {Beasley}, {Starkenburg}, {Martig}, {Battaglia}, {Christensen}, {Cole}, {de
  Boer}, \& {Wills}}]{Leaman17}
{Leaman}, R., {Mendel}, J.~T., {Wisnioski}, E., {et~al.} 2017, \mnras, 472,
  1879

\bibitem[{{Levy} {et~al.}(2018){Levy}, {Bolatto}, {Teuben}, {S{\'a}nchez},
  {Barrera-Ballesteros}, {Blitz}, {Colombo}, {Garc{\'\i}a-Benito},
  {Herrera-Camus}, {Husemann}, {Kalinova}, {Lan}, {Leung}, {Mast}, {Utomo},
  {van de Ven}, {Vogel}, \& {Wong}}]{Levy2018}
{Levy}, R.~C., {Bolatto}, A.~D., {Teuben}, P., {et~al.} 2018, \apj, 860, 92

\bibitem[{{Licquia} {et~al.}(2016){Licquia}, {Newman}, \&
  {Bershady}}]{Licquia16}
{Licquia}, T.~C., {Newman}, J.~A., \& {Bershady}, M.~A. 2016, \apj, 833, 220

\bibitem[{{Licquia} {et~al.}(2015){Licquia}, {Newman}, \&
  {Brinchmann}}]{Licquia15b}
{Licquia}, T.~C., {Newman}, J.~A., \& {Brinchmann}, J. 2015, \apj, 809, 96

\bibitem[{{Martig} {et~al.}(2014){Martig}, {Minchev}, \& {Flynn}}]{Martig14b}
{Martig}, M., {Minchev}, I., \& {Flynn}, C. 2014, \mnras, 443, 2452

\bibitem[{{Martinsson} {et~al.}(2016){Martinsson}, {Verheijen}, {Bershady},
  {Westfall}, {Andersen}, \& {Swaters}}]{Martinsson2016}
{Martinsson}, T. P.~K., {Verheijen}, M. A.~W., {Bershady}, M.~A., {et~al.}
  2016, \aap, 585, A99

\bibitem[{{Nordstr{\"o}m} {et~al.}(2004){Nordstr{\"o}m}, {Mayor}, {Andersen},
  {Holmberg}, {Pont}, {J{\o}rgensen}, {Olsen}, {Udry}, \&
  {Mowlavi}}]{Nordstrom04}
{Nordstr{\"o}m}, B., {Mayor}, M., {Andersen}, J., {et~al.} 2004, \aap, 418, 989

\bibitem[{{Pickles}(1998)}]{Pickles98}
{Pickles}, A.~J. 1998, \pasp, 110, 863

\bibitem[{{Pietrinferni} {et~al.}(2004){Pietrinferni}, {Cassisi}, {Salaris}, \&
  {Castelli}}]{BaSTI_1}
{Pietrinferni}, A., {Cassisi}, S., {Salaris}, M., \& {Castelli}, F. 2004, \apj,
  612, 168

\bibitem[{{Pinna} {et~al.}(2019){Pinna}, {Falc{\'o}n-Barroso}, {Martig},
  {Sarzi}, {Coccato}, {Iodice}, {Corsini}, {de Zeeuw}, {Gadotti}, {Leaman},
  {Lyubenova}, {McDermid}, {Minchev}, {Morelli}, {van de Ven}, \&
  {Viaene}}]{Pinna2019a}
{Pinna}, F., {Falc{\'o}n-Barroso}, J., {Martig}, M., {et~al.} 2019, \aap, 623,
  A19

\bibitem[{{Poci} {et~al.}(2019){Poci}, {McDermid}, {Zhu}, \& {van de
  Ven}}]{Poci2019}
{Poci}, A., {McDermid}, R.~M., {Zhu}, L., \& {van de Ven}, G. 2019, \mnras,
  487, 3776

\bibitem[{{Quirk} {et~al.}(2019){Quirk}, {Guhathakurta}, {Chemin}, {Dorman},
  {Gilbert}, {Seth}, {Williams}, \& {Dalcanton}}]{Quirketal2019}
{Quirk}, A., {Guhathakurta}, P., {Chemin}, L., {et~al.} 2019, \apj, 871, 11

\bibitem[{{Rix} \& {White}(1992)}]{Rix92}
{Rix}, H.-W., \& {White}, S.~D.~M. 1992, \mnras, 254, 389

\bibitem[{{Ruiz-Lara} {et~al.}(2016){Ruiz-Lara}, {Few}, {Gibson}, {P{\'e}rez},
  {Florido}, {Minchev}, \& {S{\'a}nchez-Bl{\'a}zquez}}]{RuizLara2016}
{Ruiz-Lara}, T., {Few}, C.~G., {Gibson}, B.~K., {et~al.} 2016, \aap, 586, A112

\bibitem[{{Ry{\'s}} {et~al.}(2013){Ry{\'s}}, {Falc{\'o}n-Barroso}, \& {van de
  Ven}}]{Rys2013}
{Ry{\'s}}, A., {Falc{\'o}n-Barroso}, J., \& {van de Ven}, G. 2013, \mnras, 428,
  2980

\bibitem[{{S{\'a}nchez} {et~al.}(2012){S{\'a}nchez}, {Kennicutt}, {Gil de Paz},
  {van de Ven}, {V{\'{\i}}lchez}, {Wisotzki}, {Walcher}, {Mast}, {Aguerri},
  {Albiol-P{\'e}rez}, {Alonso-Herrero}, {Alves}, {Bakos}, {Bart{\'a}kov{\'a}},
  {Bland-Hawthorn}, {Boselli}, {Bomans}, {Castillo-Morales}, {Cortijo-Ferrero},
  {de Lorenzo-C{\'a}ceres}, {Del Olmo}, {Dettmar}, {D{\'{\i}}az}, {Ellis},
  {Falc{\'o}n-Barroso}, {Flores}, {Gallazzi}, {Garc{\'{\i}}a-Lorenzo},
  {Gonz{\'a}lez Delgado}, {Gruel}, {Haines}, {Hao}, {Husemann},
  {Igl{\'e}sias-P{\'a}ramo}, {Jahnke}, {Johnson}, {Jungwiert}, {Kalinova},
  {Kehrig}, {Kupko}, {L{\'o}pez-S{\'a}nchez}, {Lyubenova}, {Marino},
  {M{\'a}rmol-Queralt{\'o}}, {M{\'a}rquez}, {Masegosa}, {Meidt},
  {Mendez-Abreu}, {Monreal-Ibero}, {Montijo}, {Mour{\~a}o}, {Palacios-Navarro},
  {Papaderos}, {Pasquali}, {Peletier}, {P{\'e}rez}, {P{\'e}rez}, {Quirrenbach},
  {Rela{\~n}o}, {Rosales-Ortega}, {Roth}, {Ruiz-Lara},
  {S{\'a}nchez-Bl{\'a}zquez}, {Sengupta}, {Singh}, {Stanishev}, {Trager},
  {Vazdekis}, {Viironen}, {Wild}, {Zibetti}, \& {Ziegler}}]{CALIFA}
{S{\'a}nchez}, S.~F., {Kennicutt}, R.~C., {Gil de Paz}, A., {et~al.} 2012,
  \aap, 538, A8

\bibitem[{{S{\'a}nchez-Bl{\'a}zquez} {et~al.}(2014){S{\'a}nchez-Bl{\'a}zquez},
  {Rosales-Ortega}, {M{\'e}ndez-Abreu}, {P{\'e}rez}, {S{\'a}nchez}, {Zibetti},
  {Aguerri}, {Bland-Hawthorn}, {Catal{\'a}n-Torrecilla}, {Cid Fernandes}, {de
  Amorim}, {de Lorenzo-Caceres}, {Falc{\'o}n-Barroso}, {Galazzi},
  {Garc{\'{\i}}a Benito}, {Gil de Paz}, {Gonz{\'a}lez Delgado}, {Husemann},
  {Iglesias-P{\'a}ramo}, {Jungwiert}, {Marino}, {M{\'a}rquez}, {Mast},
  {Mendoza}, {Moll{\'a}}, {Papaderos}, {Ruiz-Lara}, {van de Ven}, {Walcher}, \&
  {Wisotzki}}]{Sanchez-Blazquez14}
{S{\'a}nchez-Bl{\'a}zquez}, P., {Rosales-Ortega}, F.~F., {M{\'e}ndez-Abreu},
  J., {et~al.} 2014, \aap, 570, A6

\bibitem[{{Seabroke} \& {Gilmore}(2007)}]{SeabrokeGilmore2007}
{Seabroke}, G.~M., \& {Gilmore}, G. 2007, \mnras, 380, 1348

\bibitem[{{Seth} {et~al.}(2005){Seth}, {Dalcanton}, \& {de Jong}}]{Seth05a}
{Seth}, A.~C., {Dalcanton}, J.~J., \& {de Jong}, R.~S. 2005, \aj, 130, 1574

\bibitem[{{Simha} {et~al.}(2014){Simha}, {Weinberg}, {Conroy}, {Dave},
  {Fardal}, {Katz}, \& {Oppenheimer}}]{Simhaetal2014}
{Simha}, V., {Weinberg}, D.~H., {Conroy}, C., {et~al.} 2014, ArXiv e-prints,
  arXiv:1404.0402

\bibitem[{{Smee} {et~al.}(2013){Smee}, {Gunn}, {Uomoto}, {Roe}, {Schlegel},
  {Rockosi}, {Carr}, {Leger}, {Dawson}, {Olmstead}, {Brinkmann}, {Owen},
  {Barkhouser}, {Honscheid}, {Harding}, \& et~al.}]{Smee13}
{Smee}, S.~A., {Gunn}, J.~E., {Uomoto}, A., {et~al.} 2013, \aj, 146, 32

\bibitem[{{Spitzer} \& {Schwarzschild}(1951)}]{Spitzer51}
{Spitzer}, Jr., L., \& {Schwarzschild}, M. 1951, \apj, 114, 385

\bibitem[{{Spitzer} \& {Schwarzschild}(1953)}]{Spitzer53}
---. 1953, \apj, 118, 106

\bibitem[{{Statler}(1995)}]{Statler95}
{Statler}, T. 1995, \aj, 109, 1371

\bibitem[{{Str{\"o}mberg}(1925)}]{Stromberg1925}
{Str{\"o}mberg}, G. 1925, \apj, 61, 363

\bibitem[{{Tabor} {et~al.}(2017){Tabor}, {Merrifield}, {Arag{\'o}n-Salamanca},
  {Cappellari}, {Bamford}, \& {Johnston}}]{Tabor17}
{Tabor}, M., {Merrifield}, M., {Arag{\'o}n-Salamanca}, A., {et~al.} 2017,
  \mnras, 466, 2024

\bibitem[{{Toloba} {et~al.}(2011){Toloba}, {Boselli}, {Cenarro}, {Peletier},
  {Gorgas}, {Gil de Paz}, \& {Mu{\~n}oz-Mateos}}]{Toloba2011}
{Toloba}, E., {Boselli}, A., {Cenarro}, A.~J., {et~al.} 2011, \aap, 526, A114

\bibitem[{{Toth} \& {Ostriker}(1992)}]{TothOstriker1992}
{Toth}, G., \& {Ostriker}, J.~P. 1992, \apj, 389, 5

\bibitem[{{Valdes} {et~al.}(2004){Valdes}, {Gupta}, {Rose}, {Singh}, \&
  {Bell}}]{Valdesetal2004}
{Valdes}, F., {Gupta}, R., {Rose}, J.~A., {Singh}, H.~P., \& {Bell}, D.~J.
  2004, \apjs, 152, 251

\bibitem[{{Vazdekis} {et~al.}(2012){Vazdekis}, {Ricciardelli}, {Cenarro},
  {Rivero-Gonz{\'a}lez}, {D{\'{\i}}az-Garc{\'{\i}}a}, \&
  {Falc{\'o}n-Barroso}}]{Vazdekisetal2012}
{Vazdekis}, A., {Ricciardelli}, E., {Cenarro}, A.~J., {et~al.} 2012, \mnras,
  424, 157

\bibitem[{{Wake} {et~al.}(2017){Wake}, {Bundy}, {Diamond-Stanic}, {Yan},
  {Blanton}, {Bershady}, {S{\'a}nchez-Gallego}, {Drory}, {Jones}, {Kauffmann},
  {Law}, {Li}, {MacDonald}, {Masters}, {Thomas}, {Tinker}, {Weijmans}, \&
  {Brownstein}}]{Wake17}
{Wake}, D.~A., {Bundy}, K., {Diamond-Stanic}, A.~M., {et~al.} 2017, \aj, 154,
  86

\bibitem[{{Walker} {et~al.}(1996){Walker}, {Mihos}, \&
  {Hernquist}}]{Walkeretal1996}
{Walker}, I.~R., {Mihos}, J.~C., \& {Hernquist}, L. 1996, \apj, 460, 121

\bibitem[{{Weiner} {et~al.}(2006){Weiner}, {Willmer}, {Faber}, {Melbourne},
  {Kassin}, {Phillips}, {Harker}, {Metevier}, {Vogt}, \& {Koo}}]{Weiner06}
{Weiner}, B.~J., {Willmer}, C.~N.~A., {Faber}, S.~M., {et~al.} 2006, \apj, 653,
  1027

\bibitem[{{Westfall} {et~al.}(2011){Westfall}, {Bershady}, {Verheijen},
  {Andersen}, {Martinsson}, {Swaters}, \& {Schechtman-Rook}}]{Westfall11}
{Westfall}, K.~B., {Bershady}, M.~A., {Verheijen}, M. A.~W., {et~al.} 2011,
  \apj, 742, 18

\bibitem[{{Wielen}(1977)}]{Wielen77}
{Wielen}, R. 1977, \aap, 60, 263

\bibitem[{{Wisnioski} {et~al.}(2015){Wisnioski}, {F{\"o}rster Schreiber},
  {Wuyts}, {Wuyts}, {Bandara}, {Wilman}, {Genzel}, {Bender}, {Davies},
  {Fossati}, {Lang}, {Mendel}, {Beifiori}, {Brammer}, {Chan}, {Fabricius},
  {Fudamoto}, {Kulkarni}, {Kurk}, {Lutz}, {Nelson}, {Momcheva}, {Rosario},
  {Saglia}, {Seitz}, {Tacconi}, \& {van Dokkum}}]{Wisnioski15}
{Wisnioski}, E., {F{\"o}rster Schreiber}, N.~M., {Wuyts}, S., {et~al.} 2015,
  \apj, 799, 209

\bibitem[{{Yan} {et~al.}(2016{\natexlab{a}}){Yan}, {Bundy}, {Law}, {Bershady},
  {Andrews}, {Cherinka}, {Diamond-Stanic}, {Drory}, {MacDonald},
  {S{\'a}nchez-Gallego}, {Thomas}, {Wake}, {Weijmans}, {Westfall}, {Zhang},
  {Arag{\'o}n-Salamanca}, {Belfiore}, {Bizyaev}, {Blanc}, {Blanton},
  {Brownstein}, {Cappellari}, {D'Souza}, {Emsellem}, {Fu}, {Gaulme}, {Graham},
  {Goddard}, {Gunn}, {Harding}, {Jones}, {Kinemuchi}, {Li}, {Li}, {Maiolino},
  {Mao}, {Maraston}, {Masters}, {Merrifield}, {Oravetz}, {Pan}, {Parejko},
  {Sanchez}, {Schlegel}, {Simmons}, {Thanjavur}, {Tinker}, {Tremonti}, {van den
  Bosch}, \& {Zheng}}]{Yan16b}
{Yan}, R., {Bundy}, K., {Law}, D.~R., {et~al.} 2016{\natexlab{a}}, \aj, 152,
  197

\bibitem[{{Yan} {et~al.}(2016{\natexlab{b}}){Yan}, {Tremonti}, {Bershady},
  {Law}, {Schlegel}, {Bundy}, {Drory}, {MacDonald}, {Bizyaev}, {Blanc},
  {Blanton}, {Cherinka}, {Eigenbrot}, {Gunn}, {Harding}, {Hogg},
  {S{\'a}nchez-Gallego}, {S{\'a}nchez}, {Wake}, {Weijmans}, {Xiao}, \&
  {Zhang}}]{Yan16a}
{Yan}, R., {Tremonti}, C., {Bershady}, M.~A., {et~al.} 2016{\natexlab{b}}, \aj,
  151, 8

\bibitem[{{Yan} {et~al.}(2019){Yan}, {Chen}, {Lazarz}, {Bizyaev}, {Maraston},
  {Stringfellow}, {McCarthy}, {Meneses-Goytia}, {Law}, {Thomas}, {Falcon
  Barroso}, {S{\'a}nchez-Gallego}, {Schlafly}, {Zheng}, {Argudo-Fern{\'a}ndez},
  {Beaton}, {Beers}, {Bershady}, {Blanton}, {Brownstein}, {Bundy}, {Chambers},
  {Cherinka}, {De Lee}, {Drory}, {Galbany}, {Holtzman}, {Imig}, {Kaiser},
  {Kinemuchi}, {Liu}, {Luo}, {Magnier}, {Majewski}, {Nair}, {Oravetz},
  {Oravetz}, {Pan}, {Sobeck}, {Stassun}, {Talbot}, {Tremonti}, {Waters},
  {Weijmans}, {Wilhelm}, {Zasowski}, {Zhao}, \& {Zhao}}]{Yan19}
{Yan}, R., {Chen}, Y., {Lazarz}, D., {et~al.} 2019, \apj, 883, 175

\bibitem[{{Zheng} {et~al.}(2017){Zheng}, {Wang}, {Ge}, {Mao}, {Li}, {Li}, {Mo},
  {Goddard}, {Bundy}, {Li}, {Nair}, {Lin}, {Long}, {Riffel}, {Thomas},
  {Masters}, {Bizyaev}, {Brownstein}, {Zhang}, {Law}, {Drory}, {Roman Lopes},
  \& {Malanushenko}}]{Zheng2017}
{Zheng}, Z., {Wang}, H., {Ge}, J., {et~al.} 2017, \mnras, 465, 4572

\end{thebibliography}

\end{document}